\documentclass[a4paper, 11pt, twoside]{report}

\usepackage[utf8]{inputenc}
\usepackage[english.british]{babel}
\usepackage[autostyle,autopunct,english=british]{csquotes}
\usepackage{natbib}
\bibpunct{(}{)}{;}{a}{}{,} 
\usepackage{url}
\usepackage[doublespacing]{setspace}
\usepackage{fancyhdr}
\usepackage{graphicx}
\usepackage{makeidx}
\usepackage{amsmath}
\usepackage{subfigure}
\usepackage[tighter]{newpxtext}
\usepackage[upint]{newpxmath}
\usepackage[cal=pxtx,calscaled=1.0,scr=pxtx,scrscaled=1.0,frak=esstix,frakscaled=0.98,bb=txof,bbscaled=1.05]{mathalfa}
\usepackage{lscape} 
\usepackage{longtable}
\usepackage{units}
\usepackage{float}
\usepackage{rotating}
\usepackage{tabularx}
\usepackage{nicefrac}
\usepackage{parskip}
\usepackage{afterpage}
\usepackage[pdftex]{hyperref}
\usepackage{amssymb}
\usepackage[leftcaption]{sidecap}
\usepackage[hypcap, footnotesize, bf]{caption}
\usepackage[usenames,dvipsnames]{color}
\usepackage[usenames,dvipsnames,svgnames,table]{xcolor}
\usepackage{bm}
\usepackage{multirow}
\usepackage{tablefootnote}
\usepackage[hmarginratio=3:2]{geometry}
\usepackage{epigraph}
\usepackage{etoolbox}
\usepackage[toc,page]{appendix}
\usepackage{enumitem}

\newcommand{\sun}{\ensuremath{\odot}}

\DeclareUnicodeCharacter{0301}{ }

\makeatletter
\patchcmd{\epigraph}{\@epitext{#1}}{\itshape\@epitext{#1}}{}{} 
\makeatother
\hypersetup{
 colorlinks=true,
 citecolor=RoyalBlue,
 linkcolor=RoyalBlue,
 urlcolor=RoyalBlue,
 pdftitle = {Title},
 pdfauthor = {{Aaron Labdon}}
}

\fancypagestyle{plain}{%
  \fancyhf{}
  \fancyhf[HLE,HRO]{\hfill\slshape\thepage}}

\makeindex

\newcommand{\signed}{
   \begin{flushright}
   \begin{tabular}{rl}
   \\
   \\
   Signed: & \makebox[2in]{\dotfill}\\
           & {Aaron Labdon}\\
   \\
   Date:   & \makebox[1.2in]{\dotfill}\\
   \end{tabular}
   \end{flushright}
}

\newcommand{\statement}{

   \begin{quote}\small

   Submitted by Aaron Labdon to the University of Exeter as a thesis for
   the degree of Doctor of Philosophy in Physics, June, 2021. 

   This thesis is available for Library use on the understanding that it is
   copyright material and that no quotation from the thesis may be published
   without proper acknowledgement. 

   I certify that all material in this thesis which is not my own work has been
   identified and that no material has previously been submitted and approved
   for the award of a degree by this or any other University.

   \end{quote}
   \signed
}
   
\author{{Aaron Labdon} \\ \vspace{0.5in}}
\title{The Inner Astronomical Unit of Protoplanetary Disks}
\date{\statement}

\begin{document}
   \pagenumbering{alph} 
   \maketitle
   \pagenumbering{roman}
   \setcounter{page}{1}

\begin{abstract}

1st Supervisor: Prof. Stefan Kraus \qquad 2nd Supervisor: Prof. Tim Harries \\

A golden age of interferometry is upon us, allowing observations of smaller scales in greater detail than ever before. In few fields has this had the huge impact as that of planet formation and the study of young stars. State of the art high angular resolution observations provide invaluable insights into a host of physical processed from accretion and sublimation, to disk winds and other outflows.

In this thesis, I present the wide-ranging works of my PhD, encompassing both instrumentation and observational science. Instrumentational activities stem from the development of new generation baseline solutions at CHARA to the commissioning of a new observing mode on MIRC-X, allowing for the first ever J band interferometric observations of a young stellar object ever published. The science results find direct evidence of a dusty wind emanating from the innermost regions of the young object SU\,Aurigae in addition to exquisite image reconstruction revealing inclination induced asymmetries. Additionally, I find evidence of viscous heating of the inner disk of outbursting star FU\,Orionis as I derive the temperature gradient to unparalleled precision.

While it is difficult to draw one overall conclusion from the varied works of this thesis, the results described here are a testament to the uniqueness of young stellar systems and provide vital information on some the most ubiquitous processes in astrophysics. The instrumentational developments also open up exciting opportunities for future science in the ever-growing field of optical interferometry.

Copyright 2021-2023 Aaron Labdon.

\end{abstract}

   {\hypersetup{
   linkcolor=Black} 
   \tableofcontents
   \listoffigures
   \listoftables
   }

\chapter*{}

\section*{\center Declaration}

All of the work presented in this thesis was undertaken as a candidate for a research degree at the University of Exeter. Contributions by other authors have been clearly indicated when used. All original authors of illustrations have been credited. Where the work of other authors is quoted, the source has been attributed. Aside from these external contributions, this thesis is entirely my own original work.

Chapter 5 contains work published in \citet{Labdon2020b}. Chapter 6 contains work published in \citet{Labdon19}. Chapter 7 contains work soon to be published in a paper with myself as first author. Chapter 8 contains work published in \citet{Labdon20}.
   

\chapter*{Acknowledgements} \vspace{-0.75cm}
There are many people who I could thank for contributing in some way for their contributions either to the science or my mental health. Firstly, to my wonderful supervisor Stefan Kraus for his mentorship, infinite knowledge and advice, thank you for taking a chance in hiring me four years ago. Also, to the rest of the Exeter team, Claire Davies, Alex Kreplin, Narsireddy Anugu and Jacques Kluska without whom I definitely would not have made it to this stage, and to Ed Hone and Seb Zarrilli for walking the road to a PhD with me. Additional gratitude to John Monnier and Jean-Baptiste Le Bouquin for assisting me with the best technical advice I could ask for as an instrumentation newbie.  

A large thanks to my parents and grandparents who have always supported every decision I have made from almost failing to get onto an undergraduate degree program to getting the dream fellowship position. 

Finally, I will be eternally grateful to Andrew, Dan, Ethan, Ilona, Matt, Rae, and Tom whose insanity has helped keep me sane while attempting to write this thesis during a global pandemic. Without them to unwind with I would have lost my mind a while ago. May we slay more dragons, climb more mountains, sail more seas, and ride off into more sunsets in the years to come. 

\begin{flushright}
{ Aaron Labdon \\ Exeter, U.K. \\ 15$^{\rm th}$ June 2021}
\end{flushright}

   \setcounter{page}{1}
   \pagenumbering{arabic}
   
\chapter{Introduction}
\label{ch:Intro}
\epigraph{``
And all this science, I don't understand, It's just my job five days a week''}{--- Elton John, Rocketman\hphantom{spacing}} 

The story of protoplanetary systems is a story shared with ourselves. How did the Earth form? How did life evolve here? Is our story unique? To answer these questions, we must look to the stars find other such systems and other such worlds in order to capture snapshots of their formation. 

Protoplanetary disks lie at the junction between large scale star formation and planetary science. They consist of material left over from star formation, that will eventually form planets or be dissipated by stellar winds and accretion. The stages of planet formation are not well understood, with most theories failing to account for one or more selections of the now extensive known exoplanet catalogue. The only way to improve our understanding is to observe protoplanetary disks at all stages of their evolution in order to piece together the processes governing these objects.

However, observations of protoplanetary disks are not a simple undertaking as even the nearest star forming regions are still $\sim140\,\mathrm{pc}$ away. As such the angular size of these disks on sky is very small indeed. It is only in the past few decades that our instrumental capabilities have advanced far enough to observe such objects. Large 8-10\,m class telescopes such as Keck, VLT and Gemini combined with advanced adaptive optics systems have allowed us spatially resolve the delicate structures of the outer disks. At the same time the progression of interferometric techniques both in the sub-mm with ALMA and at optical wavelengths at VLTI and CHARA has allowed us to observe even the smallest scales with sub-milli-arcsecond resolution and unprecedented spectral capabilities. 

In this thesis I present the results of my PhD exploring the inner astronomical unit of protoplanetary disks. The aim of the project was originally to search for signatures of planet formation within optical interferometric data, such as gaps, rings and asymmetries. Although no direct signs of planet formation were found, several phenomena were observed and applied for the first in my work which provide a vital piece of the puzzle in the mystery that is how we can exist upon planet Earth. 

In chapter 2 I provide a detailed introduction to protoplanetary disks, from their humble beginnings in giant molecular clouds through to star, disk, and later planet formation. I outline some of the key processes taking place in protoplanetary disks from a theoretical perspective as well as the later stage evolution of disks. In chapter 3 I approach protoplanetary disks from an observational perspective, from the first studies of spectral energy distributions to the latest high angular resolution observations from cutting edge instruments. I also provide an overview of some of the key features observed and their potential connections to planet formation mechanisms. Chapter 4 serves as the final introductory chapter, providing a description of optical interferometry including its history and development into the modern artform used today. I explore the science behind the key observables used and their analysis from basic modelling to full image reconstruction techniques. 

Chapter 5 focuses on the instrumentation work undertaken during my PhD. This work is separated into two parts, the first focusing on the development of a new baseline solution for the CHARA array. In this endeavour I refined the model which predicts the location of the delay lines for a given stars position on sky in order to allow easier and more efficient observations. I succeeded in cutting the amount of time spent searching for fringes dramatically making it much easier to observe faint and low fringe contrast objects. The second instrumentation effort concerns the expansion of the operating wavelength of the MIRC-X instrument to include the J band. In this work I modelled and correct for both atmospheric chromatic dispersion and internal instrument birefringence, in addition to developing the data reduction and calibration pipeline and conducting the first science observations in this new dual JH band mode.

Chapter 6 presents  observations of the young star SU\,Aurigae using the CLIMB and CLASSIC instruments at the CHARA array in combination with archival KI and PTI observations. I present the first reconstructed image of the inner disk of SU\,Aur that confirms that the object is seen under high inclination and that we detect emission from near the dust sublimation region. In addition, I present radiative transfer modelling of the inner disk and curved rim geometry of the sublimation front. I find that a dusty disk wind is required to model a combination of interferometric and photometric observations. In chapter 7, I present follow-up observations of SU\,Aurigae from CHARA/MIRC-X in the H band which provide higher resolution and better uv coverage over a shorter observational timescale. These new observations confirm the presence of a dusty disk wind and provide an updated image clearly highlighting the strong, inclination induced, asymmetries in the disk.  

In chapter 8 I present the results of the first science observations in the MIRC-X JH band observing mode of the outbursting young star FU\,Orionis. Using the multi-wavelength observations, I derive the temperature gradient across the inner disk and find it consistent with theoretical models of a viscously heated accretion disk in contrast to the standard model of a disk heated only by reprocessed stellar radiation. This represents one of the first studies to find observation evidence for viscous heating within an accretion disk, the presence of which is a hotly debated topic. 

Finally, in chapter 9 I present some concluding remarks, including a summary of the scientific results and some of the key skills and experience learned during the completion of my PhD. I finish by providing some context on the planned future work, building upon some this work. \\

\chapter{Theory of Protoplanetary Disks}
\label{ch:TheoryPD}
\epigraph{``You're a shining star. No matter who you are. Shining bright to see. What you could truly be''}{--- Earth, Wind and Fire, Shining Star\hphantom{spacing}} 

Our understanding of protoplanetary disks and their roles in the formation of planets like our own has evolved dramatically over the past few decades, in a large part thanks to a wealth of observational data. In this chapter I will discuss the formation and evolution of protoplanetary disks, including the formation of planets and solar systems. I will then go into greater detail regarding some of the most fundamental processes at work in these disks and how our understanding has evolved over time.

\section{Star Formation \& Protoplanetary Disks}

    \subsection{Star Formation} \label{sec:starformation}

    Our inherent understanding of protoplanetary disks is based in our comprehension of their formation and its precursor, star formation. It is therefore imperative that the narrative begins with the story of how stars are born within giant molecular clouds. These interstellar clouds often many parsecs across are predominantly formed of molecular hydrogen gas, making them difficult to detect despite their size. However, small amounts of other compounds such as carbon monoxide (CO) and ammonia ($\mathrm{NH_3}$) make the mapping of these stellar nurseries possible through their emission/absorption features. Clouds are astronomical in size ranging from 10s to 100s of parsecs containing up to several million solar masses of material \citep{Murray11}. Such large clouds are not uniform entities, instead their structures are complex comprising of filaments, sheets, bubbles and irregular clumps. Filaments are large dense substructures which are ubiquitous across molecular clouds; it is primarily from these structures which the fragmentation and contraction of star formation occurs. This is evidenced in the fact that pre-stellar cores and young stellar clusters are preferentially located within filaments \citep{Schmalzl10,Konyves15}.

    The causes of the initial fragmentation of filaments is not well understood but is likely thought to be due to linear perturbations or environmental factors such as turbulence or magnetic fields \citep{Chira18}. As a filament collapses it will break into gravitationally bound cores, where if gravitational forces are sufficient, the cloud can begin to collapse. The mass above which a molecular cloud will begin to collapse is known as the Jeans mass \citep{Jeans02}, which is typically of the order of tens of thousands of solar masses. Jeans mass is reached when the internal gas pressure is no longer sufficient to maintain equilibrium with the gravitational force and is given by
    \begin{equation}\label{eq:jeansmass}
        M_J = \frac{\pi}{6} \frac{c_s^3} { G^{\frac{3}{2}} \rho^{\frac{1} {2} } } ,
    \end{equation}
    where $M_J$ is the Jeans mass, $c_s$ is the gas sound speed, $G$ is the gravitational constant and $\rho$ is the density of the cloud. For a cloud of $1000\,\mathrm{M_\odot}$ and a temperature of 20\,K, the critical density required to trigger the collapse of the cloud is $1\times10^{-25}\,\mathrm{gcm^{-3}}$. The collapse of the cloud and subsequent pre-stellar cores will continue until the internal pressure is once again sufficient to maintain equilibrium with gravity. Eventually, the density of material is such that the cores become opaque and less efficient at radiating away energy, leading to an increase in temperature, creating an object known as a stellar embryo.

    A stellar embryo will continue to contract, further raising the temperature. Eventually the gas will become hot enough for the internal pressure to support itself against the gravitational collapse. Once this hydrostatic equilibrium has been reached, the resulting object is known as the first hydrostatic core (FHSC), a molecular hydrogen object \citep{Larson69}. When the temperature of the FHSC reaches around $2000\,\mathrm{K}$ the molecular hydrogen will dislocate and later ionise creating a true protostar \citep{Hayashi66,Larson03}. Depending on the mass of material in a molecular core it can take between 1-10\,Myr for a protostar to form, while the whole protostellar phase only lasts around 500,000\,yr \citep{Dunham14}. The higher the mass of an object the shorter its lifespan at any stage of evolution.

    The small scale primordial motions of the original molecular cloud are preserved and enhanced by the conservation of angular momentum. Angular momentum can be simply calculated through 
    \begin{equation}\label{eq:angularmomentum}
        L = r^2m\omega ,
    \end{equation}
    where $L$ is angular momentum, $r$ is radius, $m$ is mass and $\omega$ is the angular velocity. If one assumes that no mass is lost, it is trivial that the same mass concentrated within a smaller radius, must have a higher angular velocity in order to conserve the angular momentum. In reality much mass and hence angular momentum is lost during the star formation phase, this is known as the spin-up of a protostar. It is from this foundation of a rotating protostar and surrounding envelope that protoplanetary disks will begin to form.

    \subsection{Formation of Protoplanetary Disks} \label{sec:diskformation}

\begin{figure*}[h!]
        \centering
        \includegraphics[scale=0.6]{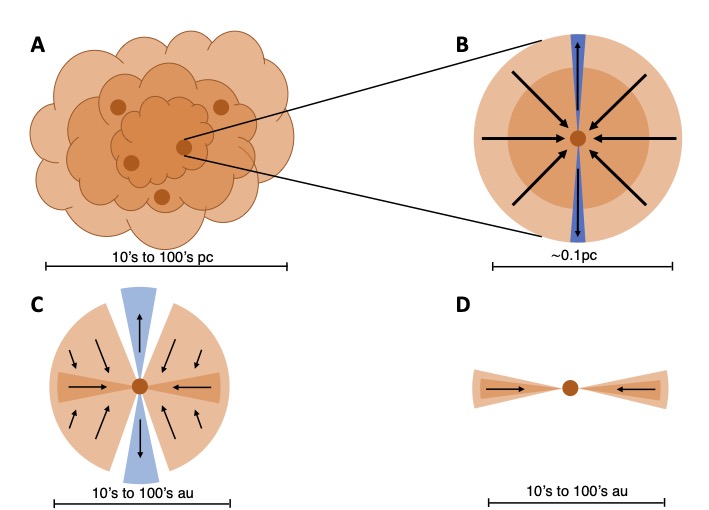}
        \caption[Schematic of disk formation from molecular clouds.]{Schematic of disk formation from molecular clouds. \textbf{A.} Large molecular cloud of uneven density distribution which will lead to the fragmentation and collapse of individual protostellar cores. \textbf{B.} A punch in of a single protostellar core, the surrounding material collapses under its own gravitational pull. Jets of material are ejected along the rotational poles of the protostar. \textbf{C.} Accretion and centripetal forces draw envelope material towards the midplane and towards the central star, flattening the envelope into a thin disk. \textbf{D.} The remaining thin disk will slowly accrete onto the central star. Diagrams are not to scale.}
        \label{fig:DiskFormToon}
    \end{figure*}

    A large amount of the original material of the molecular cloud is located within the newly formed (class 0) protostars. However, a majority of material remains within a spherical envelope surrounding the star. This material will continue its infall onto the central star in a process known as accretion (see Section\,\ref{sec:accretion} for a detailed discussion). However, in the radial direction around the protostellar equator centripetal acceleration caused by the rapid rotation is able to resist the gravitational pull of the star. In this way material in a thin radial disk will be preserved while material outside of this plane is free to collapse onto the star. This processes of disk formation is shown in Figure\,\ref{fig:DiskFormToon}.  

    This is the most simplistic disk structure, a thin disk of dust and gas aligned with the rotation of the star, which is slowly accreting onto the central object. This stage of evolution typically lasts 3-5\,Myr, with some very large disks surviving as long as 10\,Myr \citep{Russell06}. The composition of the disk is thought to be around 99\% gas and 1\% solid dusty material, a general assumption equal to the composition of the interstellar medium \citep{Bohlin78}. The evolution of disks beyond their formation is geared towards planetary formation, hence the name protoplanetary disks, the fundamentals of planet formations within disks is described in detail in the following section.

    At this stage it is useful to define the nomenclature used when referring to your star-disk systems. In the earliest stages of evolution, young systems are referred to by a class structure based on shape of their spectral energy distribution. This is described more in Section\,3.2.1. Once young stars reach a more evolved stage, the overarching term is a young stellar object or YSO which denotes a star in its early stage of evolution including protostars and pre-main-sequence stars. In turn this has several subclasses defined by the stars mass. T\,Tauri stars are the lowest mass objects, stars below $2\,\mathrm{M_\odot}$ typically fall into this class and are typically identified by their optical variability and strong chromospheric lines \citep{Joy45}. More massive objects are known as Herbig Ae/Be stars and typically exist in the $2-8\,\mathrm{M_\odot}$ mass range and are of A or B spectral type. YSOs that are significantly larger than this are simply known as massive YSOs, with only a few examples known, as their high mass means they evolve very quickly onto the main sequence. Each of these classes have a set of conditions which objects must meet in order to be included, however in reality the boundaries and definitions are often blurred such that classifications are little more than mass estimates.

\section{Planet Formation Process} \label{sec:planetformation}

    The better understanding of planet formation processes is perhaps one of the ultimate aims in the study of protoplanetary disks. The insights gained allow us to better understand the formations of our own solar system and our own planet Earth. Planet formation is thought to occur a few million years following star formation as material settles into a denser disk \citep{Briceno01,Bodenheimer02}. This timescale is constrained as material in the disk will be dissipated by outflows and accretion events over time. When it comes to the creation of planets in a disk there are two competing theories known as the disk instability and core accretion scenarios. Both pathways produce planets in significantly different ways and as such of significantly different types. However, neither can fully reproduce the observed zoo of planets and exoplanets discovered in the universe to date \citep{Matsuo07}. The true story of planet formation is likely more complex and possibly a combination of different mechanisms depending on disk masses, solar mass and type and even stellar multiplicity.

    \subsection{Core Accretion Scenario} \label{sec:CoreAccretion}

    The core accretion method of planet formation is the slow building of planets from individual dust grains mere nano-meters across to planets many thousands of kilometers across \citep{Pollack96}. Within a protoplanetary disk dust grains are constantly in motion and so collisions between grains will be relatively common. Upon colliding small grains may stick together electrostatically through the Van der Waals force to create slightly larger grains. Van der Waals force is a quantum effect caused by correlations in the fluctuating polarizations of nearby particles. This process however is relatively inefficient as colliding larger grains can also destroy one another rather than stick together \citep{Johansen14}. Even, so this initial process occurs quite early on in the planet formation process. Once a planetesimal has grown to sufficient size, it will begin to gravitationally attract material in the surrounding area, allowing it to grow rapidly though run-away accretion \citep{Rafikov03,Schlichting11}. Through the collision and accretion of grains, pebbles, and clumps \citep{Lambrechts12,Lambrechts14}, these planetesimals will grow at a steady rate of a few centimeters per year over a few million years \citep{Goldreich73}. Pebble accretion is thought to occur when small bodies experience aerodynamic drag from the gaseous disk; this reduction in velocity increases the likelihood of accretion onto nearby planetesimals \citep{Lambrechts14b}. Eventually, the core will be able to accrete non-solid material in the form of an atmosphere. The accretion of gas onto a solid core will cool the gas, causing the loss of much of the initial entropy, leading to the creation of a planets with very low initial entropy. For this reason, core accretion is sometimes known as the term cold start scenario. 

    The location of these planetesimals within the disk will determine their composition \citep{Brown50}. For example, the inner regions are too warm for volatile materials such as water and methane to condense out of their gaseous forms, hence these cores are made up of materials with higher melting points such as iron, nickel and silicates \citep{Urey51}. These heavier elements are significantly rarer in protoplanetary disks with typical ice/rock ratios, whcih in theoretical works are around around 2 \citep{Lodders03}. As such, there exists a limit to the size these inner planets can grow to, whereas planets forming further out in the disk will have access to more volatile compounds, which are in relative abundance, allowing for much larger planets to form. These larger planets can reach sufficient sizes that accretion of hydrogen and helium can occur, leading to large atmospheres such as those of Jupiter and Saturn.

    The cold start planet formation process alone does not accurately predict the positioning of the planets within our own solar system, or in many observed exoplanet populations. In order to correct this, planetary migration scenarios are proposed whereby planets change their orbits over periods of time. Perhaps the most famous scenario for describing the solar system is the 'Nice' model \citep{Gomes05,Tsiganis05,Morbidelli05}. In this model, planetary migration is driven by torques from the scattering of smaller bodies as giant planets enter resonant orbits, in particular Jupiter and Saturn. Such a model requires a wide range of vary specific starting parameters in order to recreate the known solar system, but still remains a valid theory for the movement of planets. Another model of planetary migration is the 'Grand Tack' hypothesis \citep{Walsh11}, whereby Jupiter formed at 3.5\,au before migrating inwards to 1.5\,au and then outwards to its current 5.2\,au. Such a model solves the issues with Mars being smaller than its expected mass \citep{Wetherill91} and the location and composition of the asteroid belt.

    Such migrations have also been invoked in order to understand the distribution of exoplanet populations. Hot-Jupiter planets are very large with thick gaseous atmospheres but exists very close to their host stars in a regions it would be impossible for them to have formed \citep{Wang15}. Additionally, many wide separation exoplanets have been imaged \citep{Marois08,Nguyen21}, often at separations of many 100's of au from their central star. Such planets could not form through core-accretion pathways meaning these planets either migrated to their current positions or formed through a different mechanism. 

    \subsection{Gravitational Instability Scenario}

    The disk instability scenario is an alternative planet formation model to describe the formation of giant planets on wide orbits in protoplanetary disks \citep{Kuiper51,Cameron78,Boss97}. It relies om the fact that a disk of sufficient mass will be inherently gravitationally unstable. However, a gravitationally unstable disk will not always collapse, the outcome is dependent on the cooling timescale of the disk \citep{Kratter16}. A long period cooling balances with heating caused by the dissipation of gravitational turbulence. Whereas for a short period cooling timescale the heat will be dissapated, without this extra pressure, the disk can fragment and collapse. 


    These instabilities lead to the fragmentation of the outer disk into a series of large clumps, in a similar way to the fragmentation of molecular clouds in star formation processes as described in Section\,\ref{sec:diskformation}. These clumps will then collapse directly, forming a giant planet, typically of the order $\sim5\,\mathrm{M_J}$ \citep{Kratter10}. In this way most of the initial entropy of the material is conserved hence the name hot start. 

    Disk instability planet formation cannot account for the formation of all planets, such as the inner planets of our solar system \citep{Durisen07}. However, hot start planets will be larger in radius, hotter and have higher entropy while young than those formed by core accretion. This provides observational constraints for which to search for potential young hot start planets, and several candidates have been found to date \citep{Sigurdsson03,Forrest04,Butler04}. One of the key limitations is that for a disk to be gravitationally unstable it must be very massive, indeed more massive than the majority of observed disks to date. Only a handful of disks massive enough have been found around the most massive YSOs. 

\section{Basic Disk Properties}

    Armed with a basic understanding of star, disk, and planet formation processes one can begin to explore some of the fundamental disk physics at play and consider what the observational effects of this might be. In this section, I will discuss some the key physics of protoplanetary disks and their effects on planet formation and observational astronomy. For the purposes of consistency and simplicity, this section refers to intermediate mass stellar objects with full, flared disks which are no longer highly embedded such as T\,Tauri and Herbig Ae/Be objects.  

    \subsection{Flaring}

    Protoplanetary disks are not flat, but have a vertical height due to the internal gas pressure which pushes back against the gravitational collapse of the disk. The vertical height of the disk is therefore proportional to the gas density as a function of both radius and height. For an azimuthally symmetric disk which is in hydrostatic equilibrium, the density is 
    \begin{equation}\label{eq:scaleheightdensity}
        \rho(r,z) = \frac{\Sigma(r)}{\sqrt{2\pi}H(r)} \mathrm{exp}\left(-\frac{z^2}{2H(r)^2}\right) ,
    \end{equation}
    where $\Sigma(r)$ is the surface density and $H(r)$ is the scale height of the disk at radius $r$. As such the scale height is dependent on the relationship between the temperature and surface density profiles of the disk. However, the temperature is controlled by the amount of stellar radiation impacting the disk and being absorbed. It is therefore dependent on the geometry of the disk, in particular its scale height. This coupling of geometry, temperatures and densities makes deriving an analytical disk model exceedingly difficult. 

    One of the first propositions that protoplanetary disks were not flat or even limited to a fixed opening angle was the work of \citet{Kenyon87}. \citet{Adams87} demonstrated that several YSOs with strong IR excesses could be modelled by a massive accretion disk, of which no other evidence was present. \citet{Kenyon87} were able to show that the same selection of objects could instead be modelled as a low-mass reprocessing disk which flares slightly as radial distance increases.

    A more modern and widely used model is given by \citet{Chiang97} was to approximate the disk as two layers in order to separate the flaring from the surface geometry. Their models consist of a thin, hot surface layer which is the only part of the disk directly heated by reprocessing stellar radiation. Around half of the heat from the layer is radiated outwards while the rest heats the inner layer of the disk, a cooler flared disk which is uniformly heated. This approximation allows for the derivation of an appropriate power law of $H \propto r^h$, where $h\approx1.3-1.5$ is found encompass most YSOs. This simple relation defines the range of physically valid flaring scenarios for protoplanetary disks. Iterative numerical solutions by \citet{DAlessio98} and \citet{Dullemond02} find similar results. 

    However, observations of spectral energy distributions found that this simple flared disk model overestimates the mid-infrared emission, which is thought to be due to the settling of dust grains relative to the gas height \citep{DAlessio99}.  

    \subsection{Vertical Settling}

    Within protoplanetary disks it is often easiest to think of dust and gas as being coupled. However, this is not always the case and the movement of dust grains can have a large impact on the disk structure despite only consisting of $\sim1\%$ of disk mass.

    One of the predominant motions of dust grains is the movement of larger grains from the surface layer of the disk towards the midplane. A dust grain in the surface layer of a disk will try to follow an inclined Keplerian orbit, moving in and out of the midplane as it does so. This is in contrast to the gas which is supported against gravity by its own pressure. As a consequence of this, the dust grain will experience a very strong drag force from the gas \citep{Weidenschilling77,Garaud04}. The magnitude of this effect depends on the size of the grain as a small object ($\sim0.1\,\mathrm{\mu m}$) will have a very large surface area to mass ratio and so will follow the gas motions. Whereas a large grain with a small surface area to mass ratio will not follow gas motions and will settle towards the midplane \citep{Dullemond04b}. 

    The effect of vertical settling is strongly coupled to that of grain growth. As described in section\,\ref{sec:CoreAccretion} grains will coagulate over time through the van der Waals forces and grow in size, as the grains grow they settle more towards the midplane. This increases the density of dust at the midplane which accelerates grain growth causing even deeper settling in a repetitive cycle. It is only though turbulent mixing that disks are not perfectly stratified \citep{Dullemond05}. 

    The removal of large grains from the surface layers of the disk decreases the scattering efficiency of this region, reducing the observed disk flux. This has been shown to more accurately account for observations across the NIR than non-settled disk models \citep{DAlessio99}.

    \subsection{Sublimation}

    Sublimation is simply the transition of a substances from  a solid to gaseous state without passing through a liquid state. It occurs where pressures are sufficiently low enough that the substance cannot exist as a liquid, so an increase in temperature allows for the direct transition to a gaseous state. Within protoplanetary disks this process occurs in the innermost regions where temperatures are sufficient to sublimate the dust grains in the disk. This gives a minimum inner radius at which a dusty disk can exist, depending on the dust grain composition \citep{Natta01,Dullemond01}. Such a limit is not imposed on the gaseous disk component, which is instead truncated through magnetospheric or tidal interactions. 

    Dust grains are commonly modelled as graphite or astronomical silicates (such as olivine) \citep{Mathis77,Draine84}. At one atmosphere of pressure the sublimation temperature of graphite is around $4000\,\mathrm{K}$, this is much lower in the low pressures of protoplanetary disks and is typically found to be in the range of $1200-2000\,\mathrm{K}$. This is further complicated by grain sizes and local gas density dependencies. Changing the dust grain size has the effect of changing the fundamental optical properties of the grain. In this way, the scattering efficiency will change, a larger grain is more efficient at scattering incident radiation and can cool more efficiently, allowing them to survive in higher temperature regions. The gas density dependency is more complicated. It can be understood by considering that sublimation is the process of breaking the equilibrium between gas pressure and surface tension of the grains. The surface tension of a grain attempts to maintain the solid form, while the gas pressure attempts to pull it apart, sublimating the grain. The higher the local gas density the more energy is required sublimate the grains as the gas pressure is lower. The sublimation temperature of common disk compounds such as graphites and silicates as a function of gas density within protoplanetary disks is given by \citet{Pollack94} where they find the sublimation temperature varies as a rough power law
    \begin{equation}\label{eq:sublimation}
        T_{\mathrm{sub}} = G\rho^\gamma_g(r,z) ,
    \end{equation}
    where $T_{sub}$ is the sublimation temperature, $G=2000$, $\gamma = 1.95\times10^{-2}$ and $r$ and $z$ are the radial and vertical distance through a disk respectively \citep{Isella05}. The treatment of sublimation physics within disks can dramatically affect the shape and scale of the inner regions, observational consequences of this are explored in Section\,3.4.1

\section{Viscosity and Accretion Processes} \label{sec:accretion}

    Viscosity and accretion are perhaps the two most ubiquitous processes in astrophysics and can be found on all scales from supermassive blackholes to circum-planetary disks. Our understanding of these processes is limited as theoretical and observational works do not always agree, but over the past few decades we have still made great strides to further our comprehension of these most fundamental of effects. 

    \subsection{Viscosity and Viscous Heating}

    Viscosity is simply the measure of the internal friction that is present between adjacent layers of fluid undergoing relative motion. Due to large amounts of shear in disk structures, viscosity should be an important factor. Indeed, this is the case as material at a greater radial distance will lag behind in terms of rotational velocity \citep{Lynden74,Frank02}. Assuming a Keplerian system, the radial velocity can be defined as
    \begin{equation}\label{eq:Viscous}
      \upsilon_r = -\frac{3}{r^{1/2}\Sigma}\frac{\delta}{\delta r} \left( \nu \Sigma r^{1/2} \right) ,
    \end{equation}
    where $\upsilon_r$ is the radial velocity at radius $r$, $\Sigma$ is the surface density of the disk and $\nu$ is the disk viscosity. It is important to note that this generalisation only applies to thin disks. \citet{Shakura73} showed that a thin disk implies Keplerian physics as the radial pressure gradient through the disk become unimportant, a thick disk would not be Keplerian. The radial dependence of rotational velocity causes a transfer of angular momentum from the fast-inner disk to the slower outer disk, this loss of angular momentum in the inner disk causes material to spiral inwards. This infall of material is known as accretion and is one of the most ubiquitous processes in astrophysics. It can be demonstrated that viscosity is proportional to the mass flow rate in the steady state solution
    \begin{equation}\label{eq:MdotViscous}
        \nu \Sigma = \frac{\dot{m}}{3\pi} \left[ 1-\left( \frac{r_\star}{r} \right) ^{1/2} \right] ,
    \end{equation}
    where $\dot{m}$ is the rate of mass flow in the disk measured in $\mathrm{M_\odot yr^{-1}}$. From this it follows that viscosity determines the rate of radial inflow (accretion), and hence gravitational potential energy release. It is common to describe viscosity as an $\alpha\mathrm{-viscosity}$ as prescribed by \citet{Shakura73} in the form
    \begin{equation}\label{eq:alphaviscosity}
        \nu = \alpha c_s H ,
    \end{equation}
    where $c_s$ is the local mean sound speed and $H$ is the scale height of the disk and $\alpha$ is a free parameter between zero (no accretion) and approximately one specifying the local rate at which angular momentum is transported. Later work has since constrained this value further to $10^{-2}$ to $10^{-3}$ on large scales in protoplanetary disks \citep{Hartmann98,Johansen14}. Gratifyingly, many of the properties of steady thin discs turn out to have rather weak dependences on $\alpha$ as there is much disagreement between theoretical and observational models as to the value of $\alpha$ by as much as an order of magnitude \citep{King07}. 

    From the nature of viscosity, it naturally follows that fluids can be viscously heated. Viscous heating represents the effect of an irreversible process by means of which the work done by a fluid on adjacent layers due to the action of shear forces is transformed into heat. Circumstellar disks as described by \citet{Adams87} are heated only by the reprocessing (absorption and emission) of stellar radiation, however viscous heating allows a disk to be heated internally which can dramatically change the temperature structure of the disk. A viscously heated disk will create a stronger temperature gradient across the inner regions where temperature falls off rapidly with radial distance.

    As mentioned viscosity arises from the relative motions between layers of a fluid, the causes of such relative motions has been the subject of much discussion in literature. One of the preeminent theories for the source of viscosity is that of magnetorotational instabilities (MRI) described by \citet{Balbus91} and is often called the Balbus–Hawley instability. In this scenario the disk material contains mobile electrical charges which are subject to the influence of the magnetic field \citep{King07}. This magnetised material will feel the effect of the Lorentz force in which charge carriers orbit magnetic field lines. If the fluid is in a state of differential rotation about a fixed origin, this Lorentz force can be surprisingly disruptive, even if the magnetic field is very weak. This holds true for disk geometries where the angular rotational velocity decreases with radial distance as shown in equation \ref{eq:Viscous}. This effect is inherently destabalising and results in the turbulent motions within the disk, creating the necessary conditions for viscous heating.

    Various other mechanisms have also been proposed as causes of viscous heating over the past few decades, including thermal convection \citep{Lin80,Ruden86}, shear instabilities \citep{Dubrulle93} and gravitational instabilities \citep{Tomley91,Laughlin94}. However, MRI remains the most plausible explanation for the presence of viscous heating. 

    \subsection{Fundamentals of Accretion}

\begin{figure*}[h!]
        \centering
        \includegraphics[scale=0.6]{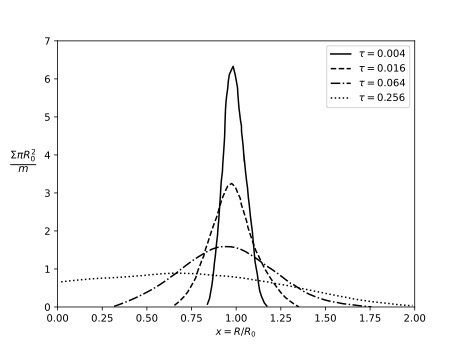}
        \caption[Effect of viscous spreading on a ring of material]{The evolution of a ring of mass $m$ and surface density $\Sigma$ as a function of the dimensionless radius $x=R/R_0$, where $R_0$ is the initial radius, and of dimensionless time $\tau = 12\nu t/R_0^2$. Recreated from \citet{Pringle81}. } 
        \label{fig:ViscousSpread}
    \end{figure*}

    As described in the previous section, the radial dependence of rotation velocity in combination with disk viscousity causes the transfer of momentum outwards through the disk. This transfer of angular momentum causes material to spread in a process known as viscous spreading. Considering an initial ring of material undergoing viscous spreading, most of the material will move inwards losing energy, but a tail of matter also moves outwards gaining angular momentum. This spread over time is demonstrated in Figure\,\ref{fig:ViscousSpread} where an initially thin ring is spread over a large radius with the majority of mass moving inwards. It is this movement of material known as accretion which is the one of the primary mass loss mechanisms in early disks. However, viscosity driven accretion through the disk is only the first stage, the process of moving material from the inner disk onto the central star is most often driven by the magnetic field of the central star.

\begin{figure*}[h!]
        \centering
        \includegraphics[scale=0.6]{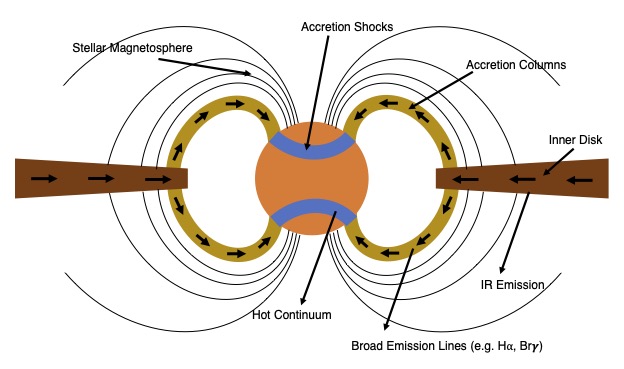}
        \caption[Schematic of magnetospheric accretion columns originating from the inner disk]{Schematic of magnetospheric accretion columns originating from the inner disk, travelling along stellar magnetic field lines and terminating at bright accretion shocks on the stellar photosphere. The origins of various emission types are also shown. Adapted from \citet{Hartmann16}.}
        \label{fig:MagnetoAccretion}
    \end{figure*}

    A young star with a sufficiently strong magnetic field will truncate the disk as the magnetic stress is sufficient to remove the excess angular momentum of the nearly Keplerian flow over a narrow transition zone. The radius of this truncation can be predicted by
    \begin{equation}\label{eq:amagnetictruncation}
        R_{\mathrm{trunc}} = \beta \mu_\star^{4/7} \left( 2GM_\star \right)^{-1/7} \dot{M}^{-2/7},
    \end{equation}
    assuming the gas is efficiently coupled to the magnetic field, where $R_{mathrm{trunc}}$ is the magnetic truncation radius, $\mu_\star$ is the stellar dipole moment and $\beta$ is a parameter $\leq1$ (with $\beta=1$ corresponding to the classical Alfvén radius for spherical accretion) \citep{Koenigl91}. $R_{\mathrm{trunc}}$ is usually confined to a few stellar radii for a moderate accretion rate of $10^{-7}$ to $10^{-9}\,\mathrm{M_\odot yr^{-1}}$. Interior to this radius disk material will flow along magnetic field lines out of the disk plane towards the dipoles of the star, giving rise to magnetospheric accretion columns \citep{Alencar12}. A schematic of magnetospheric accretion is shown in Figure\,\ref{fig:MagnetoAccretion}

    The infalling gas hits the star along two circular rings of (cylindrical) radius $R_{\mathrm{acc}} = \eta \left(R_\star/R_{\mathrm{trunc}}\right)^{-1/2}R_\star$ and width $\Delta_{\mathrm{acc}} \approx 0.5 \delta R_{\mathrm{acc}}$,  where the parameter $\eta$ accounts for the effect of screening currents in the disk \citep{Koenigl91}. (If the magnetic dipole axis is not aligned with the rotation axes of the disk and the star, then the accreted matter occupies only a portion of each ring.) This ring comprises of accretion shocks where the gas collides with the stellar surface at speed of $200 - 600\,\mathrm{kms^{-1}}$ \citep{Sacco10}. The strength of these shocks is such that material in accretion columns can be heated to temperature exceeding several MK \citep{Gullbring94,Lamzin98} allowing them to emit highly energetic radiation such as X-rays. 

    Magnetospheric column accretion is supported by observational evidence in the form of inverse P-Cygni profiles in the hydrogen Balmer lines \citep{Alencar12}. P-Cygni profiles occur when the same molecular line is seen in both absorption and emission at different redshifts. Material moving towards an observer will absorb radiation, creating a blue-shifted absorption trough while material moving away will only be seen through emission creating a red-shifted emission peak. However, an inverse P-Cygni profile denotes the opposite, red-shifted material is closer to the observer and so is seen in absorption while blue shifted material is only seen in emission. A standard P-Cygni profile describes typical ejection scenarios such as shell ejection, an inverse P-Cygni profile describes infalling material. By modelling the expected profiles of hydrogen lines it was found that magnetospheric accretion could recreate the observations \citep{Hartmann94,Muzerolle01}.

\section{Outflows and Mass Loss}
    
    Within disk systems material does not only move inwards, indeed a significant fraction of the mass and angular momentum loss is from the outward movement of material. This allows stars to reach their final masses and rotation rates, which are known to be significantly lower that the total mass and angular momentum contained with the initial molecular cloud core. 
    
    \subsection{Jets}

    One of the key aspects of the star formation process outlined in Figure\,\ref{fig:DiskFormToon} is that a fraction of the infalling material is ejected along the polar direction, perpendicular to the disk equatorial plane. These jets are thought to be launched centrifugally from the disk surfaces via the stresses of open magnetic field lines that thread the disk \citep{Blandford82}. The full range of the jet launching region is not fully constrained, but observations by \citet{Coffey04} and \citet{Coffey07} place the region between $0.2 - 0.5\,\mathrm{au}$ from the rotational axis. Launched material is superheated by its proximity to the central star and becomes a narrow jet of ionised gas. Once ejected these jets are highly collimated with an opening angle of only a few degrees and travelling at supersonic speeds in the range $150-400\,\mathrm{kms^{-1}}$ \citep{Cabrit07}. Such jets have been observed at every stage of the star formation process from very young protostars to more evolved transitional disks with very low accretion rates \citep{Rodriguez14}. 

    In dense star forming regions, protostellar jets will often collide with nearby clouds of dust and gas. The ionised nature of the jets and their fast velocities causes shock waves through the surrounding nebulosity emitting radiation. Such objects are known as Herbig-Haro (HH) objects after the first people to directly study these phenomenon \citep{Herbig50,Herbig51,Haro52,Haro53,Ambartsumian57}. HH objects represent the first observation and classification of jets from young stars owning to their bright nebulous emission.

    \subsection{Dusty Disk Winds} \label{sec:DiskWinds}

\begin{figure*}[h!]
        \centering
        \includegraphics[scale=0.6]{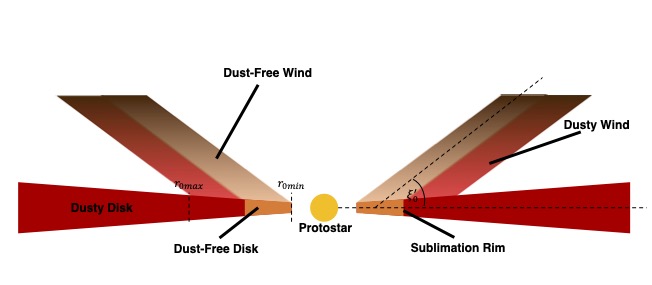}
        \caption[Schematic of a disk with associated dusty disk wind]{Schematic of a disk with associated dusty disk wind. The dusty disk/wind areas are shown in red, while the dusty free disk/wind areas are in orange. Disk winds below the disk have been omitted for clarity.}
        \label{fig:DustyDiskWind}
    \end{figure*}

    Magnetohydrodynamic disk winds describe the process by which material is removed from the disk surface along magnetic field lines \citep{Blandford82,Pudritz83}. This mechanism is based on the presence of a large-scale, ordered magnetic field which threads the disk. The field could originate from the interstellar field that permeates the molecular cloud and is dragged by in-falling gas into the disk. The magnetic field strength required to drive these winds is of the order of kGauss. Fields of this strength have been shown to be present in T~Tauri stars ($2.35\pm0.15\,\mathrm{kG}$ for T~Tau) by \citet{Guenther99} and \citet{Krull99}. In this model material is flung out along magnetic field lines highly inclined to the disk surface. The high magnetic pressure gradient above the disk surface accelerates the material which is then collimated through the azimuthal and poloidal field components \citep{Bans12}. These centrifugally driven winds are highly efficient at distributing density above the plane of the disk, carrying angular momentum away from the disk surface. Dust grains coupled to the gas will stream into the wind, the distribution of dust grains within the wind is not well characterised, but is likely dominated by smaller grains which are better coupled to the gas. \citet{Bans12} assume a simple single grain size of $1\,\mathrm{\mu m}$ for both disk and wind models in their radiative transfer work. It is for this reason that these winds are often referred to as dusty disk winds, schematic of the dusty disk wind is shown in Figure\,\ref{fig:DustyDiskWind}.

    A modern iteration of magnetospheric disk winds is given by \citet{Bans12}. In this scenario the wind is launched from the disk surface in a region between an inner and outer radius, located at a maximum of twice the sublimation radius. This is because FUV radiation that plays a key role (through photoevaporation) in the mass loading of gas into the wind at the inner launching region, at more distant regions the disk is shielded from FUV radiation by its own wind. The gas density structure is given as
    \begin{equation} \label{eq:2}
    \rho = \rho_1\bigg(\frac{r_0}{r_1}\bigg)^{-3/2} \eta(\chi),
    \end{equation}
    where $\rho$ is the density along a flow line at a distance $r_0$ from the symmetry axis, that is, along this disk surface. $\rho_1$ denotes the density at the fiducial radius of $r_1$, which is given as $1$\,au while $\chi$ is related to the cylindrical coordinate $z$ through
    \begin{equation}\label{eq:3}
       \chi = z/r
    \end{equation}
    $\eta(\chi)$ is obtained from the solution of the MHD wind equations. The into-wind mass transfer is linked to the density through
    \begin{multline}
     \rho = 1.064\times10^{-15}\Big(\frac{\dot{M}_{\mathrm{out}}}{10^{-7}M_\odot yr^{-1}}\Big)\Big(\frac{M_*}{0.5M_\odot}\Big)^{-1/2}  \\ \times \frac{1}{\mathrm{ln}(r_{\mathrm{0max}}/r_{\mathrm{0min}})\psi_0(1-h_0\xi_0^{'})} \mathrm{~g~cm^{-3}},
    \end{multline}
    where $\dot{M}_{\mathrm{out}}$ is the mass outflow rate, $M_*$ is the stellar mass, $h_0$ is the disk scale height and $\psi_0$ is the ratio of vertical speed to Keplerian speed at the disk's surface. For a full derivation of the equation and constants, see \citet{Blandford82} and \citep{Safier93}. $\xi_0^{'}$ is related to the angle $(\theta_0)$ at which the poloidal component of the magnetic field threads the disk, defining the opening angle of the disk wind \citep{Safier93}
    \begin{equation}
    \xi_0^{'} = \mathrm{tan}(\theta_0).
    \end{equation}
    This rate generally controls the magnitude of the NIR excess added by the dusty wind. It was this addition to the NIR excess which first inspired the derivation of this process. Many circumstellar disks exhibit a stronger NIR excess than can be recreated using basic disk models. This model has been shown to successfully account for the NIR excess of the SED and the basic visibility features of AB\,Aur, MWC\,275 and RY\,Tau \citep{Konigl11,Petrov19}.

    A similar magnetically driven model that has been invoked to explain the NIR excesses of YSOs is that of a magnetically-supported dust atmosphere to the inner disk \citep{Turner14b}. A buoyant magnetic pressure adds to the internal gas pressure to create a dusty 'atmosphere' across the inner disk. This is thought to be confined only to the inner regions where the high temperatures ensure good magnetic coupling. Unlike the dusty disk wind, the material here is not outflowing, but is suspended by magnetic pressure. The material here is optically thick near the base of the atmosphere and has been shown to double the fraction of stellar luminosity reprocessed by the disk.

    \subsection{Photoevaporation}

    Photoevaporation is the process by which energetic radiation, usually in the form of far-UV photons, ionises and disperses gas within the surface layers of protoplanetary disks. Sufficiently energetic UV radiation will excite Hydrogen and Helium molecules within the disk, causing them to photodissociate and ionise \citep{Alexander06a}. The ionisation of gas will increase the local thermal energy which counters the gravitational energy of the disk mass. Beyond some critical radius in the disk the local thermal energy will break equilibrium, allowing the gas to escape as a wind \citep{Alexander06b}. This radius is known as the Gravitational Radius and is given by
    \begin{equation}\label{eq:GravyRadius}
       R_g = \frac{G M_\star}{c_s^2} = 8.9 \left( \frac{M_\star}{1M_\odot} \right) \mathrm{au},
    \end{equation}
    where $R_g$ is the gravitational radius, $G$ is the gravitational constant, $M_\star$ is the mass of central star and $c_s$ is the sound speed of the gas (typically $\sim10\,\mathrm{kms^{-1}}$). Such an effect is particularly effective at removing material from the disk in the later stages of disk evolution. In young disks the accretion rate is significantly larger than photoevaporative winds, however as the accretion rate drops over time, the effect becomes more significant as it deprives the disk of material resupply inside of $R_g$. At this point, the inner disk drains on its own, short, viscous time-scale, giving a dispersal time much shorter than the disk lifetime. This is thought to be one of the potential mechanisms behind transitional disks, see section\,\ref{sec:transitional} for a more detailed exploration of disk evolution at this point.

\section{Disk Evolution}

    The evolution of disks beyond their initial formation is one which appears to diverge even between objects of similar masses and spectral types. As a consequence the exact timeline of disk evolution is not well understood, neither are the effects of the evolutionary stages on the efficiency of planet formation. In this section, I will discuss some of the key evolutionary stages which some disks have been observed to go through and how they might impact the planet formation process. 

    \subsection{Variable Accretion \& Outbursting Events}

    Classically, it was not thought that accretion varied much over time, with only a slow decrease as the disk material is depleted due to outflows, planet formation and accretion processes. This occurs on long timescales over several million years of disk evolution. However, our view of accretion changed dramatically with the characterisation of the star FU\,Orionis \citep{Herbig66}. In 1937 FU\,Ori suddenly increased in luminosity by $\sim6^m$ over a matter of a few days and has remained at a significantly increased brightness ever since, only decreasing by around $0.025^m$ per year. A change in luminosity of such a magnitude and speed had never been observed before outside of supernovae, particularly in a usually quiescent star. However, since this discovery over a dozen similar objects have been observed undergoing a similar event \citep{Herbig77,Audard14}. This class of star have been dubbed FUors after the archetypal FU\,Ori. 

    The causes of such a marked increase in the brightness of FUors was the subject of much speculation until \citet{Hartmann85} argued that a rapid accretion rate onto the central star could be the cause. In this scenario, a large amount of material from of the disk falls onto the central star, leading to a large increase in the accretion rate over a very short time scale. The expected molecular lines of such a scenario matched high spectral resolution observations of several FUor stars \citep{Zhu09}, and this scenario has since become widely accepted and the driving force behind outbursting events. 

    The question of the driving force behind the change in accretion rate however, is still an unanswered question. Various scenarios have been proposed to explain the massive increase in accretion rate seen in these objects. \citet{Vorobyov05,Vorobyov06} propose gravitational instabilities on large scales cause the disk to fragment and for clumps of material to fall onto the central star. \citet{Bell94} suggest that a thermal instability in the very inner regions ($<0.1\,\mathrm{au}$) could be sufficient to cause outbursts of this magnitude. \citet{Bonnell92} propose that a binary companion on a highly eccentric orbit could perturb the disk and cause repeated outburst of accretion. Similarly, \citet{Reipurth04} suggest that FUor stars are newly created binary systems, where the two stars become bound following the breakup of larger multiple systems. Such a scenario leads to the ejection of companions and the rapid infall of material. However, given the limited number of known FUors there is little consensus on the FUor outburst triggering mechanism.

\begin{figure*}[h!]
        \centering
        \includegraphics[scale=0.6]{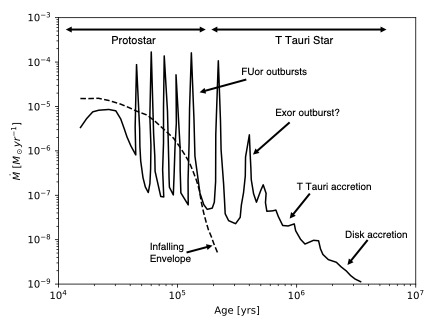}
        \caption[Illustrative sketch of accretion rate over time for a low mass young star through the protostellar and T Tauri stars]{Recreation of graphic from \citet{Hartmann96}. Model of accretion rate over time for a low mass young star through the protostellar and T Tauri star phases. They propose that most stars undergo multiple FUor type outbursts every $\sim$100,000 years until the surrounding envelope can no longer replenish the disk with new material. At which point accretion settles into a steady state scenario. 
        }
        \label{fig:VariableAccretion}
    \end{figure*}

    The prospect of variable accretion also goes some way to solving other astrophysical problems, notably, the young star luminosity problem. YSOs are known to be 10-100 times less luminous than expected from steady-state accretion scenarios \citep{Kenyon90}. Particularly given the accretion rates of the order $10^{-7}$ to $10^{-8}\,\mathrm{M_\odot yr^{-1}}$ observed around many YSOs. However, if FUor outbursting events are not limited to a subset of stars and variable accretion is a more ubiquitous process, this goes some way to explaining the observed luminosity deficit. The possibility of most YSOs undergoing multiple outbursting events during their early evolution is explored extensively by \citet{Kenyon95} and \citet{Evans09}. Figure\,\ref{fig:VariableAccretion} shows a model accretion scenario that is thought to be typical for many YSOs. FUor outbursts occur every $\sim$100,000 years over the early stages of disk evolution while the disk is replenished by the wider envelope \citep{Audard14}. As the envelope depletes, outbursts become smaller and less frequent. As the disk ages and the envelope and later the disk itself are depleted, accretion stabilises into a steady state scenario.  

    A secondary class of star which is also thought to exhibit variable accretion are EX\,Lupus objects or EXors, which exhibit similar outbursts to their FUor counterparts. These outbursts are typically weaker and shorter, only increasing in brightness by $1-4^m$ and lasting $10-100$ days \citep{Audard14}. In some cases they also occur more frequently, being separated by only a matter of months. Similarly to FUor objects, the cause of these outbursts is thought to be an enhanced rate of accretion onto the stellar surface. 

    \subsection{Transitional Disks} \label{sec:transitional}

\begin{figure*}[h!]
        \centering
        \includegraphics[scale=0.6]{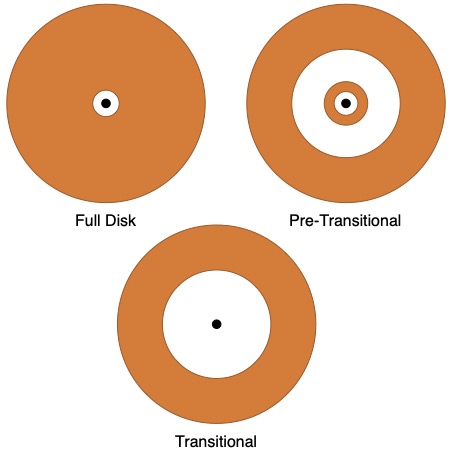}
        \caption[Evolution of a protoplanetary disk through the transitional stages]{Evolution of a protoplanetary disk through the transitional stages. From a full, optically thick flared disk, to a planet induced gap in the inner disk known as the pre-transitional stage, to a fully cleared inner disk known as transitional. Beyond these stages the disk evolves into a planetary/debris system.}
        \label{fig:Transitional}
    \end{figure*}

    Transitional disks are so called because they link together two key stages in planet formation. They provide the bridge between optically thick flared disks and dissipated planetary/debris systems \citep{Espaillat14}. They were first identified by their distinct lack of NIR excess, while simultaneously exhibiting strong MIR and FIR excesses \citep{Strom89}. Such observations indicate the absence of an inner disk, or at least the absence of a dust component to the inner disk. In addition to transitional disks, a sub-class known as pre-transitional has been observed which exhibit an inner gap as opposed to an a hole \citep{Brown07}. Figure\,\ref{fig:Transitional} shows the difference in disk geometries between full, pre-transitional and transitional disks.

    Currently dozens of transitional disks have been discovered indicating that this is a ubiquitous and vital stage of disk evolution. The exact causes of the absence of an inner disk is not well understood, however the strongest suggestion is that this is the natural consequence of core accretion planet formation \citep{Paardekooper04,Zhu11}. Standard planet formation models predict that a protoplanet will clear a region of the disk through accretion and tidal forces. Indeed several transitional or pre-transitional disks have reported examples of protoplanet candidates \citep[e.g. PDS\,70;][]{Keppler18,Haffert19}. These clearings detected in the dust disk often coincide with gas cavities in the same region, as expected if cleared by a forming planet \citep{vanderMarel18}. The combination of these dust cavities with the presence of continuing gas accretion onto the central star is a challenge to theories of disk clearing. Even if companions are not responsible for the clearings seen in all (pre-)transitional disks, these objects still have the potential to inform our understanding of how disks dissipate, primarily by providing constraints for disk clearing models involving photoevaporation and grain growth \citep{Dullemond05,Alexander07}.

    \subsection{Debris Disks} \label{sec:debris}

    A very late stage of evolution is a class of disks known as debris disks. Debris disks are gas depleted, disks of circumstellar dust which have been found around many older stars (typically $>10\,\mathrm{Myr}$) such as Beta Pictoris, Fomalhaut and 51 Ophuichi \citep{Smith84,Mennesson13,Stark09}. These are thought to have evolved is one of two ways. Firstly, they could be natural progression of disk evolution, where accretion and radiation pressure removes almost all the gas from a protoplanetary disk, leaving behind only a dusty remnant. However, it is thought that the lifetime of dust grains is significantly shorter that the ages of debris disk systems owing to the Poynting-Robertson effect, radiation pressure and partial collisions. Therefore, this dust likely formed after the protoplanetary disk stage, from the collision of planetesimals in an extrasolar analogue to the Kuiper belt within our own solar system \citep{Wyatt08}. These collisions cause the breakdown of large scale objects to mm sized grains in what can be seen as a reversal of the planet formation process.


\chapter{General Observations of Protoplanetary Disks}
\label{ch:ObsPD}
\epigraph{``And everything under the sun is in tune, But the sun is eclipsed by the moon''}{--- Pink Floyd, Eclipse\hphantom{spacing}}

Our understanding of protoplanetary disk has evolved with the capabilities of our observatories. From initial hints at the presence of disks from photometric infrared excess to exquisite high resolution imaging, the past few decades have opened doors to previously undreamt of realms of astrophysics. In this section I summarise the observational techniques used to explore protoplanetary disks and scientific insights gleaned from them. 

\pagebreak 

\section{Spectral Energy Distributions}

The small angular size of protoplanetary disks has historically made spatially resolved observations difficult. It is only over the past few decades with the building of larger aperture telescopes and the development of interferometry that we have been able to directly study the structures of these disks. As an alternative astronomers have typically relied on indirect methods to study disks, the most prominent of these being the interpretation of spectral energy distributions (SEDs).

\begin{figure*}[h!]
	\centering
	\includegraphics[scale=0.45]{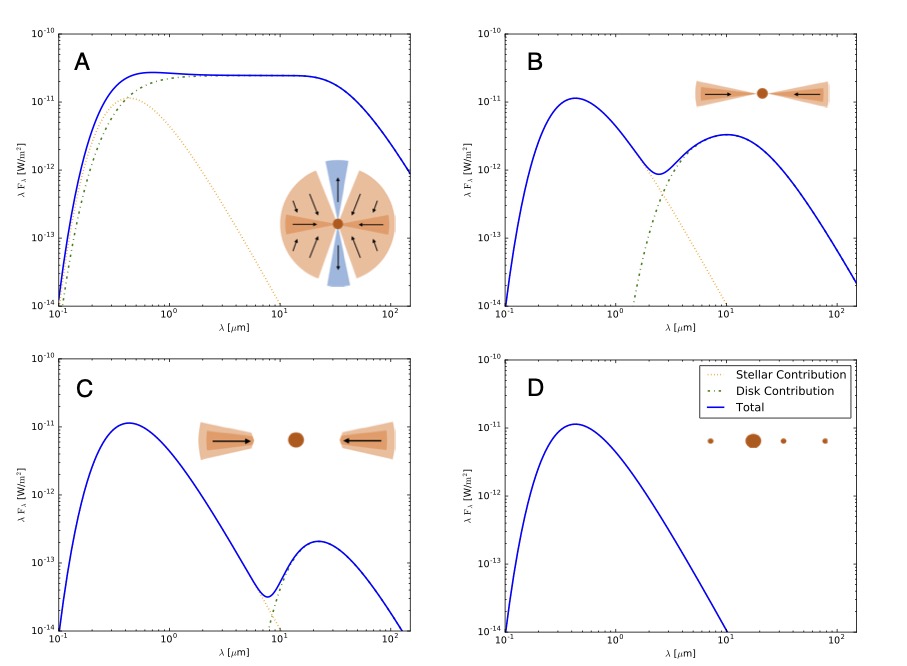}
	\caption[Evolution of the spectral energy distribution as this disk evolves]{Evolution of the spectral energy distribution. \textbf{A.} A protostar still enclosed within its protostellar envelope. The envelope contributes significantly to the SED across all wavelengths owing to its almost total obscuration of the central star (Obscuration not included in diagram). \textbf{B.} A full protoplanetary disk contributes little to the SED across the optical, but provides a strong excess across the infrared and beyond. \textbf{C.} A transition disk, where the inner disk has been cleared by accretion and radiation pressure. The lack of hot material means the NIR excess is non-existent, but the outer disk still contributes to longer wavelengths of the SED. \textbf{D.} The disk material has been completely removed or used up in planet formation, the only SED contribution comes from the central star.}
	\label{fig:SEDevolution}
\end{figure*}

An SED is simply a plot of the measured energy against the frequency or wavelength of light emitted. The shape and magnitude of the SED encodes the geometry of the source and is a powerful diagnostic tool for astronomers. The most basic form of an SED is that of a blackbody which can be used to approximate the photosphere of a single star. Additional material surrounds the star provides additional flux to the SED in the form of an excess to the blackbody curve. Figure\,\ref{fig:SEDevolution} shows the expected evolution of an SED over the lifetime of a solar-mass young star from being completely enclosed within its envelope to the full planetary system. It shows that disk material close to the central star is hotter and so contributes to shorter wavelength excess across the infrared, which is the first region to be cleared by stellar radiation. The longer wavelength sub-mm and radio emission is dominated by colder material further from the central star, which persists longer against radiation pressure. 

The different stages of SED evolution allow us to split objects into a loose class system based upon the slope of the spectral index \citep{Lada84}. The spectral index is typically measured at $2-2.4\,\mathrm{\mu m}$ and is a measure of the strength of infrared excess over the blackbody of central star. A class I object has a very strong IR excess extending to shorter wavelengths owing to the surrounding protostellar envelope remnants resulting in reprocessed light from the shock fronts that develop as material falls onto the protostar and disk. A class II object is where a strong IR excess is still present, but contributes little to shorter wavelengths. These objects can also exhibit soft X-ray and UV excesses from the shock of accreting material onto the stellar surface. Finally, class III objects refer to an object with little-to-no IR excess, these are typically debris disk or main-sequence stars. The small amount of excess present likely originates from the secondary generation dust particles found there. 

Subsequent to the three stage classification developed by \citet{Lada84} two additional classes have been added. Class 0 objects represent the very earliest stages of collapse of a cloud core and are characterised by no infrared emission owning to colder temperatures and lack of fully formed star \citep{Andre02}. A flat-spectrum (FS) source is a halfway point between class I and II objects \citep{Greene94}.

	\subsection{Characteristics of SEDs}

	As discussed, the wavelength of emission corresponds to the temperature of the emitting region meaning a great deal of additional detail can be gauged from the shape of the spectral energy distribution. However, the interpretation of the SED must be taken in the context of the disk structure, in particular the inclination of the disk. A disk of moderate to high inclination will be self-shadowed by its own flaring, altering the regions of disk that are seen by an observer. A disk of sufficiently high inclination will also obscure the central star due to extreme line-of-sight extinction. Only with some understanding of the viewing geometry can an SED be understood. The different regions of an SED correspond to different regions of the disk and different geometric and compositional aspects. An example SED of the young star SU\,Aurigae is shown in Figure\,\ref{fig:SEDexample}, the regions A through E are described here:

	\begin{figure*}[h!]
		\centering
		\includegraphics[scale=0.45]{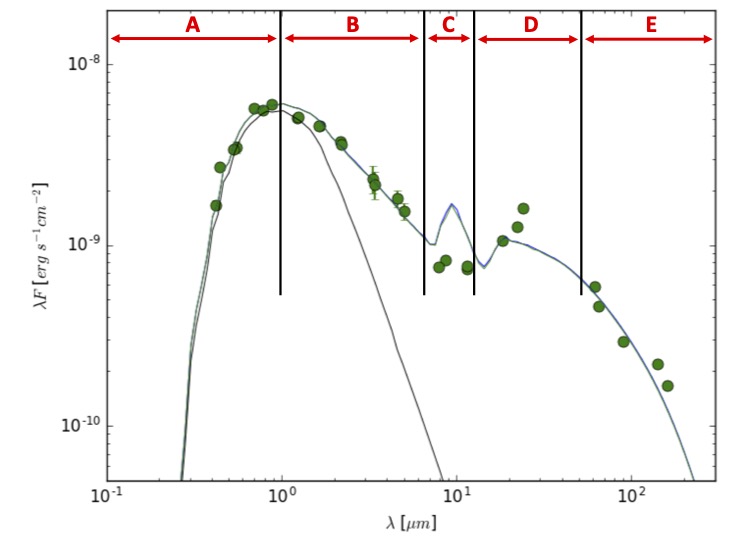}
		\caption[Spectral energy distribution of the young star-disk system SU\,Aurigae]{Spectral energy distribution of the young star-disk system SU\,Aurigae. The green points represent the photometric observations taken with a range of instruments. The solid black body is that of the stellar photosphere. \textbf{Region A} is the UV optical region of the spectrum. \textbf{Region B} is the NIR excess, \textbf{Region C} is the Silicate feature, \textbf{Region D} is the multi-colour region and \textbf{Region E} is the millimetre slope. }
		\label{fig:SEDexample}
	\end{figure*}

	\begin{enumerate}[label=\Alph*]
		\item[A.]

		Short wavelength radiation across the UV and optical parts of the spectrum is dominated by the stellar photosphere. There is almost no contribution from the disk itself owing to its relatively cold temperature (typically $\sim 1500\,\mathrm{K}$). An exception to this is in objects with a very high accretion rate such as FUor and EXor objects where accretion shocks and material interior to the sublimation rim will contribute significantly to this regime. 

		\item[B.]

		In full disks the NIR ($1\leq\,\lambda\,\leq5\,\mathrm{\mu m}$) region is in a strong excess over stellar emission owing to the presence of hot dust close to the central star. Much of this emission is thought to arise from the sublimation rim of the disk, the structure of which is described in Section\,1.3.3. The gradient of this region can be used to estimate the radial temperature gradient of the disk and the magnitude of the excess can indicate the presence of other processes such as disk winds and puffed up sublimation rims. 

		\item[C.]

		One of the most prominent features is known as the $10\,\mathrm{\mu m}$ silicate emission feature which is present in many SEDs, in particular this feature is attributed to stretching of the Si-O bonds in silicates. The variation in emission strength and shape can be well explained by the dominant grain size in the upper layers of a disk \citep{Przygodda03,Olofsson09}. Smaller sub-micron sized grains result in a larger, more pointed silicate feature; inversely larger grains, several microns in size create a weaker, flatter silicate feature \citep{Kessler06}. As such, the size and shape can be used as an indicator for grain growth within a protoplanetary disk, and important stage in planet formation. Studies such as \citet{Kessler06} and \citet{Lommen10} track the growth of dust grains across a wide range of disk types and ages and find that the shape and size of the silicate feature correlates well with the expected grain size. 

		\item[D.]

		This region of the SED typically lies between $30\leq\,\lambda\,\leq300\,\mathrm{\mu m}$, corresponding to slightly cooler regions of the disk. At these radii the disk is still optically thick allowing the dust to contribute to continuum emission. This region will typically also be the turn-over point of the SED, this is where the coolest material of disk is reached meaning the SED begins to turn down. 

		\item[E.]

		Another region of the SED that can provide useful information is the shape at millimetre-plus wavelengths, the so called millimetre slope. Emission from these regions can provide broad estimates of the dust mass of the disk and size of the grains \citep{Wood02}. these however, still require assumptions about grain opacity and temperature.
	\end{enumerate}

	\subsection{Interstellar Extinction} 

	One of the practicalities needed for interpretation of SEDs is the understanding of interstellar extinction or reddening. As radiation passes from its source to the observer it will be absorbed and scattered by dust and gas. This reprocessing is more efficient at shorter wavelengths, hence blue light is scattered and absorbed more leading to a reddening of the spectra.
	The degree of this reddening is determined by the distance to an object and the presence on any additional material between the object and observer. Extinction can be measured through
	\begin{equation} \label{eq:extinction}
  R = \frac{A_V}{E_{B-V}},
  \end{equation}
  were R is the ratio of total to selective extinction in the V band and typically lies between $2.2$ and $5.8$, but is generally accepted to be around $3.1$. $A_V$ is the total extinction in the V band and $E_{B-V}$ is the colour excess calculated across the B and V bands \citep{Fitzpatrick99}. 

\begin{figure*}[h!]
		\centering
		\includegraphics[scale=0.45]{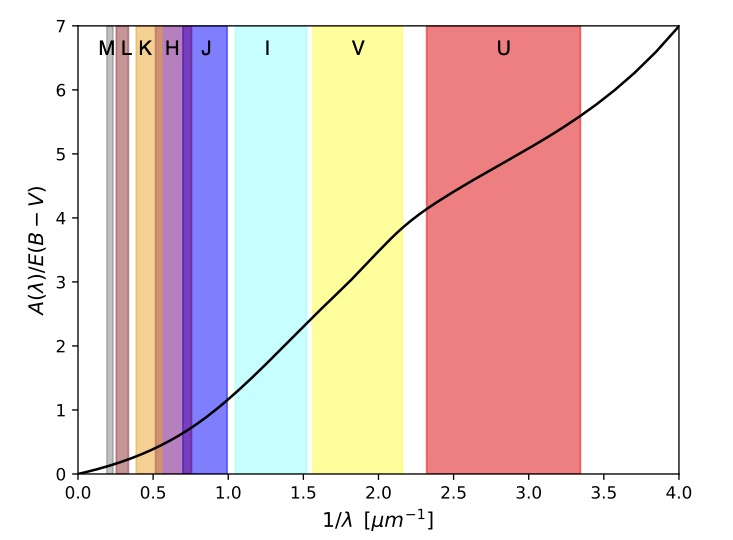}
		\caption[Wavelength dependence of extinction across the visible and NIR wavebands]{Wavelength dependence of extinction across the visible and NIR wavebands for an extinction ratio of $R=3.1$ and $E_{B-V}=1$. Coloured regions correspond to different photometric bands as defined in the Johnson system.}
		\label{fig:Extinction_Curve}
	\end{figure*}

  However, extinction is spectrally and spatially variable, making correcting for its effects very difficult. The spatial dependence of extinction arises from the differing composition and density of the interstellar medium (ISM), often in the local environment of star forming regions. Differences in the optical properties of the dust grains, such as their size and composition can affect the degree of observed reddening. The spatial dependence of extinction is characterised by the R term of Equation\,\ref{eq:extinction} where the value of $3.1$ is the global mean, and is applicable to most YSOs. The spectral dependence of extinction is reliant on a knowledge of the spatial dependence, as different dust grains exhibit different optical properties. \citet{Fitzpatrick99} provide model extinction curves for a variety of values of R, allowing the extinction to be derived in any wavelength given only the total extinction in a single band. An example extinction curve for $R = 3.1$ for a star at a distance of $157\,\mathrm{pc}$ with an $A_V = 0.5$ is shown in Figure\,\ref{fig:Extinction_Curve}. 

  By correcting for the effects of extinction, we can explore an SED as if the ISM were not present, allowing us to interpret astrophysical objects as they truly are. It is common statistical practice to apply this correction to the SED model rather than the photometric data points which comprise the model.

\section{Spectroscopic Observations}

	The natural evolution of studying a spectral energy distribution is to study the finer details of the spectra, in particular the emission and absorption lines of elements and compounds found within. Radiation from the central star will be reprocessed by molecules within the gas and dust of a disk, the wavelength of this interaction is highly specific to the chemistry of the molecule. Hence an over or under brightness at a particular wavelength of the spectrum is a signpost for the presence of a particular molecule. In addition, the absorption or emission peaks may be shifted and broadened by motions relative to the observer, allowing for the tracing of disk kinematics; including orbital rotation, ejection outflows and accretion. These aspects make spectroscopic studies of protoplanetary disks a very powerful technique. In this section, I will examine some of the most common spectral lines associated with the inner regions of protoplanetary disks and their ramifications on disk physics. 

	Perhaps the most common and numerous of the lines are those resulting from molecular hydrogen, of these numerous series arise from the inner regions of protoplanetary disks. Primarily the Paschen ($0.82-1.87\,\mathrm{\mu m}$), Brackett ($1.45-4.51\,\mathrm{\mu m}$) and Pfund lines ($2.30-7.46\,\mathrm{\mu m}$) emit across the infrared. Of these, one of the commonly studied is the Brackett-$\gamma$ (Br$\gamma$) molecular line corresponding to gas temperatures between $6000$ and $12,000\,\mathrm{K}$. The processes associated with Br$\gamma$ emission are thought to be mass accretion and ejection scenarios. Work by \citet{Kraus08} found that the radial size of the emitting region relative to the continuum size was significantly different between the objects in their sample, indicating that the physical mechanisms traced by the Br$\gamma$ line are varied for different objects.

	The coupling of rotational and vibrational excitations in carbon-monoxide (CO) creates a distinct emission feature. This is the CO bandhead so called as it appears in bands in low-resolution spectra and is found in almost a quarter of YSOs \citep{Carr89}. The first overtone of this emission is found at around $2.3\,\mathrm{\mu m}$ and is associated with temperatures of $\sim3000\,\mathrm{K}$, conditions which match those found in the very inner disk. The CO profiles observed in most YSOs are consistent with gas motions in the innermost regions, in particular accretion processes \citep{Bik04,Ilee14}. In particular, \citet{Martin97} demonstrate that observed CO profiles can be recreated through accretion funnels where material flows from the disk to the star along magnetic field lines. An alternative scenario proposed to create the conditions needed for CO bandheads is that of disk wind where large amounts of hot gas are in motion \citep{Chandler93,Chandler94}.

	Polycyclic aromatic hydrocarbons (PAHs) profoundly influence the thermal budget and chemistry of protoplanetary disks \citep{Meeus01,Seok17}. PAH in the surface layers of the disk are directly exposed to stellar radiation and so are photoionized and provide photoelectrons allowing for the heating of gas within the disk \citep{Kamp04}. PAHs also have large surface areas for chemical reactions to occur, which can significantly alter the carbon chemistry of the disk \citep{Habart04}. Thus the understanding of PAHs and their presence within protoplanetary disks is vital to understand the diverse nature of disks and the formation of planets within them. 

	In addition to NIR spectroscopic observations, much work has been done to constrain the chemistry of disks are longer wavelgnths using specialised instruments such as Spitzer and mm-wavelength observatories such as ALMA. \citet{Pontoppidan14} review the contributions of volatiles in protoplanetary disks and highluight detections of $\mathrm{H_2O}$, $\mathrm{CO_2}$, HCN, $\mathrm{SiO_4}$ and many other molecules in the inner regions of protoplanetary disks of many stars \citep{Lahuis06,Gibb07}. A large scale survey of disks by \citet{Pascucci09} of over 60 disks around low mass stars found that a significant proportion contained organics, notably HCN in $\sim30\%$ of disks. The wide variety of volatiles found in protoplanetary disks provide the building blocks for complex planets, including those capable of sustaining life.


\section{Spatially Resolving Protoplanetary Disks}

	While the SED and spectroscopic studies of protoplanetary disks are immensely useful, only a limited amount of spatial information can be obtained. In order to gain a full appreciation for the structure, the disk must be spatially resolved. This raises a host of challenges, both instrumentational and analytical, which must be overcome. The difficulty of imaging protoplanetary disks is one of resolution, the typical small angular size of these disks on sky has meant that imaging has been historically impossible owing to the size of telescopes. The resolution of the current largest group based telescope is only $\sim32\,\mathrm{mas}$ in the H\,band, insufficient to resolve small scales of inner disks which are typically only a few mas in diameter. However, over the last few decades with observatories such as Keck, Hubble, VLT, and ALMA providing larger and larger telescopes in addition to advanced adaptive optics we have been able to explore these disks in unprecedented detail. 

	\subsection{Scattered Light Observations}

	  As mentioned previously it is thought that most protoplanetary disks exhibit a flared structure which allows radiation from the central star to interact with almost all of the disks surface. At the disk surface small dust grains are well coupled to the gas component and so are resistant to vertical settling. These small grains are very efficient at scattering infrared photons. As the disk is optically thick infrared radiation cannot penetrate deep into the disk, therefore this scattered light is an excellent probe of gaseous phase structures in the disk surface. 

	  \begin{figure*}[h!]
			\centering
			\includegraphics[scale=0.45]{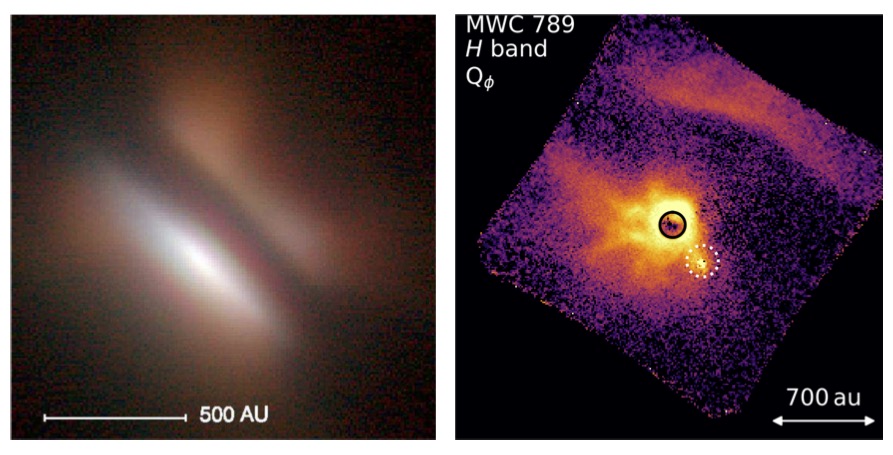}
			\caption[Example scattered light observation of protoplanetary disks]{Scattered light observation of protoplanetary disks. LEFT: Edge on disk around PDS\,144N as seen with Keck NIRC2 instrument \citep{Perrin06}. The disk inclined at almost $90^\circ$ meaning only the flared outer disk is visible, while the optically thick inner disk is obscured. RIGHT: Circumstellar disk of MWC\,789 as seen with GPI using the polarimetric differential imaging technique, shown is the radial stokes parameter $Q_\phi$ \citep{Laws20}.}
			\label{fig:ScatteredLight}
		\end{figure*}

		Figure\,\ref{fig:ScatteredLight} shows scattered light images of the disks around PDS\,144N and MWC\,789 from \citet{Perrin06} and \citet{Laws20} respectively. PDS\,144N is almost completely edge-on  such that the central star is completely obscured by the optically thick midplane of the disk, only the surface layers of the flared disk are visible to the observer. On the other hand MWC\,789 is more face-on and so more details of the radial disk structure are visible. In the north-west a large disconnected arc of material and a stellar companion was detected in the south-west. Other outflows and arms can be seen along the eastern edge of the disk. As the star is not obscured by the disk a coronagraph is required to block light from the central star. 

		There are several different methods with which we can observe protoplanetary disks in scattered light. The primary issue of imaging disks is that light from the star is often significantly brighter than the scattered light from the disk, which makes observing features within the disk very difficult. The aim of imaging disks is therefore to remove the contaminating photon noise from the central star. In the first instance this is done through the use of a coronagraph, a physical block in the telescope which blocks the starlight from passing to the detector. Originally designed to obscure the solar photosphere to allow for viewing of the solar corona, stellar coronography is now widely used in the imaging of disks and planets. However, there is often a significant amount of photon noise which leaks around the coronagraph, which must be removed through other techniques. Perhaps the most widely used of these is the process of differential imaging, of which there are two methods, Polarised Differential Imaging (PDI) and Angular Differential Imaging (ADI).

		Polarised differential imaging relies on the assumption that photons from the central star are unpolarised while any dust scattering interaction will polarise the photons. In such a way the polarisation signal will originate only from this disk and can be used to create an image of the disk \citep{Mauron98}. This technique is very powerful for imaging faint disk features as the central star is almost completely subtracted from the image in comparison to total intensity images \citep{Avenhaus18,Laws20}. However, care must be taken with PDI as the polarisation is heavily dependent on the dust grain size distribution, composition and the angle of incident radiation. Hence the disk composition must be at least partially understood to fully interpret PDI images. 

		Angular differential imaging involves the capturing of multiple short exposures of an object as it moves across the sky without correcting for changes in the field of view. In between different exposures any asymmetric features will rotate in frame, while centro-symmetric features such as the PSF will appear not to rotate. As such the images can be combined in such a way that the central PSF is subtracted from the final image, leaving just a disk image \citep{Marois06}.

		The now widespread use of direct imaging in the study of protoplanetary disks has allowed us to examine disk structures in unprecedented detail. Through this we have revealed a host disk gaps, rings, spiral arms and many other features many of which are direct consequences of active planet formation processes \citep{Avenhaus18}. However, even these observations are limited in their resolution ($\sim34\,\mathrm{mas}$ for an 8\,m telescope) and cannot resolve detail in the innermost regions of protoplanetary disks, to explore these regions, other techniques must be applied. 

	\subsection{Sub-mm \& Radio Observations}

		In addition to direct imaging with optical telescopes the development of long wavelength sub-mm and radio telescope arrays has opened new doors in high resolution imaging over the past couple of decades. Perhaps the most prolific facility is the Atacama Large Millimetre Array (ALMA) boasting the highest resolution \citep{Andrews18}.

		Rather than relying on the resolving power of a single dish telescope, these observatories make use of arrays of smaller aperture scopes in a method known as very long baseline interferometry. For a discussion of the basics of interferometry at optical wavelengths see Chapter\,4. The resolving power of interferometric arrays is based on the distances between the antenna rather than the aperture of an individual disk. The multiple antenna signals are combined and processed to simulate an individual radio telescope with an effective aperture up to 15\,km in diameter at ALMA.

		\begin{figure*}[h!]
			\centering
			\includegraphics[scale=0.4]{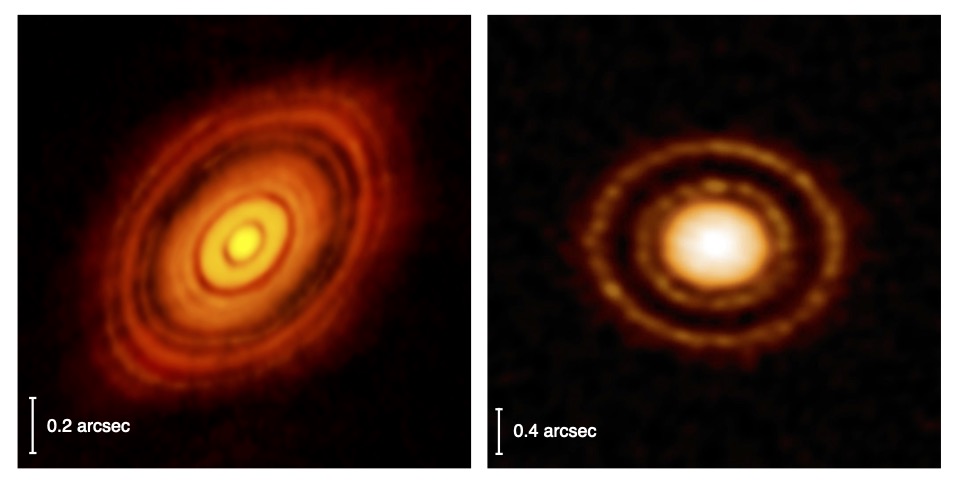}
			\caption[Example ALMA images of two protoplanetary disks.]{Example ALMA images of two protoplanetary disks. LEFT: multiple concentric rings of HL\,Tau from \citet{ALMA15}. RIGHT: Gapped disk structure of AS\,209 from \citet{Fedele18}. ALMA images trace the the dust and pebble content of the disk at mm continuum.}
			\label{fig:ALMAImages}
		\end{figure*}

		Many of the most striking images of protoplanetary disks have been obtained with ALMA such as those shown in Figure\,\ref{fig:ALMAImages}. In addition to disk images, ALMA also works at high spectral resolutions, resolving common lines such as C-O transitions. This allows for the study of the gaseous disk, in particular disk kinematics, inflows and outflows \citep{Andrews18,Guzman18}. Observational studies have unveiled a plethora of different disk features sometimes unknown and unexpected from theory.

\section{Observed Structures in Protoplanetary Disks}

		As we have explored, the structures of protoplanetary disks can be inferred indirectly from analysis of SEDs and spectroscopic observations, or observed directly in scattered light imaging. From these observations we have discovered a host of different structures and substructures with the disk population, in this section I will summarise of the most important features relevant to the innermost regions of disks and planet formation in these regions. 

		\subsection{Sublimation Fronts}

		High resolution optical interferometric observations of full protoplanetary disks are often dominated by flux from around the sublimation radius. The majority of disks cannot be modelled as wide extended components, but rather as thin rings around the central star. This was found to be the case by \cite{Lazareff17} who conducted a survey of 51 Herbig AeBe stars using the PIONIER instrument at the VLTI. It was found that over half of the disks could be successfully modelled using a ring-like structure. This is due to the high optical depth of the circumstellar material, where most of the radiation is absorbed, scattered or re-emitted. As a result most of the disk flux originates from the very inner-rim of the disk despite the radial extent of material. 

		\cite{Lazareff17} fit the observational data using simple geometric models, such as Gaussian and ring distributions. They find that, in the majority of cases, ring-like distributions provides significantly better $\chi^2$ values that elliptical Gaussian. In addition, when a ring shape is preferred, it is a wide ring, with $w>0.4\,\mathrm{mas}$, where $w$ is the half-light radius. Additional, they constrain the sublimation temperature to $1800\,\mathrm{K}$ using radiative transfer modelling, larger than the $1500\,\mathrm{K}$ predicted. As explanation for this is not immediately offered. 

		The presence of an illuminated inner-rim and a shadowed outer disk has been shown to be present in a number of objects. However, the exact shape of the inner-rim is still open for debate. The initial question: \textit{Is the rim sharp of diffuse?} is addressed by \cite{Kama09}. It was concluded that the inner-rim is not infinitely sharp, but a diffuse region exists that can extend to cover a significant fraction of the rim radius. This region is created through a combination of sublimation, condensation and backwarming effects. Backwarming occurs when a dust grain cannot cool into empty space in every direction, so its temperature increases. 

		The shape of the inner rim has traditionally been defined as a vertical wall \citep{Chiang97,DAlessio98,DAlessio99}. A good example of such a study is \citet{Akeson05}, who obtain PTI observations of three YSOs and model the inner disk with more advance radiative transfer models to fit both visibilities and SEDs simultaneously. They are able to reproduce observations using a flared disk with a vertically flat inner wall. However, they also require the presence of material interior to the dust sublimation rim to explain their observations. This material could exist in the form of a hot gaseous disk or very refractive dust grain species such as Ca-Al-Ti bearing compounds which are well characterised from chondrite meteroites within our own solar system \citep{Lodders03}.

		However, this has been shown to be a poor fit to many observations, as the strong asymmetries predicted by flat rim models have never been observed in inclined objects \citep{monnier06}. This lead to the consideration of older models of sublimation rims; \cite{Dullemond02} proposed several different structures dependent on their density distribution as a function of radius. For the case where optical depth is high, the structure is similar to that of a simple 'disk with a hole'. At large radii that disk adopts a highly flared shape, while nearer to the sublimation radius the dust becomes superheated and so becomes 'puffed-up' with the rest of the disk in shadow if sufficiently optically thick. The degree of shadowing depends upon the position of the flaring and the optical thickness along the different disk axes. These structures provide a detailed and complex way of modelling disk geometries, one that requires sophisticated radiative transfer modelling, that can produce much better fits to the observed visibilities than basic geometric modelling. 

		A commonly used prescription for inner-rim morphology is that of \cite{Isella05}. They argue that the shape of the inner rim is controlled by a very large vertical density gradient, through the dependence of grain evaporation temperature on gas pressure density as described in section\,2.3.1. This results in a curved rim, rather than a vertical wall. The exact shape of the inner-rim can still vary slightly from object to object, however, most modern interpretations are based upon a curved and puffed-up rim which casts the extended disk into shadow. However, to further complicate matters, as more disks are observed more discontinuous, asymmetric inner-rims are discovered, and modelling those brings its own set of challenges.

		\subsection{Rings and Gaps}

    Gaps within the radial structure of protoplanetary disks are the best signature available for the detection of planets within the disk beyond direct imaging. They can form at any position within the disk, including within the inner astronomical unit, and can vary greatly in size.  

    If a planet forms within a circumstellar disk, it is reasonable to assume that tidal forces can clear the orbit of the planet. For such a phenomenon to occur, that Roche radius of the planet must be comparable to the thickness of the gas disk, equating to typically masses somewhere between Saturn and Jupiter \citep{Goldreich80}. Indeed one of the definitions of a planet by the IAU is that its orbit has been cleared of debris. \cite{jang08} initially the shape of a dimple in the disk created by gravitational perturbations of a planet, however, they did not anticipate that tidal forces could clear a large gap in the disk, as shown in the hydrodynamical simulations of \cite{Bate03}. These gaps were much larger than the dimples and if well resolved allow for determination of the mass of the planet. However, care must be taken when interpreting disk gaps as planet induced as it has been shown that magneto-rotational or gravitational instabilities can also create gaps under the correct circumstances \citep{Johansen09,Takahashi16}. 

    These gaps invariably create a second rim, the far edge of which can become thermally heated and puffed up. In flared disks, this creates a second ring of bright flux which can be observed in the same way as inner-rim, through interferometry \citep{Varniere06}. Alternatively for a flatter disk, the far edge of the gap will remain in the shadow of the inner-rim and so not be visible \citep{Wolf02}. 

    With recent improvements in the spatial resolution of instruments, an increasing number of circumstellar disks have been shown to contain gaps in their morphology. Perhaps the most famous example is that of HL\,Tau \citep{ALMA15} shown in the left of Figure\,\ref{fig:ALMAImages}. HL\,Tau demonstrates a remarkable series of bright and dark concentric rings though to be induced by the planet formation process. \cite{Calvet02} discuss the observation of gap within a highly evolved disk that is theorised to be caused by a planet within $4\,\mathrm{au}$ of the central star. Another example is provided by \citet{Andrews16} in the form of ALMA images of the YSO TW\,Hya where similar concentric rings are found between 1-6\,au. Various scenarios are proposed for the occurrence of these rings, some occur at temperatures that may be associated with the condensation fronts of major volatile species \citep{Zhang15,Okuzumi16}, while others may be the consequence of magnetised disk evolution. However, a particularly dark annulus located at 1\,au  could indicate interactions between the disk and young planets.   
		


		\subsection{Asymmetries}

		Dramatic asymmetric features have been captured across a wide range of protoplanetary disks. The most spectacular demonstrations of asymmetric features are found in the outer disk when imaged directly or with sub-mm interferometry \citep{Garufi17,Avenhaus18,vanderMarel21}. Asymmetries are found in the form of spiral arms, arcs, or large blobs of material, which have a variety of possible causes. 

		Several objects have been observed with string asymmetries thought to arise from dust trapping mechanisms. Examples of this include Oph\,IRS\,48 and V1247\,Orionis \citep[][respectively]{vanderMarel13,Kraus17} where dust trapping leads to the formation of a huge 'blob' or arc like features on one side of the disk. These are thought to form when dust is trapped within local gas pressure maxima in both the radial and azimuthal direction. The high local gas pressure prevents both the inward drift of dust and potentially destructive collisions, such an effect is more efficient for larger grains as the small grains are better coupled to the gas. The origin of these local pressure maxima is not fully understood, but potential causes include the appearance of a 'dead zone' within the disk \citep{Regaly12} or vortices created by a substellar companion/planet \citep{Pinilla12}. The efficiency of the dust trapping in these regions is such that it has been touted as a potential planet formation mechanism. 

		Another commonly observed feature in the outer disk is one or more spiral arms such as HD\,142527, HD\,34700A, AB\,Aur and SU\,Aur \citep[][respectively]{Avenhaus17,Monnier19,Boccaletti20,Ginski21}. Spiral arms a known to be more common in larger Herbig\,AeBe stars than their lower mass T\,Tauri counterparts which fuels the theories regard spiral arm formation \citep{Avenhaus18}. Herbig disks tend to be more massive, raising the possibility of gravitational instabilities, Herbigs also have a high binarity fraction meaning the dynamical influence of a secondary component must be considered \citep{Casassus15,Perez15}. Most of the observed disks are not thought to be massive enough to experience gravitational instabilities, making this an unlikely scenario \citep{Vorobyov10}. 

    In addition to outer disk studies, inner disk asymmetries have been observed in several disks to date, looking back at the survey of Herbig AeBe stars by \cite{Lazareff17}. They find numerous candidates with asymmetric disks from geometric fitting, which are modelled as simple asymmetric, skewed rings. However, no specific explanations are offered for the presence of these asymmetries as more information is required on specific objects. 

		Work by \cite{Kluska16} reveals an extreme case of disk asymmetry in the material around MWC\,158. A large asymmetry on one side of the disk is shown to move across the different epochs of interferometric data, through a combination of image reconstruction and geometric fitting. The causes of this asymmetry are explained in a number of ways. The most obvious cause is a close companion in orbit around the star. A binary system is shown to successfully reproduce the strong closure phase signal of the data, however, none of the geometric binary fits alone were able to reproduce the data, and fitting of circular Keplerian orbits proved unsuccessful. 

		Other possible explanations are also proposed, several years earlier \cite{Fernandes09} discussed two possible scenarios for MWC\,158. Firstly, the formation of a one-armed spiral due to matter ejection from a hot-spot on the stellar surface \citep{Kervella14} is shown to sufficiently explain the shifted emission features and the changes in radial velocities seen in the photospheric lines. Secondly, a shell ejection episode that can carry dust and gas away from the central star is capable of reproducing the photometric variability observed in the different epochs of observations \citep{Fernandes11,Fernandes12}. 

		The presence of disk asymmetries has been shown to be caused by a variety of interesting factors, that can greatly affect disk morphology. Perhaps the most interesting, also the easiest to explain, is that companions within the disk have the potential to create many of the observed structures.

\chapter{Introduction to Interferometry}
\label{ch:Interferometry}
\epigraph{``It is obvious from what has gone before that it would be hopeless to attempt to solve the question of the motion of the solar system by observations of optical phenomena at the surface of the earth.''}{--- A. Michelson \& E. Morley\hphantom{spacing}} 

Interferometry has truly entered its golden age over the past few decades. We have witnessed the dramatic transformation of what was a fringe observational technique to a mainstream method of discovery. Interferometry provides us with the highest possible resolution view of the universe, which no single telescope can ever hope to rival. In recent years this has culminated in the first image of a super massive black hole in another galaxy taken by the event horizon telescope \citep{EHT19}. An interferometric array spanning the whole Earth.

\section{Origins of Interferometry}

The property of light to interfere with itself was first scientifically described by Thomas Young in the late 18th, early 19th century, by resurrecting the century old theory that light is indeed a wave. Young was able to describe the double-slit phenomenon whereby light from a single source passes through two narrow slits and is projected onto a screen and a series of light and dark bands appear \citep{Young02}. Light from the two slits will interfere either constructively when in phase or destructively when out of phase, producing an interference pattern of bright and dark fringes on the screen. The spacing and contrast of these fringes can be described solely by the characteristics of the source and slits. It was from this very simple experiment that interferometry was born.

\begin{figure*}[h!]
    \centering
    \includegraphics[scale=0.5]{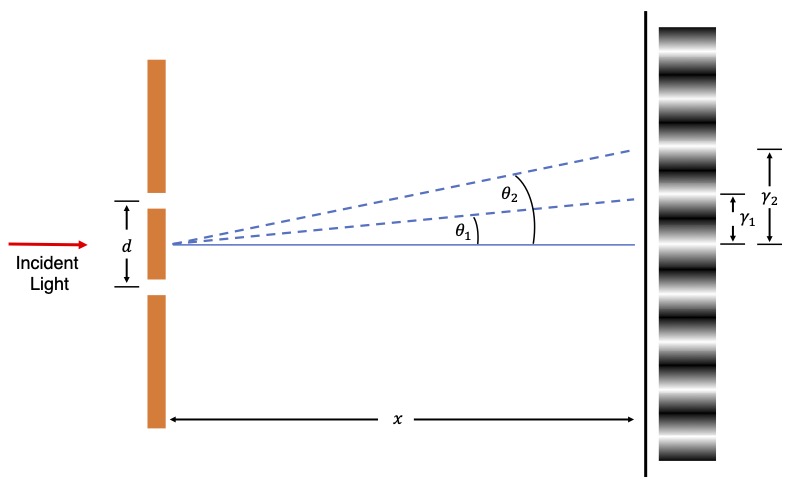}
    \caption[Young's double slit experimental setup]{Young's double slit experimental setup. Light is incident on the double slit from the left of the setup, after passing through the slits, the two beams interfere to produce a fringe pattern on the distant screen.}
    \label{fig:Youngs}
\end{figure*}

The simple setup of Young's double slit experiment is shown in Figure\,\ref{fig:Youngs} where coherent light of wavelength $\lambda$ is incident upon a double slit of separation $d$. At a distance of $x$ is a screen upon which an interference pattern is formed with bright constructive peaks at angle $\theta_m$ from the normal, where $m$ is the mode of the peak. The relationship can be described as $d \mathrm{sin}(\theta)  = m\lambda$, this can be expanded to include the separation between the peaks as $\Delta \gamma = x\lambda/d$ and the small angle separation can be induced.

One of the earliest uses of interferometry in astronomy, lies within attempts to discover the now absurd luminiferous aether. The aether was the postulated medium for the propagation of light since the 17th century, where it was invoked to explain the movement of light through apparently empty space. The aether was theorised to be directional, hence light travelling in different directions would exhibit a phase shift. It was on this basis which Albert Michelson and Edward Morley designed an experiment to detect the luminiferous aether in 1887. The idea was simple, to compare a source of light with itself after being sent in different directions via a beam splitter and series of mirrors. The recombining of the two beams of light would create an interference pattern of bright and dark fringes. The presence of the luminiferous aether 'wind' would cause a change in the phase of the beams leading to a shift in the fringe pattern. No such shift in the fringe pattern was observed, and in later years the theory of a luminiferous aether was completely discredited. However, the foundations for the first interferometer had been laid. 

\begin{figure*}[h!]
        \centering
        \includegraphics[scale=0.62]{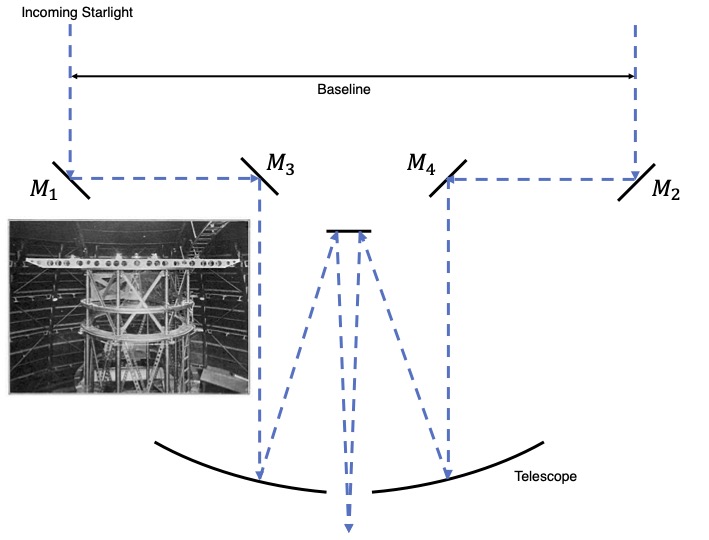}
        \caption[Schematic of Michelson's Stellar interferometer]{Schematic of Michelson's Stellar interferometer. The variable distance between the M1 and M2 mirrors in the baseline length of the system, which feeds light via M3 and M4 onto the primary telescope mirror where the light forms an interference pattern at the focal length. \textbf{Insert:} Stellar interferometer in place at the Mt Wilson observatory Hooker telescope \citep{hale1922new}.}
        \label{fig:StellarInterferometer}
    \end{figure*}

The first observations interfering light from celestial objects were conducted in 1891 and used to measure the diameter of Jupiter’s moons \citep{Michelson91}; only a few years later did Michelson, following the suggestion of Hippolyte Fizeau, propose a stellar interferometer in order to determine the diameter of stars \citep{Michelson21}. The first such design was installed at the 100-inch telescope at Mount Wilson observatory in California/USA. The geometry of the system consisted of four mirrors as shown in Figure\,\ref{fig:StellarInterferometer}, M1 and M2 collect light from the primary telescope mirror and are separated by a variable length $d$. The M3 and M4 mirrors redirect the light onto a detector. In this way light from same star can create an interference pattern. The movable primary mirrors allow for the baseline to be increased or decreased in order to find the point of maximum or minimum fringe contrast. By finding the baseline separation at which the fringes disappeared completely Michelson was able to calculate the diameter of Betelgeuse to be $18\pm2\,\mathrm{mas}$ \citep{Michelson21}. His measurement was surprisingly accurate given the only detector of fringe contrast was the human eye with modern estimates ranging from $42$ to $56\,\mathrm{mas}$ depending on observing wavelength \citep{Dolan16}. 

The major advantage of  astronomical interferometry is the very high achievable resolution. For a single dish telescope, the maximum resolution is defined by the Rayleigh criterion which states $R\propto \frac{\lambda}{D}$ where $D$ is the diameter of the telescope aperture. For the single dish telescope, two point sources are described as resolved when the principle diffraction maximum of the first point source coincides with the principle diffraction minimum of the second. In interferometry, the same two point sources are considered resolved when their fringe contrast is equivalent to zero, this occurs at $\frac{\lambda}{2B}$ where $B$ is the separation between the two telescopes. At this separation the two fringe packets created will be exactly out of phase, creating destructive interference. As such, the resolution of an interferometric array is independent of the apertures of the individual and instead depends of the distance between telescopes. This allows astronomer to achieve unprecedentedly high angular resolution. 

The first successful multi-aperture stellar interferometer was the Interféromètre à 2 Télescopes (I2T) \citep{Labeyrie75}. Initially located at Nice observatory and later moved to CERGA in southern France, it consisted of 2 small aperture telescopes separated by a 12\,m baseline. Fringe tracking was done manually using a micrometer screw to form fringes on a photon-counting television camera. First fringes were obtained on Vega in 1974 \citep{Labeyrie75} and has since been used for a wide array of science \citep[see][and reference therein]{Koechlin88}. However, its primary purpose was for the design and implementation of new technologies and techniques for creating modern interferometric arrays.

\begin{figure*}[h!]
    \centering
    \includegraphics[scale=0.5]{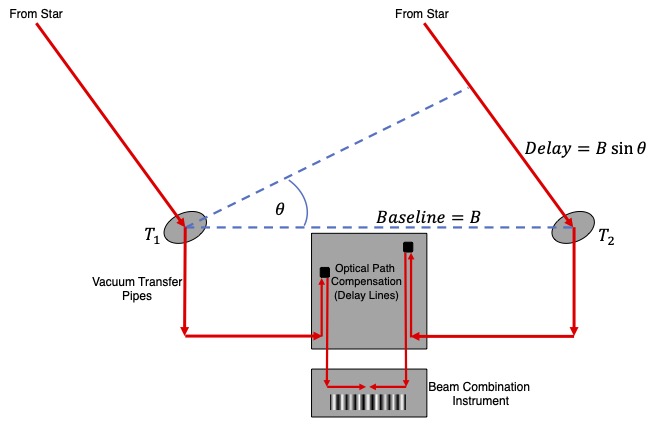}
    \caption[A schematic of a modern two telescope interferometer]{A schematic of a modern two telescope interferometer. Due to the large separation of the telescopes, light will be delayed arriving at one telescope with respect to others. This delay must be corrected in the optical path compensation laboratory before the light is passed to the beam combination instrument.}
    \label{fig:ModernInter}
\end{figure*}

A modern Michelson interferometer is significantly more complex than Michelson’s original design despite using the same basic principles. Figure\,\ref{fig:ModernInter} is a schematic of a simple two telescope interferometer. Light is collected from the two telescopes and brought into a central facility via vacuum pipes in order to remove the effects of atmospheric diffraction. Due to the large separation of the telescopes there is a significant delay in the time taken for light to arrive at one telescope with respect to another, making it impossible to form interferometric fringes. Such a delay must be compensated before the beams are combined, this is done through delay lines. Delay lines consist of mirrors mounted on carts which can move up and down rails to increase or decrease the optical path length of an individual telescope. The light is then passed into a beam combination instrument which interferes the light creating fringes which are read on a detector. 

In addition to Michelson stellar interferometers and a small number of observatories use a technique known as Fizeau interferometry, which does not require corrective delay lines. These facilities are more similar in design to the original Michelson-Morley experiment. Light is collected from multiple telescopes at a sparse aperture telescope which serves as the beam combiner which focus the combined light onto a detector. An example of a Fizeau interferometer is the Large Binocular Telescope Interferometer (LBTI) \citep{Rothberg20}.

The setup of a beam combiner instrument varies dramatically between different instruments which are capable of different spectral resolutions, wavelength modes and sensitivities. A brief summary of the current generation of facilities and instruments is provided in the following section.

\section{Overview of Interferometric Facilities} \label{sec:Facilities}

    Since the first observations using the Mt Wilson stellar interferometer multiple facilities dedicated to optical interferometry have sprung up around the world. The two most prolific interferometric sites are CHARA and VLTI, the key characteristics of which I discuss in this section. In addition, I discuss the KI and PTI sites which have direct relevance to the work in this thesis. This section is dedicated to optical observatories, for a summary on radio/sub-mm interferometric sites see \citet{Omont03,Sargent04}.

    \subsection{CHARA}

    On the same site as Michelson's original stellar interferometer now stands the CHARA (Centre for High Angular Resolution Astronomy) array \citep{Brummelaar05}. CHARA consists of six telescopes of 1 metre in diameter in a fixed Y-shaped configuration. This provides 15 baselines ranging from 34 - 331 meters in length and up to 20 possible closure phase triangles (see Section\,\ref{sec:ClosurePhases}). Light is passed from the telescopes to the centre beam combination facility with delay lines 40 metres long, allowing up to 80 metre of delay to be added. Such delay lines are particularly short for the long baselines of CHARA, but this is addressed using fixed delay mirrors known as PoPs (Pipes of Pan) which can easily add large amounts of delay in vacuum. 

    There are currently 5 instruments located within the beam combining laboratory operating across the visible and NIR. 

    \begin{itemize}

    \item CLASSIC and CLIMB are open air, broadband instruments optimised for sensitivity across the J, H and K band. CLASSIC is the original two-beam combiner at the array, while CLIMB can combine light from up to 3 telescopes \citep{Brummelaar13}. 

    \item Precision Astronomical Visible Observations (PAVO) \citep{Ireland08} is a 3-telescope visible beam combiner. PAVO spatially modulates fringes in a pupil plane before dispersing them with an integrated field unit, this allows PAVO to utilise the full multi-r0 aperture of the CHARA array over a standard $630-950\,\mathrm{nm}$ bandwidth. PAVO typically relies on fringe tracking from CLIMB in the NIR, allowing for high precision sensitivity-optimised observations. 

    \item The Visible spEctroGraph and polArimeter (VEGA) \citep{Mourard09} is a 3 telescope combiner designed for high resolution spectro-interferometry and can provide spectral resolutions of R = 6000-30,0000 across the visible wavelengths of 480-850 nm. The short operating wavelength of VEGA also makes it the highest spatial resolution of the array, down to 0.3 mas on the longest baseline. The successor SPICA (Stellar Parameters and Images with a Cophased Array) will combine light from all 6-telescopes while fringe tracking in the NIR.

    \item Finally, MIRC-X (Michigan InfraRed Combiner-eXeter) \citep{Anugu20} is currently the only 6 telescope infrared beam combiner in the world, allowing it to make use of all 15 baselines and 20 closure phase triangles simultaneously. MIRC-X operates in the H band with low/medium spectral resolutions between R = 22-190 and is equipped with a revolutionary sub-electron noise and fast-frame rate camera allowing for very precise and sensitive observations. A deeper discussion of MIRC-X and associated engineering activities can be found it Chapter\,5.

    \end{itemize}

    \subsection{VLTI}

    The other primary interferometric facility is located at the European Southern Observatories Very Large Telescope (VLT) at Paranal, Chile. The array consists of multiple telescopes used for a wide variety of astronomical purposes, just one of which is interferometry. Light can be collected from either the 4 static 8.2m unitary telescopes (UTs) or 4 movable 1.8m auxiliary telescopes (ATs) allowing for optimisation of either sensitivity or uv coverage. What the VLTI lacks in number and length of baselines (maximum baseline length of 110m) it makes up for in sensitivity with significantly better limiting magnitudes than CHARA instrument. This is primarily owing to larger telescopes and a fully operation adaptive optics system.

    There are currently 3 instruments in the beam combination laboratory operating at different wavelengths and spectral modes. PIONIER (Precision Integrated-Optics Near-infrared Imaging ExpeRiment) \citep{JBLB2011} is a H band 4 telescope combiner which operates at low spectral resolutions providing good sensitivity and very precise visibility and closure phase measurements. Secondly, GRAVITY \citep{GRAVITY17} is a K band instrument operating at both low and high spectral resolutions. It can provide exquisite precision, with relative astrometric accuracy down to $10\,\mathrm{\mu mas}$ and is capable of resolving spectral lines for spectro-interferometric observations. GRAVITY is equipped with a built-in fringe tracking system making it extremely sensitive, ideal for observing very faint targets. Finally, MATISSE \citep{Lopez14} is a MIR multi-wavelength instrument which is capable of L, M and N band observations allowing for maximum angular resolution ($\lambda/2B$) between 3.5 and 8mas depending on wavelength. MATISSE is also capable of both high and low spectral resolution observations, allowing for both faint target work and spectro interferometry. Work is underway to allow for GRAVITY to be used as a fringe tracker for MATISSE to allow for even fainter observations in the MIR in a project known as GRAV4MAT \citep{Gonte16}. 

    \subsection{Other Facilities}

        In addition to CHARA and VLTI there are several historic, current, and future interferometric sites around the world including IOTA \citep{Carleton94}, SUSI \citep{Davis99}, NPOI \citep{vanBelle20}, and MROI \citep{Buscher13} amongst others. However, I will only discuss the sites relevant to the scientific work contained in this thesis, the Keck interferometer and Palomar Testbed Interferometer.

        \subsubsection{PTI}

        The Palomar Testbed Interferometer (PTI) \citep{Colavita99} was originally intended as a test site for the Keck interferometer, but became a widely used facility in its own right. PTI consisted of three 40cm telescopes which could be combined pairwise on baselines up to 110m in length. It operated in both the H and K bands with no spectral resolution. First fringes were obtained in 1995, and the site remained in operation until 2008, many of the learnings and techniques were later adopted in the construction of the Keck interferometer. 

        \subsubsection{KI}

        The Keck observatory at the summit of Mauna Kea Hawaii consists of two of the largest telescopes in the world, each with 10m apertures. Operational from 2003 to 2012, the Keck Interferometer (KI) \citep{Colavita13,Eisner14} combined light from the two telescopes situated 84m apart in a central laboratory. A variety of instruments allowed KI to operate in H, K and L band modes with an additional nulling interferometry instrument. The sheer size of the telescope apertures allowed for high sensitivity observations of faint objects.

\section{Observables in Interferometry}

    The process of extracting observables from interferometric data is significantly more complex than most other areas of astronomy. One cannot simply measure the flux in a few pixels or ogle a directly imaged nebula, an often-painstaking processes must be endured to extract any useful information for the interference fringes available.  

\begin{figure*}[h!]
        \centering
        \includegraphics[scale=0.48]{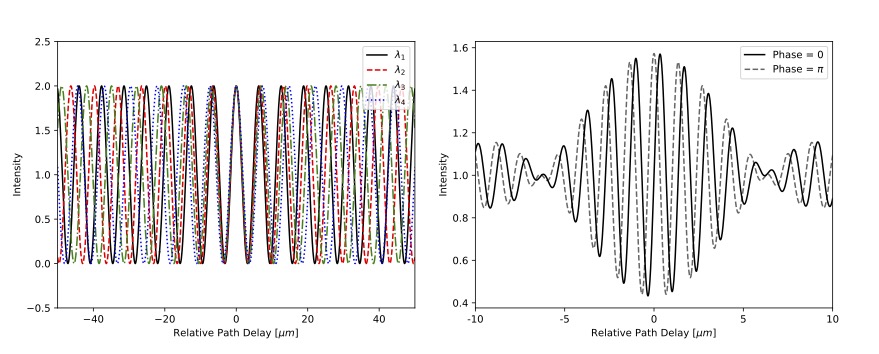}
        \caption[Fringe pattern for monochromatic and polychromatic sources]{LEFT: Individual monochromatic responses for a simple two element interferometer. The different colors represent arbitrary, individual, but adjacent wavelengths. RIGHT: The resulting fringe pattern for a polychromatic source centered on $0.75\,\mathrm{\mu m}$ with a bandpass of $0.1\,\mathrm{\mu m}$ and a visibility amplitude of 0.5. Solid line exhibits zero phase measurement, dashed line is shifted by $\pi$ radians. Recreated from \citet{Thiebaut17}.}
        \label{fig:Phases}
    \end{figure*}

    There are two key aspects of an observed interference pattern, the fringe amplitude and fringe phase. The fringe amplitude is simply a measure of the fringe contrast between the light and dark regions and is described as: 

    \begin{equation}\label{eq:Vis}
    \centering
        \mathcal{V_{\mathrm{amp}}} = \frac{I_{\mathrm{min}}-I_{\mathrm{max}}}{I_{\mathrm{max}}+I_{\mathrm{min}}} ,
    \end{equation}

    where $\mathcal{V_{\mathrm{amp}}}$ is the visibility amplitude (also known as the Michelson visibility) and $I_{\mathrm{min}}$ and $I_{\mathrm{max}}$ are the minimum and maximum intensity of the interference pattern. From this equation it follows that visibility is normalised between 0 and 1 and a decreased fringe contrast corresponds to a lower visibility and greater fringe contract to a higher visibility. The physical interpretation of visibilities is described below in section\,\ref{sec:Visibilites}. In Figure\,\ref{fig:Phases}, the fringes on the left have a visibility amplitude equal to one.

    Fringe phase is a measure of the location of the central fringe with respect to the location of zero optical path difference (OPD), The phase offset is measured in radians, with a phase of $2\pi\,\mathrm{radians}$ corresponding to a whole fringe period. The right panel of Figure\,\ref{fig:Phases} depicts a fringe panel of zero phase as a solid line as the central fringe located at the position of zero OPD, the dashed line has been shifted by $\pi$ radians. 

    \subsection{Complex Visibilities} \label{sec:Visibilites}

    Visibilities are the most basic measurement available to any interferometer. An interference fringe has two properties a visibility amplitude and a visibility phase, both contained within a complex visibility measurement. As described above visibility amplitude is simply a measure of the fringe contrast between the bright and dark regions of the interference pattern between two telescopes.

    However, this description of visibilities does not describe the brightness of the astronomical object being observed. In order to explore this the Van Cittert-Zernike theorem \citep{Cittert34,Zernike38} provides a physical description of visibilities. In short, it states that the Fourier transform of the intensity distribution (object geometry) of a distant, incoherent source is equal to its complex visibility. Mathematically this corresponds to
    \begin{equation}\label{eq:VanCZ}
        I(\alpha,\beta) = \iint_{-\infty}^\infty\mathcal{V}(u,v)\mathrm{exp}(2\pi i(\alpha u+\beta v))~dudv ,
    \end{equation}
    where $\alpha$ and $\beta$ represent the angular coordinates on the sky and $u$ and $v$ are coordinates in Fourier space describing the baseline $\alpha$ and $\beta$, and therefore the spatial frequencies of the brightness distribution. One can see that this equation can very simple be re-written as the Fourier transform:
    \begin{equation}\label{eq:fourier}
        \mathcal{V}(u,v) = \mathcal{F}(I(\alpha,\beta)) ,
    \end{equation}
    This simple relation allows for information about the intensity distribution of the target to be directly extracted from visibility measurements. 

    The difficulty with visibility measurements lies in their analysis, as understanding measurements taken in Fourier space is not an intuitive process. It is only though inverse modelling or more complex image reconstruction that an appreciation can be gained for the original intensity distribution of the object. A description of modelling techniques, including visibility modelling can be found in Section\,\ref{sec:ModellingDescription}.
    
    \subsection{Closure Phases} \label{sec:ClosurePhases}

    While the measure of fringe phase is mathematically relatively simple, the measurement of phases in astronomy is a notoriously difficult endeavor owing to the presence of a turbulent atmosphere, mirror reflectivity, local scintillation and other effects. Turbulence above any individual telescope, in addition to differing optical systems will introduce a phase delay that is specific to that telescope. As such the phases measured by two separate telescopes will be significantly different, even if the telescopes are only separated by a few meters. This will manifest as a phase shift in the measured phase of a detected fringe. 

    However, \cite{Jennison58} introduced a method of calculating a reliable phase measurement based on a closed triangle of telescopes. Given three telescopes, A, B and C, where a phase delay is introduced above telescope B. This will induce a phase shift in the fringes between telescopes A-B and an equal but opposite phase shift between B-C. Hence, one can sum up the phase differences of A-B, B-C and C-A. In this way, an atmospheric phase delay above any single telescope will be subtracted from the final phase measurement. This closed triangle of telescopes leads to the name of closure phases. The closure phase $\Phi_{ABC}$ can therefore be written as 

    \begin{equation}\label{eq:CloPha}
      \Phi_{ABC} = \phi_{AB}+\phi_{BC}+\phi_{CA} ,
    \end{equation}

    where $\phi_{AB}$ represents the measured Fourier phase for the baseline connecting telescopes A and B. Closure phases were first carried out at optical wavelengths in aperture masking experiments \citep{Baldwin86,Haniff87,Readhead88}. The first facility to apply closure phase techniques to optical interferometry was the Cambridge Optical Aperture Synthesis Telescope (COAST) facility in 1996 \citep{Baldwin96}. Since then closure phase measurements have become standard practice for many interferometric facilities. 

    These closure phases encode the spatial structure of the source, in particular deviations from point symmetry. A resolved point source or other point symmetric object will have closure phases equivalent to $0^\circ$ or $180^\circ$. In addition, an unresolved object cannot exhibit non-zero closure phases. Any deviation from 0 or 180 degrees implies an asymmetric intensity distribution. Closure phases are therefore immensely powerful observables for precision astrometry and imaging techniques. 
    
    \subsection{Differential Phases}

    Another method of extracting phase information from interference is the derivation of differential phases. Differential phases are simply a measure of the change in the object phase between adjacent spectral channels and generally rely on the presence of emission/absorption lines within the spectrum of the object. When looking at individual emission lines, the line width is very narrow compared to the bandwidth of the emission frequency. As such, the phase error introduced by the atmosphere is expected to be nearly constant across the spectral line \citep{Monnier07}. Thus, it is possible to measure the visibility phase of a spectral line relative to the continuum by subtracting the continuum phase measured on either side of the spectral line. 

    Differential phases can be a powerful tool in interferometry as they allow for the study of kinematics within the objects, by selecting spectral lines that trace such processes. As an example, the $Br\gamma$ emission line in the K band is thought to inflow/outflow processes in protoplanetary disks \citep[see][]{Hone17,Kraus12b}.  

    While differential visibilities can be very useful, they rely on the presence of strong, emission or absorption lines. In order to spectrally resolves such lines, high spectral resolution observations are required which limits the sensitivity of interferometry.

\section{Geometric Modelling} \label{sec:ModellingDescription}

    The understanding of interferometric variables is not intuitive in the way that many astronomical variables are. The ability to visualise geometric structures in their Fourier space is reserved only for those who have spent many years in interferometric studies. The analysis of the data is therefore intrinsically inverse, such that models are adapted in order to fit the data. Visibilities are often expressed as function of baseline length or a combination of baseline and observational wavelength known as spatial frequency, given by $SF=B/\lambda$, were $SF$ is the spatial frequency. These visibility profiles can provide direct information such as the characteristic size and basic morphology of an object. In order to interpret visibility profiles quantitatively, analytic models must be applied to model simple structures such as point sources, ring and Gaussian brightness distribution. The brightness distribution of these models can then be Fourier transformed in order to extract visibilities as shown in Equation\,\ref{eq:VanCZ}. 

    It is often necessary to combine two or more such models in order to effectively model an object, for example a point source and ring in order to model a star and disk system.  In this case, the linearity property of Fourier transforms can be employed to compute model visibilities:
    \begin{equation}\label{eq:MultiVis}
        \mathcal{F}\left( \sum_{j=1...n} \left(I_j\left(x,y\right)\right)\right) =  \sum_{j=1...n} \left( \mathcal{F} \left(I_j\left(x,y\right)\right)\right),
    \end{equation}
    Therefore, the visibility of a complex brightness distribution is the normalised sum of the visibilities of the individual components that make up that brightness distribution. In addition to the multi-component nature, most objects are also inclined with respect to the observer, meaning the orientation must be considered. These projection effects must be accounted for by de-projecting the baselines. A complete description of geometric modelling of visibilities is given by \citet{Berger03b}, which also contains mathematical expressions for commonly used models.

\begin{figure*}[h!]
        \centering
        \includegraphics[scale=0.48]{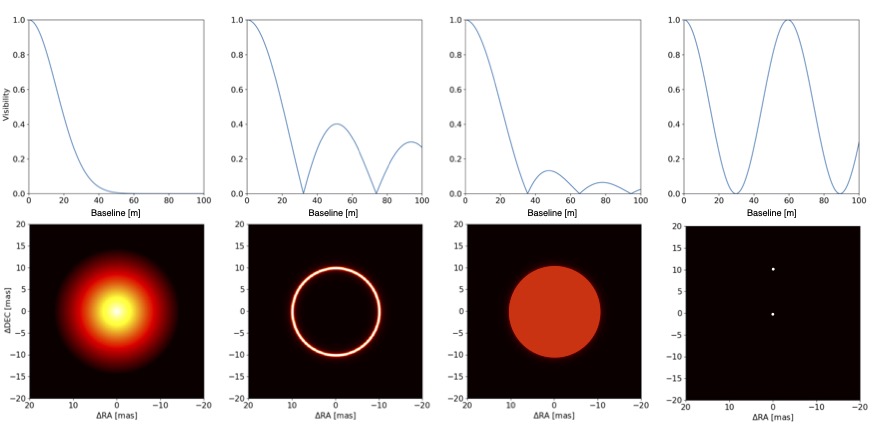}
        \caption[Visibility profiles and associated brightness distributions for simple geometric models]{Visibility profiles and associated brightness distributions for simple geometric models. From left to right: A Gaussian with a full-width-half-maximum of 10\,mas, a ring with diameter of 10\,mas, a uniform disk of diameter 10\,mas and a binary of separation 10\,mas of equal flux contribution. The visibility profile over the first 30\,m is almost identical between models, it is only at longer baselines the difference can be discerned. None of the models suffer from any projection effects, with $0^\circ$ inclinations and position angles. The baseline for all visibility calculations in the north-south direction. Brightness distribution colours are normalised flux and are not to scale between models.}
        \label{fig:GeoModels}
    \end{figure*}

    Geometric modelling is a very powerful technique that allows an observer to get an appreciation for object morphology from relatively few data points. The more complex the geometry of a system the greater the number of observables required to disentangle analytic models. As shown in Figure\,\ref{fig:GeoModels} the defining feature between different model’s visibility profiles is only apparent at longer baselines, while projection effects require a more expanse uv coverage to fully interpret. 

    Geometric modelling is particularly useful for the interpretation of smaller data sets of simple objects. However, for studying finer details and more complex datasets, such modelling is not always sufficient as the models lack robust 3-dimensional information. In order to study greater details, either more complex models are needed through radiative transfer work or model independent processes such as image reconstruction must be employed.  

\section{Image Reconstruction}

    The reconstruction of an interferometric image is not as simple as performing an inverse Fourier transform on the data. The observables of optical interferometry provide an incomplete sampling of the uv plane owing to the discrete possible baselines and position angles. The voids of the uv plane must therefore be filled in with assumptions. Additional problems arise from the lack of Fourier phase information, with only closure phases available. As such, image reconstruction becomes a complex compromise between fitting available data and keeping the image as regularised or simple as possible. However, used in a careful and considered way, image reconstruction is a powerful, model independent analytical tool. In this section I shall describe the basic process of image reconstruction and outline some of the commonly used algorithms. 

    \subsection{Basic Principles}

    Because of the incomplete sampling of the uv plane, the measured data can be thought of as the true, complete distribution $\mathcal{V}(u,v)$ multiplied by some sampling function $S(u,v)$. To this we can apply the convolution theorem to state that the Fourier transform of the sampled distribution (also known as the dirty map $I_D(x,y)$) is equal to the convolution of the Fourier transform of the true source distribution image $I(x,y)$ with the Fourier transform of the sampling function (also known as the dirty beam $B_D(x,y)$). For the cases where the full complex visibility is known:
    \begin{equation}\label{eq:Dirty}
        I_D(x,y) = I(x,y) \otimes B_D(x,y) \,\mathcal{F} \left( V(u,v) \times S(u,v) \right)  ,
    \end{equation}
    where $\otimes$ represents convolution and $\mathcal{F}$ indicates the Fourier transform. The process of obtaining an approximation of the true source distribution image therefore becomes a problem of deconvolution of the dirty map by the dirty beam. However in optical interferometry, the full Fourier phase is not a known quantity, as we are limited to only closure phase measurements. Image reconstruction algorithms can be designed following the same inverse problem approach.

    The primary factor hampering deconvolution is the sparse sampling of spatial frequencies meaning that fitting data alone does not define a unique image solution.  Such a problem can be solved by an inverse problem method by imposing a priori constraints to select the most likely image among all those which are consistent with the data. The requirements for such priors are that they must smoothly interpolate across the uv plane while avoiding the highest frequencies beyond the diffraction limit. These prior constraints can be monitored within the term $f_{\mathrm{prior}}(x)$ of image $x$ which measures the agreement between the image and the priors, the lower the value the better the fit. The term $f_{\mathrm{prior}}(x)$ is often called regularisation, the two most commonly used algorithms are \textit{MEM} and \textit{CLEAN}

    Maximum entropy methods (MEM) are based on the idea of obtaining the least informative image which is consistent with the data. This is the same as maximising the entropy within the image in order to maintaining simplicity. This amounts to a minimising of the term $f_{\mathrm{prior}} = -S_{\mathrm{entropy}}(x)$ where the entropy $S_{\mathrm{entropy}}(x)$ measures the informational content of an image $x$. One of the most common expressions considered for this is 
    \begin{equation}\label{eq:MEM}
        f_{prior}(x) = \sum_j \left[ x_j \mathrm{log}\left( x_j/\bar{x}_j \right) - x_j + \bar{x}_j \right]   ,
    \end{equation}
    where $\bar{x}$ is some default image, either a flat, a previous reconstruction or a lower resolution image of the same set. The subscript $j$ refers to the $j-th$ pixel in the image \citep{Thiebaut17}. In order to enforce some correlation between close pixels in the image, some averaging or smoothing linear operator is usually chose to prevent truly maximum entropy. 

    The \textit{CLEAN} algorithm is another method employed which favours images with a limited number of significant pixels to attempt to preserve simplicity \citep{Hogbom74}. This method is iterative and follows the prescription: Given a dataset as a dirty image, the location of a point source which best fits the data is found. The model image is then updated by a fraction of this intensity. This fraction is then subtracted from the dirty image and the procedure is repeated with this updated dirty image. When the level of the residuals in the image become smaller that a given noise threshold the model image is convolved with a Gaussian PSF to set the resolution of the uv plane. Once most point sources have been removed, the residual dirty image is essentially due to the remaining extended source. Adding the residual dirty image to the clean image produces a final image consisting of compact sources plus smooth extended components. 

    \subsection{Reconstruction Algorithms}

    Over the past few decades several algorithms have been developed to reconstruct images from interferometric data. Notably \textit{MiRA} \citep{Thiebaut13} and \textit{BSMEN} \citep{Buscher94}. \textit{MiRA} or the Multi-aperture Image Reconstruction Algorithm uses the minimisation of both $f_{\mathrm{prior}}(x)$ and $f_{\mathrm{data}}(x)$, where $f_{data}(x)$ measures the agreement between the model and the data, making use of a very wide range of regularisation techniques including MEM, smoothness and total variation. As \textit{MiRA} is purely based on an inverse problem approach it can cope with incomplete data sets, such as those missing phase information.

    \textit{BSMEM} is the Bi-spectrum maximum entropy method and uses MEM to regularise the problem of image reconstruction to the complex bispectrum. The strength of BSMEM is that it makes no attempt to convert data into complex visibilities. As such it can handle any type of missing data, including closure phases. 

    In order to autonomise the process of reconstruction as much as possible, the different algorithms have been implemented into pipelines. One such pipeline is the Poly-Chromatic Image reconstruction Pipeline (PIRP), as outlined by \cite{Kluska14}. PIRP primarily makes use of MiRA for image retrieval, but adds the Semi-Parametric Approach for image Reconstruction of Chromatic Objects (SPARCO) which parameterises the chromaticity of the stellar emission from the interferometric data across the spectrum to reconstruct images based on different temperatures producing different wavelength. It is shown that using SPARCO improves the fit of the image to the data by a factor of 9 using this method \citep{Kluska14}, thus proving it is highly effective at accurately reproducing the data.

\newcommand\blfootnote[1]{%
  \begingroup
  \renewcommand\thefootnote{}\footnote{#1}%
  \addtocounter{footnote}{-1}%
  \endgroup
}
\chapter{Engineering Activities with MIRCX}\blfootnote{Parts of this chapter are published in \citet{Labdon2020b}}
\label{ch:Engineering}

A significant portion of my work during the course of this PhD has been dedicated to instrumentation activities in conjunction with the both the CHARA array and the MIRC-X instrument. My instrumentation work can be split into two projects. Firstly, to expand the operational wavelength of MIRC-X to include the J band in addition to the H band. The first scientific results of this work are described in section Chapter\,8. In this section, I shall describe this instrumentation work. Secondly, the development of a next generation baseline solution for CHARA in order to provide more efficient observations for all instruments.

\section{J band Interferometry with MIRC-X}\label{sec:J_intro}  

	The Michigan Infrared Combiner (MIRC) was a six-telescope infrared beam combiner at the CHARA telescope array \citep{Monnier04}, the world’s largest baseline interferometer in the visible/infrared, located at the Mount Wilson Observatory in California. In the summers of 2017 and 2018 MIRC underwent significant upgrades, with the newly commissioned instrument being named MIRC-X. The commissioning of MIRC-X occurred in two phases, the first of these phases involved the implementation of an ultra-fast, ultra-low read noise near-infrared camera \citep{Anugu18} and an overhaul of the MIRC control software. The second phase involved the replacement of the optical fibres and beam combination element in addition to the commissioning of polarisation controllers. A detailed description of the upgraded MIRC-X instrument can be found in \citet{Kraus18,Anugu20}.

	Over the past 15 years MIRC/MIRC-X has produced many outstanding results in a variety of areas of astrophysics including the imaging of stellar surfaces of rapid rotators, eclipsing binary systems and star spots on the surfaces of highly magnetically active stars. \citep{Monnier12,Roettenbacher17,Schaefer19,Chiavassa20}. In more recent years, the upgrades have allowed for observations of young stellar objects (YSO) with all six telescopes for the first time \citep{Kraus20,Labdon20} and for high precision single-field astrometry for the detection of Hot Jupiter objects in binary systems \citep{Gardner20}.

	Historically MIRC-X has only operated in the H band, however, one of the ultimate aims of the upgrades was to conduct dual H and J band observations. The J band ($1.1-1.4\,\mu m$) is a relatively untapped resource in long-baseline interferometry, but one with huge potential for scientific discoveries. It allows access to the photosphere in giant and super-giant stars relatively free from opacities of molecular bands. In addition, the J band traces the warmest and smallest scales of protoplanetary disks, where accretion and viscous heating take place. Finally, the J band allows for higher angular resolution observations than near and mid-IR (as $R=\lambda/B$) allowing us to probe the smallest scales of astrophysical objects.
	 
	In this section I present a summary of the work done to commission the J band for observations at the MIRC-X instrument. In section \ref{sec:J_LDCs} I discuss the J band specific hardware improvements required, while in sections \ref{sec:J_software} I summarise the corresponding software developments undertaken in this endeavor. In section \ref{sec:J_pipeline} I discuss the changes and updates needed in order to reduce and calibrate the first scientific observations.

	\subsection{MIRC-X Filters}\label{sec:J_filters}

		Beyond the major instrument upgrades in 2017 and 2018 mentioned in the introduction, various other hardware requirements were needed in order to conduct J band observations. In this section I discuss this work and the design decisions involved. Operating in standard observing mode MIRC-X employs an H band filter with central wavelength $\lambda = 1.61\,\mu m$ and bandwidth $\Delta\lambda = 0.27\,\mu m$. However, in order to observe in the J band a new filter which extends to a lower wavelength regime must be employed. However, this is hampered owing to the presence of a metrology laser at CHARA. This laser operates at $1.319\,\mu m$ and is responsible for providing accurate metrology of the delay line cart position, a vital role at any optical interferometer. This laser wavelength lies in the upper J band and so will saturate the detector in addition to contaminating stellar flux.

\begin{figure} [ht]
		   \begin{center}
			 \begin{tabular}{c} 
		   \includegraphics[scale=0.4]{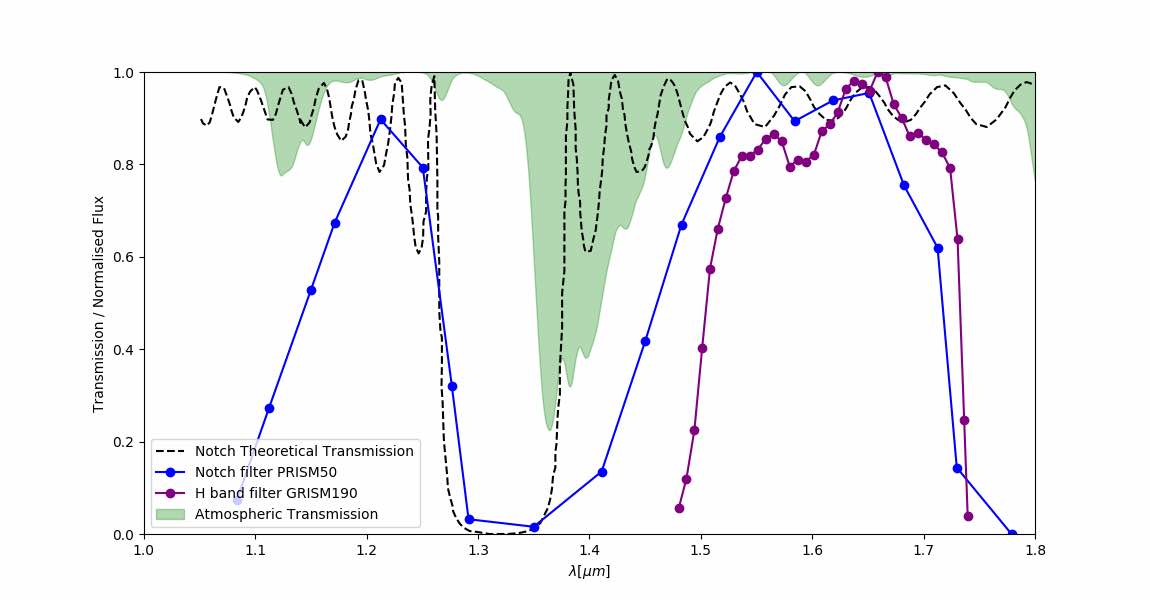}
		   \end{tabular}
		   \end{center}
		   \caption[Theoretical and measured transmission of the notch filter across the H and J bands]{ \label{fig:Atmosphere} Theoretical and measured transmission of the notch filter across the H and J bands. The dashed black curve represents the theoretical transmission curve of the filter. The solid green area is the atmospheric transmission. The blue line with points is the normalised flux measured using the MIRC-X detector and the CHARA STS through the notch filter and the PRISM50. The solid purple line with points is the normalised flux measured using the MIRC-X detector and the CHARA STS through the H band filter and the GRISM190, the current highest spectral mode of MIRC-X.   }
		 \end{figure}

		This was overcome with addition of a notch or band-stop filter. These filters reject/attenuate signals in a specific wavelength band and pass the signals above and below this band. In this way I can reject light from the CHARA metrology laser and accept stellar contributions from the H band J bands above and below this respectively. The theoretical and measured characteristics of the notch filter I employed is shown in Figure\,\ref{fig:Atmosphere}. The rejection factor at the metrology laser wavelength was found to be sufficient as no laser signal is seen on the detector, indicating it to be below the level of the background noise.

	\subsection{Longitudinal Dispersion Compensation}\label{sec:J_LDCs}

		Long-baseline optical interferometry poses unique engineering challenges. One of the most demanding aspects is the requirement for optical path length compensation. The path length of the various light beams are not identical for each telescope, owing to the position of the scope and position of the astronomical object on the sky. As such, the path difference much be corrected in order to ensure coherence for interferometric fringes. At CHARA this is done using two methods, the first is to employ fixed amounts of delay which can be moved in and out of the light path using retractable mirrors. These are known as Pipes of Pan or PoPs. The second method is to employ carts affixed to delay lines which track the geometric phase difference as a stars moves across the sky. The difficulty with the second method is that it is very difficult to achieve in a vacuum, due to the size and number of moving parts. This introduces a large amount of air into the system, potentially up to $80\,m$ difference in air path between the beams. This air leads to atmospheric, chromatic dispersion. 

		In standard single band observing, this does not pose a problem due to the small wavelength range. However, in dual band observations this become a major problem. If delay line carts are tracking H band fringes, the J band light will be incoherent due to chromatic dispersion and fringes will be severely reduced contrast to the point of being un-observable. Therefore a method of correcting for atmospheric dispersion must be employed. At CHARA, this is in the form of Longitudinal Dispersion Compensators (LDCs). 

\begin{figure} [ht]
		   \begin{center}
			 \begin{tabular}{c} 
		   \includegraphics[scale=0.45]{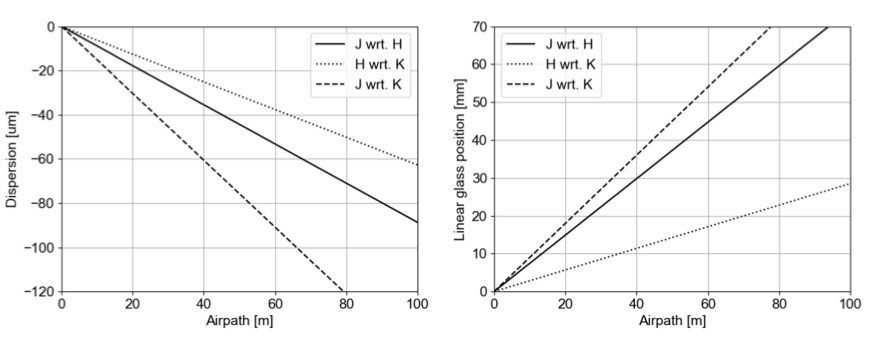}
		   \end{tabular}
		   \end{center}
		   \caption[Modelled atmospheric dispersion across the NIR and LDC correction required to compensate dispersion] { \label{fig:AirDisp} LEFT: Theoretical atmospheric dispersion between the NIR bands covered by MIRC-X and the upcoming MYSTIC K band instrument. Group delay measurements are used an analogy for chromatic dispersion. The J band is taken as $1.2\,\mu m$, the H band as $1.55\,\mu m$ and the K band as $2.2\,\mu m$. The solid line represents the dispersion in the J band with respect to the H, the dotted line the dispersion in H with respect to K and the dashed the dispersion in J with respect to K. RIGHT: The linear LDC glass movement required to correct the dispersion between the J, H and K bands. The wavelengths adopted and line meaning are the same as shown on the left.  }
		\end{figure}

		The LDCs consist of wedges of glass that can move in and out of the beam to increase or decrease the thickness of glass in the beam path. The characteristics of the LDCs are described in great detail in \citet{Berger03} when the LDCs were commissioned in order to enable simultaneous near-infrared (NIR) observations with the CLIMB instrument and visible observations with PAVO \citep{Ireland08}. Despite not being designed for use between the H and J bands, the choice of Schott-10 glass proved ideal for use with MIRC-X. Shown in Figure\,\ref{fig:AirDisp} is the expected dispersion between the NIR wavebands and the expected linear movement of the glass to successfully correct for this. The thickness of glass required for a given air path can be calculated using
		\begin{equation}
			\label{eq:glass_per_m}
			\Delta L = \Delta x \times \frac{n_a(\sigma_1) - n_a(\sigma_2) + \sigma_1 n_a^{'}(\sigma_1) - \sigma_2 n_a^{'}(\sigma_2)} {n_g(\sigma_1) - n_g(\sigma_2) + \sigma_1 n_g^{'}(\sigma_1) - \sigma_2 n_g^{'}(\sigma_2)} \, .
		\end{equation}
		
		Here, $\Delta L$ is the thickness of glass, $\Delta x$ is the air path in meters, $n_a$ is the refractive index of air and $n_g$ is the refractive index of the LDC glass. The $n_a^{'}$ and $n_g^{'}$ terms are the derivatives of the air and glass refractive indicies respectively. $\sigma_1$ and $\sigma_2$ refer to the wavenumbers of the two observing wavelengths \citep{Berger03}. In our model the refractive index of air was calculated using a composite of \citet{Ciddor96} and \citet{Mathar07} models in order to cover the full NIR. Atmospheric conditions, such as temperature and humidity were taken to be average values for the CHARA laboratory. 

		Overall the LDCs are capable of correcting for atmospheric longitudinal dispersion between the H and J bands, making simultaneous observations possible.

	\subsection{Software control} \label{sec:J_software}

		A variety of software developments were required in order to conduct these observations. In addition to changes to the real-time operating code of MIRC-X/CHARA, the reduction pipeline had to be modified, as outlined below. Also, I independently verified the data products by calibrating known calibrator stars and ensuring the stability of the transfer function.

		\begin{figure} [ht]
		  \begin{center}
			\begin{tabular}{c} 
		   \includegraphics[scale=0.25]{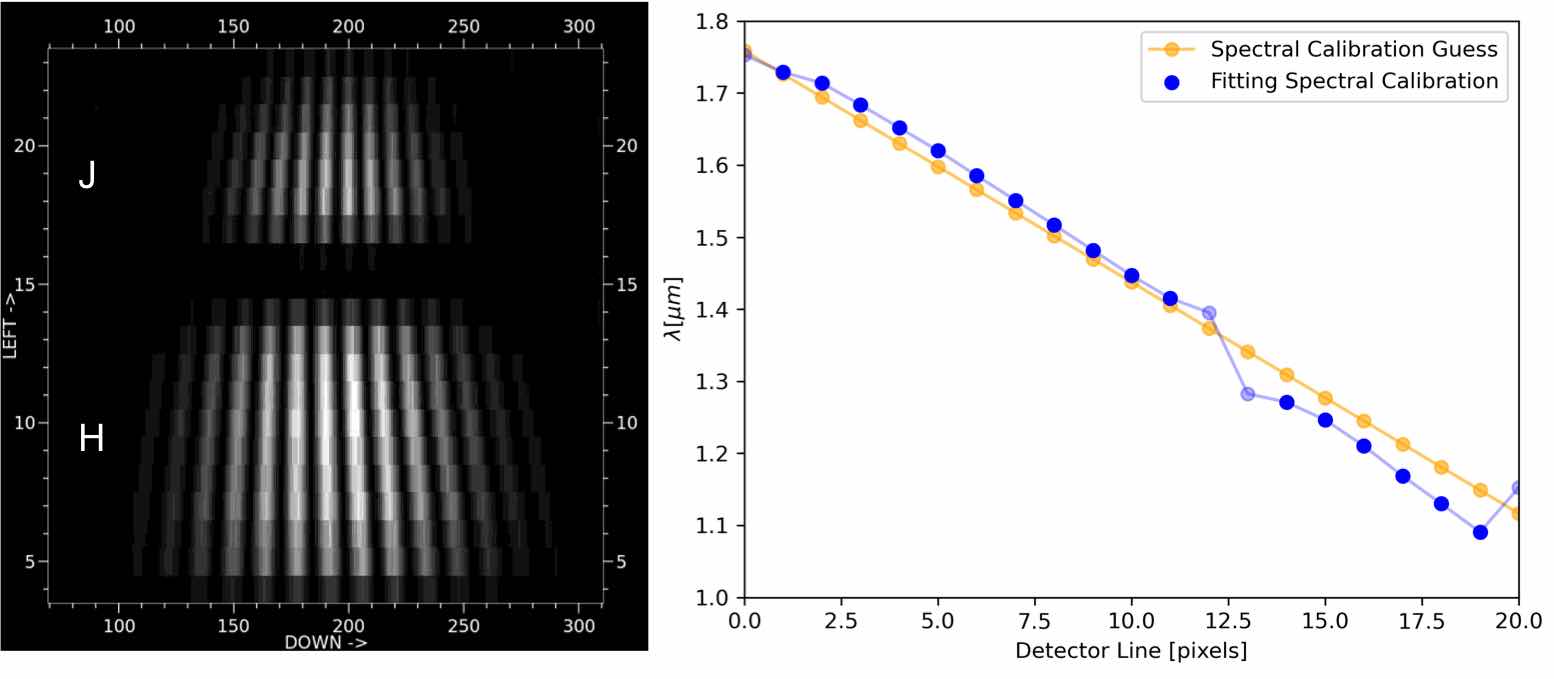}
		   \end{tabular}
		   \end{center}
		   \caption[MIRC-X J+H band observing window and resultant spectral calibration of data]{ \label{fig:speccal} LEFT: MIRC-X detector window showing laboratory induced fringes along a single baseline. The wavelength increases towards the bottom, meaning the J band is on top. The notch filter rejection window is clearly visible between the H and J bands at $\sim1.3\,\mathrm{\mu m}$. RIGHT: Spectral calibration of the MIRC-X simulated data. The initial guess of the spectral calibration is shown in pale orange and the fitted calibration is shown in blue. The translucent blue points represent those that were rejected by the pipeline due to low flux.}
		\end{figure}

		The control of the LDCs is based upon that developed by \citet{Berger03} for controlling dispersion between the visible and NIR. This software was adapted and expanded by adopting the dispersion models described in section\,\ref{sec:J_LDCs}. The basic function of the software is to receive the delay line cart positions in order to calculate the total amount of atmosphere in the system. Using equation\,\ref{eq:glass_per_m} it  calculates the thickness of glass required to correct for this dispersion, which is then translated into the linear movement for the LDC. This is done at a rate of one LDC per second, meaning a full movement cycle of the LDCs is 6 seconds. Theoretically a fast update rate is possible given the technological constraints of the network and translation stages. However, $1\,\mathrm{s}$ rate is sufficient given the slow rate of change of the total air path length in the system.

	\subsection{Data Reduction and Calibration Pipeline}\label{sec:J_pipeline}

		The data reduction pipeline for MIRC-X is already well established and publicly available \citep{Anugu20}. It has a proven track record through several publications. However, it was not designed to reduce J band data and so several changes had to be made to the existing code. 

		The first change to the data reduction pipeline consisted of adjusting the fringes and photometry windows to accurately detect and account for the J band flux. Previously the flux window had been detected by fitting a simple Gaussian to the detector flux, however, due to the presence of the notch filter the dual-band data is double-peaked. Figure\,\ref{fig:speccal} shows the detector window of MIRC-X covering both the H and J bands, the spectral range of the J band is just over half that of the H band.


		The second change comes from the spectral calibration, that is the process of determining the wavelength of each spectral channel. The wavelength is assigned by fitting the frequency of the fringe patterns, based on an initial guess. The initial guess is based on the known spectral dispersion of the prism used, in this case $R = 50$. Figure\,\ref{fig:speccal} shows the spectral calibration result obtained from the MIRC-X data reduction pipeline for artificially produced interference fringes. Both the initial guess and the final fitted spectral calibration are shown.  

		\begin{figure} [ht]
		  \begin{center}
			\begin{tabular}{c} 
		   \includegraphics[scale=0.38]{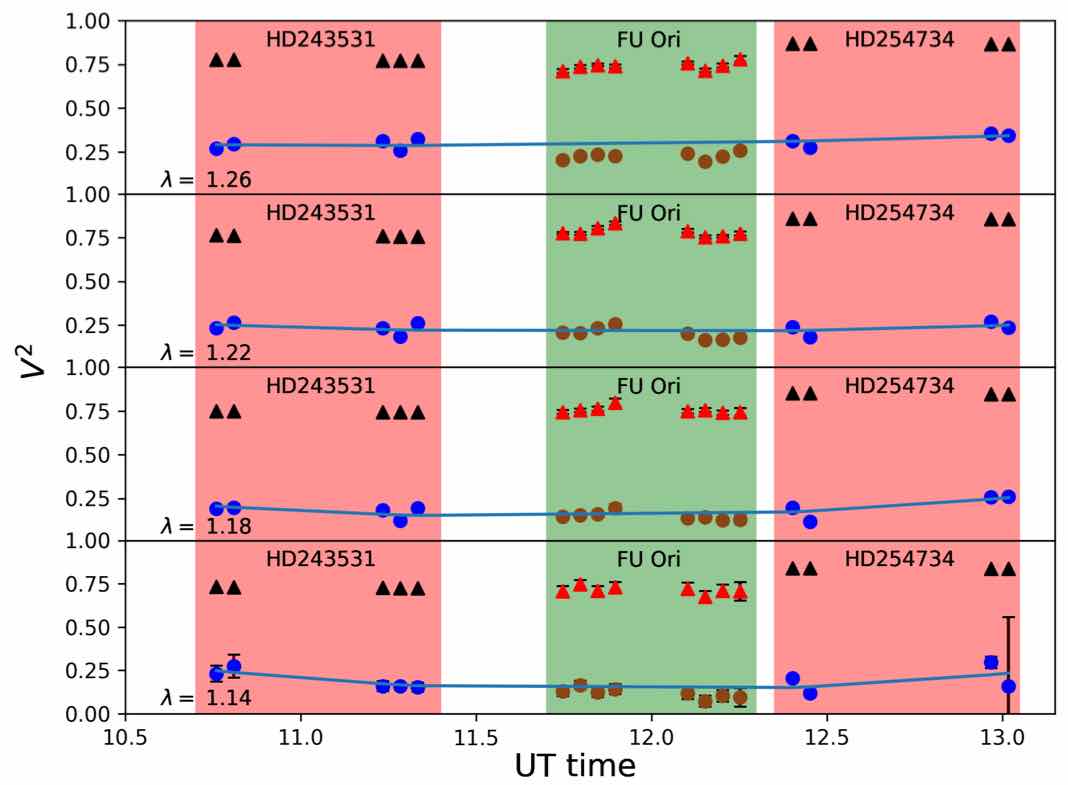}
		   \end{tabular}
		   \end{center}
		   \caption[Calibration transfer function for 2019 November 19]{ \label{fig:TFucntion} Calibration transfer function for 2019 November 19. The red-shaded areas are those representing calibrators and green are those representing science observations of FU\,Ori \citep{Labdon20}. these are along a single baseline (S1-S2) for the four J band spectral channels. The blue and brown points are the raw, uncalibrated squared-visibilities for the calibrators and science target respectively. The black and red triangles are the calibrated squared-visibilities for the calibrators and science target respectively. The blue represents the transfer function across the calibrators. }
		\end{figure}

		In order to ensure the validity of the data products, careful calibration tests were undertaken. The calibration procedure was unchanged from the standard MIRC-X pipeline as described in great detail elsewhere \citep{Anugu20}. Figure\,\ref{fig:TFucntion} shows the transfer function on a single baseline across the three hour observing period on the night of 2019 November 19. The 4 wavelength channels are those across the J band only, as the calibration of the H band channels has already been verified in previous papers. The linearity of the transfer function shows the stability of the observing across the 2.5 hour period. Large jumps in the transfer function on short timescales would indicate instability/inaccuracies in the dispersion control subsystems.

	\subsection{Impact and Next Steps} \label{sec:ImpactSteps}

		The impact of this work has been to demonstrate that dual H and J band observations are possible with MIRC-X and can produce scientifically valid, valuable results. I have shown that atmospheric chromatic dispersion can be corrected for using longitudinal dispersion compensators and the results can be reduced and calibrated effectively. This work has paved the way for the H+J observing mode to be commissioned for public use in the near future.

		There are some additional steps in the commissioning of the J band observing mode as one available to the community. Firstly, in early 2021 new LDCs were installed at CHARA in order to support the upcoming SPICA instrument and its integration with MIRC-X as a fringe tracker. Though the effect of these new LDCs has been modelled it has yet to be tested in the J band mode, these observations are planned for summer 2021. Additionally, there willl be an installation of new optical fibres into the MIRC-X instrument. These fibres have been created so as to minimise the differences in their lengths in order to minimise the effect of internal instrumental birefringence. The new fibres will also come with a new V-groove with optimised fibre spacing, allowing us to achieve 6-telescope J+H band observations for the first time. They have a factor 2 better length equalisation with a planned install at MIRC-X in summer 2022. The data reduction pipeline is believed to be ready for deployment. However, further tests are required to validate the data products following the deployment of new LDCs and instrumental fibres. 

		My contributions to this work involve the modelling of atmospheric dispersion and measuring of internal birefringence. In addition, the LDC correction models were developed by myself, the control of the LDCs and MIRC-X integration was done in collaboration with Jean-Baptiste Le Bouquin and Narsireddy Anugu. The observations were proposed and conducted by myself with more planned this summer. Following the first science observations, the data were reduced, calibrated and verified by myself. Additional guidance and advice was provided in collaboration with the MIRC-X consortium: John D. Monnier, Stefan Kraus, Jean-Baptiste Le Bouquin, Narsireddy Anugu, Tyler Gardener, Cyprien Lanthermann, Claire L. Davies, Benjamin Setterholm \& Jacob Ennis.

\section{Baseline Solution Modelling}\label{sec:Baseline_Intro}

	The baseline solution of an interferometer is a model which predicts the position of the delay line carts for a given pointing on sky. As described in Chapter\,4 in order to achieve coherence of the incoming wavefronts across all telescopes in an array, the delay lines must be positioned with micrometer level precision. A baseline solution requires precision measurements of the entire optical system, including telescope positions and differential light paths. The measurement of these variables to the required accuracy is too difficult a task to undertake directly, they can however be obtained indirectly. By recording the actual delay line position when fringe tracking for multiple positions on sky, a series of simultaneous equations can be built up. These equations can then be solved for their unknown components, the exact geometry of the system. There is never a universal solution for all observations due to changing optics, temperatures, and  turbulence. For this reason the parameters are fit to minimise the residuals between the predicted fringe position and actual fringe position. 

	The consequences of a poor baseline solution can have dire consequences for observers. The solution determines the distance over which a user must 'scan', manually moving the delay lines to find fringes. A good baseline solution means the prediction of the delay line positions is good and fringes can be found more efficiently. It is also vitally important for observations of faint and low visibility objects where fringes may not be found in the first scan.  

	\subsection{The Difficulty of Baseline Solutions at CHARA}\label{sec:Baseline_Problems}

		Given a simple interferometer with fixed beam paths, the construction of a baseline solution is a mathematically relatively simple to parameterise problem, requiring only the modelling of 4 parameters per telescope. The X, Y and Z positions each telescope relative to a reference telescope and the differential light paths between each scope and the reference. The contribution of the telescope positions to the optical path difference (OPD) is given by
		\begin{equation}\label{eq:T1Y}
	  	\centering
	  	\Delta OPD = + T1_X\,\mathrm{cos}(E) \mathrm{sin}(A) ,
	  \end{equation}
	  \begin{equation}\label{eq:T1X}
	  	\centering
	  	\Delta OPD = - T1_Y\,\mathrm{cos}(E) \mathrm{cos}(A) ,
	  \end{equation}
	  \begin{equation}\label{eq:T1Z}
	  	\centering
	  	\Delta OPD = - T1_Z\,\mathrm{sin}(E)  ,
	  \end{equation}
		Where $\Delta OPD$ is the optical path difference, $T1_{X,Y,Z}$ are the coordinates of telescope 1 and $E$ and $A$ are the elevation of and azimuth of the observation respectively. This approach results in a low number of degrees of freedom within the model, meaning few observations are required to constrain the model to a good degree of precision.

		However, the CHARA array represents a unique challenge in baseline solution modelling. In addition to the fundamental parameters already described, there are additional parameters to consider. Firstly, the beam order of the CHARA array can and does change frequently. That is different telescopes will pass light through different laboratory beams on different nights, depending on the setup. Each beam within the lab will have a different length air path. This introduces 6 additional parameters to the model, one for each of the beam positions. Secondly, CHARA employs Pipes of Pan (PoPs) in order to extend the range of delay of the delay lines. These are fixed amounts of delay that can moved in or out of the light path of each telescope. At CHARA there are 5 independent PoPs per telescope, this results in 30 additional parameters in the modelling. Overall the number of degrees of freedom within a CHARA baseline solution is much greater than that of simpler arrays. To compute the baseline solution commercial software known as \textit{iphase}\footnote{http://www.tpointsw.uk/} was used by CHARA, this software used every individual pair of telescopes, beams and PoPs as a degree of freedom in order to compute the x, y and z coordinates of each telescope and the light path length of each scope. It did not have the ability to calculate PoP or lab beam lengths or to model any changes over time. As such the software was severely limiting.

\begin{figure*}[h!]
		  \centering
		  \includegraphics[scale=0.8]{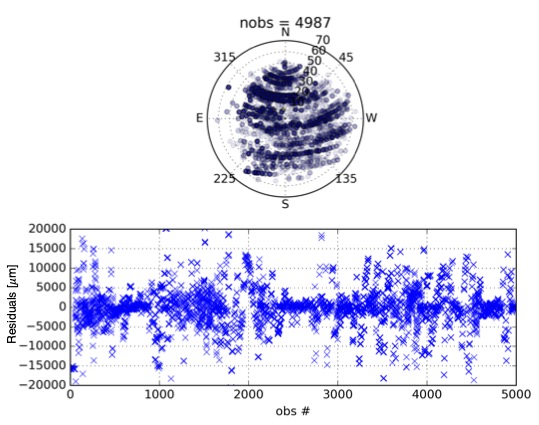}
		  \caption[Graphical representation of the CHARA baseline solution computed in 2017]{Graphical representation of the CHARA baseline solution computed in 2017. Almost 5000 observations taken over a year using multiple instruments operating at different wavelengths. \textbf{TOP:} Sky map of the locations of the observations, opacity represents the normalised magnitude of the residuals, high opacity equates to greater residuals. \textbf{BOTTOM:} fit residual against observation number, which roughly corresponds to observation date. Total rms of the fit is $4591\,\mu m$.  }
		  \label{fig:OldSolution}
		\end{figure*}

		Historically the baseline solution of CHARA has been poor for several reasons. Firstly, it was created using observations from only 2 or 3 telescope combiners, providing 1 or 3 baselines per observation, and not the 6 telescope combiner MIRC-X providing 15 baselines. As a result there were significantly fewer linked observations available to constrain the model, creating considerable discrepancies. Following from this, a significant number of observation were required in order to produce a complete model, often a year of more of observations were combined. Over such long periods of time, there are inevitably changes in the geometry of the array structure from both thermal expansion and contraction on seasonal timescales and geographical changes in a seismically active area. Figure\,\ref{fig:OldSolution} shows the problems with such a dataset when attempting to compute a new solution, many of the fit residuals are several mm in magnitude and vary widely based on position on sky. This leads to large offsets in the expected cart position and fast drifts in offsets over time. Clearly a better approach was required.

	\subsection{A New Generation Model for CHARA}\label{sec:Baseline_New}

		The development of a new baseline solution for CHARA was a two-step process. Firstly, the recording of an additional, improved dataset and secondly, the creation of a new, more intuitive fitting tool. In terms of data collection, the key to an improved solution is multiple telescope data taken over a short time period. To this end MIRC-X data is the ideal instrument. As such, software updates were implemented to allow the efficient and convenient recording of MIRC-X delay line offsets. In addition, to allow the quality control of data additional information was recorded including the name of instrument, operational wavelength and the date/time of observation. The lack of this data prior to the project made analysis of any potential issues very difficult. In addition, I proposed for, and obtained dedicated observing time over 2 nights in November 2018. During this time I conducted extensive observations with MIRC-X with the aim of completing a good sky coverage with as many telescope, beam and PoP configurations as possible. In this way long term changes in array geometry and laboratory setups were minimised to allow for a consistent dataset to be recorded.

		The origin of the data used to compute the baseline solution was not the only problem, indeed the computation was problematic as explained in the previous section. In order to overcome the shortfalls of \textit{iphase}, a new software was developed called the BaselinE SOlution Tool (BESOT). Written in Python as a separate, alternative code to \textit{iphase} by myself and Jean-Baptiste Le Bouquin. This tool allows for intuitive degrees of freedom that reflect specific array parameters rather than a degree of freedom for each individual configuration. For example PoP 1 of telescope S2 is a degree of freedom that is assigned to all observations containing that specific PoP. In this way there are only 66 degrees if freedom to solve for rather than the hundreds to thousands using the 'iphase method'. This allows for all parameters to be fitted simultaneously and a vastly improved solution to be obtained. 

		The newly developed BESOT also allowed for an in depth look at the stars with make up the observations to allow me to explore potential correlations between stellar parameters and fitting residuals. Of particular interest were stars of high or poorly constrained proper motions. The computations within the CHARA pointing model are based on Hipparcos measured positions and proper motions, as opposed to the more recent and far more accurate GAIA values. If the proper motion of a star was poorly constrained by Hipparcos the recorded stellar position in the baseline solution may be wrong compared to the actual position on sky. In addition to proper motions, I also checked for correlations with stellar classification and parallax and found no significant correlations. 

		\begin{figure*}[h!]
		  \centering
		  \includegraphics[scale=0.8]{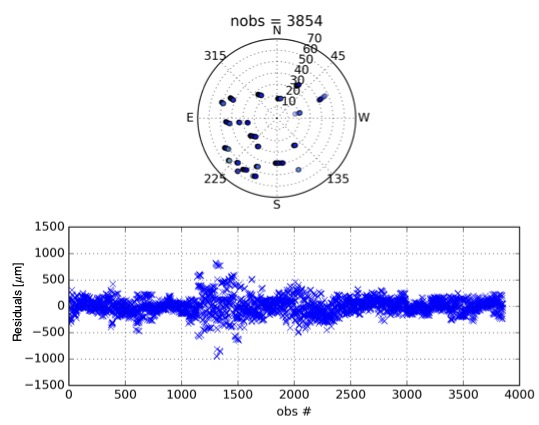}
		  \caption[Graphical representation of the CHARA baseline solution computed in 2019]{Graphical representation of the CHARA baseline solution computed in 2019. Almost 4000 observations taken over 2 nights using the MIRC-X instrument. \textbf{TOP:} Sky map of the locations of the observations, opacity represents the normalised magnitude of the residuals, high opacity equates to greater residuals. \textbf{BOTTOM:} fit residual against observation number, which roughly corresponds to observation date. Total rms of the fit is $134\,\mu m$.  }
		  \label{fig:NewSolution}
		\end{figure*}

	\subsection{Tracking Long Term Drifts}\label{sec:J_BaselineLongDrifts}

\begin{figure*}[h!]
		  \centering
		  \includegraphics[scale=0.8]{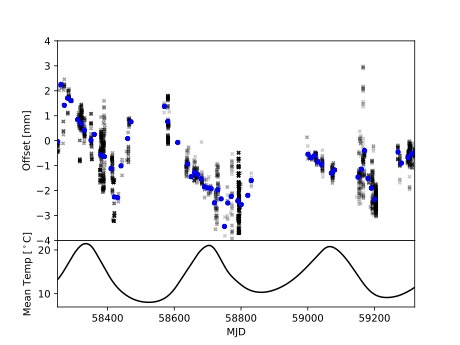}
		  \caption[Long term drifts in offsets computed for standard baseline solution]{TOP: Delay line offsets from 3 years of observations with the MIRC-X instrument at CHARA on the S1-W2 baseline. Computed by retroactively applying the most recent baseline solution to all data in order to show only variations with time. Black crosses are the raw data, blue circles are data binned into 10 day increments. BOTTOM: 10 day moving average of mean nightly temperature measured at the E1 telescope of CHARA.}
		  \label{fig:DATA_S1W2}
		\end{figure*}

		It was discovered early in the process that using offset data taken over a long period of time resulted in a poor fit to the solution parameters. This raised the idea of long term temporal changes in the array geometry, the question remained whether these changes were random or structured trends. The extent of these trends over the past 3 years is shown in Figure\,\ref{fig:DATA_S1W2}. These offsets are computed from observations corrected to assume the same baseline solution was in place across the whole 3 years. In this way the only cause of changes to the level of the offsets are physical changes at the array. 

		The scale of changes over several months is of the order of $\sim6\,\mathrm{mm}$ and appears to weakly correlate to the seasonal temperature changes. As temperatures decrease over the latter half of the year the offsets move in a negative direction, this cycle roughly repeats annually but is difficult to model given the limited data available. Lower nightly temperatures appear to signify an over-estimation of the delay line position, meaning negative delay line offsets must be applied by the observer. Inversely, higher temperatures signify an under-estimation of the delay line positions. The correlation with temperature indicates that thermal contraction/expansion is the cause. However, narrowing down the part(s) of the array which are expanding and contracting on long time scales is a near impossible task and beyond the scope of this work. Knowing the timescales and magnitudes associated with these trends is enough to plan effective countermeasures.

	\subsection{Impact of Development at CHARA}\label{sec:Development_Impact}

		The impact of my intervention will be lasting at CHARA as the benefits of an improved baseline solution have now been proven. Prior to this work searching for fringes manually by applying offsets to the delay lines was a labourious task with offsets frequently reported up to $\pm15\,\mathrm{mm}$ from their predicted position. Since the developments described in this chapter these offsets have been significantly reduced to $\pm0.3\,\mathrm{mm}$ immediately following the deployment of a new solution, with low degradation over time as described in Section\,\ref{sec:J_BaselineLongDrifts}. This substantial improvement allows for highly efficient observing in addition to making it possible to observe faint and low visibility objects easier. 

		The recognition of long term drifts within the solution on the scale of weeks-months was fully recognised during this undertaking. It is now routine operation that a new, complete, solution is computed at least twice a year and that smaller adjustments are made during the interim periods as the needs of observers dictates. This ensures the continuity of a good baseline solution and is a vast improvement over previous efforts in which a single solution was computed each year. 

		This work was undertaken in collaboration with Jean-Baptiste Le Bouquin who co-developed the new baseline solution tool BESOT\footnote{https://gitlab.com/alabdon/baseline-solution-multi-tool} and assisted with interpretation. Also involved in the undertaking was Theo ten Brummelaar, who as director of CHARA was able to provide assistance in recording MIRC-X offsets and in the operation of \textit{iphase}. In addition, he developed additional \textit{iphase} modules to validify the results obtained from BESOT. I served as project coordinator leading the push for a better baseline solution, co-authoring and later solo-developing the BESOT program. In addition, I was responsible for the analysis of archived data to identify problem data sets and the analysis and tracking of long term drifts in the solution.

\chapter{Dusty disk winds at the sublimation rim of the highly inclined, low mass young stellar object SU Aurigae}\blfootnote{Large parts of this chapter are published in \citet{Labdon19}}
\label{ch:DustyDisk}

T\,Tauri stars are low-mass young stars whose disks provide the setting for planet formation. Despite this, their structure is poorly understood. I present infrared interferometric observations of the SU Aurigae circumstellar environment that offer resolution that is three times higher and a better baseline position angle coverage than previous observations. In this section I aimed to investigate the characteristics of the circumstellar material around SU\,Aur and constrain the disk geometry. To this end, the CHARA array offers unique opportunities for long baseline observations, with baselines up to $331$\,m. Using the CLIMB three-telescope combiner in the K-band allows us to measure visibilities as well as closure phase. I undertook image reconstruction for model-independent analysis, and fitted geometric models such as Gaussian and ring distributions. Additionally, the fitting of radiative transfer models constrain the physical parameters of the disk. For the first time, a dusty disk wind was introduced to the radiative transfer code TORUS to model protoplanetary disks. This implementation was motivated by theoretical models of dusty disk winds, where magnetic field lines drive dust above the disk plane close to the sublimation zone. Image reconstruction reveals an inclined disk with slight asymmetry along its minor-axis, likely due to inclination effects obscuring the inner disk rim through absorption of incident star light on the near-side and thermal re-emission and scattering of the far-side. Geometric modelling of a skewed ring finds the inner rim at $0.17\pm0.02~\mathrm{au}$ with an inclination of $50.9\pm1.0^\circ$ and minor axis position angle $60.8\pm1.2^\circ$. Radiative transfer modelling shows a flared disk with an inner radius at $0.18$\,au which implies a grain size of $0.4\,\mu$m assuming astronomical silicates and a scale height of $15.0$\,au at $100$\,au. Among the tested radiative transfer models, only the dusty disk wind successfully accounts for the K-band excess by introducing dust above the mid-plane. It is shown that a dusty disk wind model is a viable scenario to explain these interferometric observations and the spectral energy distribution of SU Aurigae, with enough hot dust to reproduce the observed NIR excess.

\section{Introduction} \label{sec:intro}

    SU\,Aurigae (SU\,Aur) is a $5.18\pm0.13\,\mathrm{Myr}$ old \citep[from isochrone fitting,][]{Bochanski18} low mass pre-main-sequence star in the Upper Sco star forming region at a distance of $158^{+1.49}_{-1.48}\,\mathrm{pc}$, obtained from GAIA DR2 parallax measurements \citep{GAIA2par}. As a G2-type star it has a similar effective temperature to the sun \citep{DeWarf}, but a much higher bolometric luminosity at $12.06\,\mathrm{L_\odot}$ \citep[calculated from GAIA DR2,][]{GAIA2phot} putting it in the sub-giant star class. The stellar parameters adopted are listed in Table~\ref{table:Stellar_SU1}. SU\,Aur is known to be variable in the V\,band, varying up to $0.5$\,mag over a several day cycle \citep{Unruh04}. On the other hand variability in the K\,band is minimal \citep{Akeson05} allowing us to assume the flux contribution from the star is constant across different epochs of observations. As such, any variation in the visibility is likely geometric. 

    \begin{table}
        \caption[Stellar parameters of SU\,Aurigae]{\label{table:Stellar_SU1}Stellar parameters of SU\,Aurigae. (1) \citet{Gaia}; (2)~\citet{DeWarf}; (3)~\citet{Cutri03}}
        \centering
        \begin{tabular}{c c c} 
            \hline
            \noalign{\smallskip}
            Parameter  &  Value &   Reference   \\ [0.5ex]
            \hline
            \noalign{\smallskip}
            RA (J2000) &  $04~55~59.39$   & (1)  \\
            DEC (J2000) & $+30~34~01.50$  & (1)          \\
            Mass & $2.0~M_\mathrm{\odot}$ & (2) \\ 
            Sp. Type  & G2~IIIe & (2)  \\
            Distance & $157.68^{+1.49}_{-1.48}~\mathrm{pc}$ & (1)  \\
            Radius & $3.5~R_\mathrm{\odot}$ & (2) \\
            $K_{\mathrm{mag}} $ & $5.99$ & (3) \\
            $T_\mathrm{{eff}}$ & $5860~\mathrm{K}$ & (2)  \\ [1ex] 
            \hline
        \end{tabular}

    \end{table}
    
    Spectroscopic and photometric monitoring of SU\,Aur by \citet{Petrov19} has revealed that a dusty disk wind is the potential source of the photometric variability in both SU\,Aur and RY\,Tau, caused by changes in circumstellar extinction as opposed to chromospheric variability. The characteristic time of change in the disk wind outflow velocity and the stellar brightness indicate that the obscuring dust is located close to the sublimation rim of the disk, in agreement with previous theoretical disk wind models \citep{Bans12,Konigl11}. 
    
    Interferometric observations in the K band carried out by \citet{Akeson05} using the Palomar Testbed Interferometer (PTI), a three-telescope interferometer with baselines up to $110$\,m, found a disk inclined at $62^{\circ+4}_{-8}$ with a minor axis position angle of $24\pm23^\circ$ and a sublimation rim at $0.21$\,au. The disk was modelled as a flared disk with a vertical inner wall of emission using radiative transfer to fit both interferometric and photometric data. However, optically thick gas close to the central star was needed to fit the SED. The disk geometry is in agreement with values from \citet{Eisner14} using the Keck interferometer with values derived from an upper limit to the fit of the Br$\gamma$ emission. They find an upper limit on the disk inclination of $50^\circ$ and minor axis position angle of $50^\circ$. The difference in the derived position angles in previous studies is likely due to the lack of Br$\gamma$ emission and poor signal to noise of \citet{Eisner14} and the limited baseline range of \citet{Akeson05}. Hence, new observations with significantly better uv coverage on much longer baselines were needed. Additionally, polarimetric imaging campaigns, undertaken by \citet{deLeon15} and \citet{Jeffers14} revealed the presence of tails protruding from the disk at both H-band and visible wavelengths, likely associated with an extended reflection nebula and a possible undetected brown dwarf companion encounter. A companion search was undertaken by the SEEDS (Strategic Exploration of Exoplanets and Disks with Subaru) imaging survey and ruled out the presence of a companion down to $10$ Jupiter masses at separations down to $15\,\mathrm{au}$, contradicting the potential brown dwarf encounter theory of \citet{deLeon15}.     
    
    In this chapter I present the lowest mass YSO to be studied with very long baseline ($>110$\,m) NIR interferometry to date and the first study to probe the detailed rim structure of SU\,Aur with interferometric observation on baselines up to $331$\,m. Three different modelling methodologies were applied: (i) Image reconstruction was used to obtain a model-independent representation of the data and to derive the basic object morphology. (ii) Following this geometric model fitting allowed me to gain an appreciation for the viewing geometry of the disk by fitting Gaussian and ring models to the data. (iii) Finally, I combined interferometry and photometry to derive physical parameters with radiative transfer analysis, where the particular focus was on the physical characteristics of the inner rim.
    
    The observations are described in section \ref{lab:observations},  with image reconstruction detailed in section \ref{ImRecSU}. The geometrical model fitting approach is described in section \ref{RAPIDO} and radiative transfer modelling and SED fitting are discussed in section \ref{MCMC}. A discussion and analysis of the results can be found in section \ref{Diss}, followed by concluding remarks made in section \ref{Conc}.

\section{Observations} \label{lab:observations}
\begin{figure}[h!]
        \centering
        \includegraphics[scale=0.7]{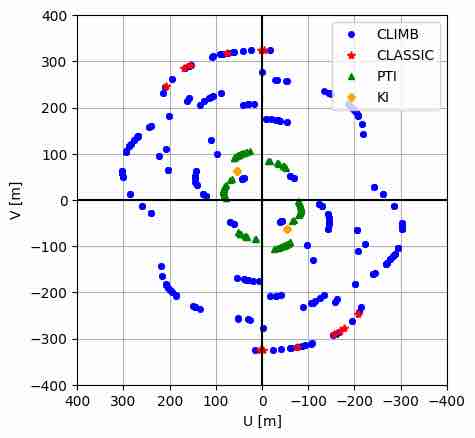}
        \caption[Coverage of the uv plane of the SU\,Aur interferometric observations]{Coverage of the uv plane of the interferometric observations. Blue points represent observations  using the CLIMB instrument, while red stars represent the CLASSIC instrument at the CHARA array. Yellow triangles represent observations using the Keck interferometer in 2011 and green triangles using the PTI with dates between 1999 and 2004.}
        \label{fig:uvplane_SU1}
    \end{figure}

    The CHARA array is a Y-shaped interferometric facility that comprises six $1\,$m telescopes. It is located at the Mount Wilson Observatory, California, and offers operational baselines between $34$ and $331\,$m \citep{Brummelaar05}. The CLIMB instrument, a three-telescope beam combiner \citep{Brummelaar13}, was used to obtain observations in the near-infrared K-band ($\lambda=2.13\,\mu m, \Delta\lambda=0.35\,\mu m$) between October 2010 and November 2014. I obtained 28 independent measurements of SU\,Aur, using seven different two-telescope configurations with maximum physical baseline of $331\,$m corresponding to a resolution of $\lambda/(2B) = 0.70\,\mathrm{mas}$ [milliarcseconds], where $\lambda$ is the observing wavelength and $B$ is the projected baseline. In addition, a small number of observations were taken in 2009 using the two-telescope CLASSIC beam combiner \citep{Brummelaar13}, also at CHARA in the K-band along the longest ($331\,$m) projected baseline. Details of the observations, and the calibrator(s) observed for the target during each observing session, are summarised in Table~\ref{table:LOG_SU1}. The uv\,plane coverage that I achieved for the target is displayed in Figure~\ref{fig:uvplane_SU1}. The data covers a relatively wide range of baseline lengths and position angles, making the data set suitable for image reconstruction.

    The CLIMB and CLASSIC data were reduced using pipelines developed at the University of Michigan \citep{Davies18}. This is much better suited to recovering faint fringes from low visibility data than the standard CHARA reduction pipeline of \citep{Brummelaar12}. The measured visibilities and closure phases were calibrated using interferometric calibrator stars observed alongside the target. Their adopted uniform diameters (UDs) were obtained from JMMC SearchCal \citep{Bonneau06, Bonneau11}, when available, or gcWeb\footnote{http://nexsciweb.ipac.caltech.edu/gcWeb/gcWeb.jsp} and are listed in Table\,\ref{table:LOG_SU1}. 

    \begin{landscape}
        \renewcommand*{\arraystretch}{0.6}
        \begin{longtable}{c c c c c } 
        \caption[SU\,Aur observing log from 1999 to 2014 from the CHARA, KI and PTI interferometers.]{Observing log from 1999 to 2014 from the CHARA, KI and PTI interferometers. Baselines involving the S2 station (CLIMB/CHARA) on 2012-12-03 produced no (or very faint) interference fringes and so were not suitable for data reduction. All uniform disk (UD) diameters quoted obtained from \citet{Bourges14}.
        *{Calibrator found to be a binary, solution used shown in Section~\ref{appendix1}.}
        }
        \label{table:LOG_SU1}
        \endfirsthead
        \endhead
        \endlastfoot
            \hline
            \noalign{\smallskip}
            Date  &  Beam Combiner &   Stations  & Pointings & Calibrator (UD [mas]) \\ [0.5ex]
            \hline
            \noalign{\smallskip}
            2010-10-02 &  CHARA/CLIMB   & S1-E1-W1   & 2 & HD 29867 ($0.280\pm0.007$), HD 34499 ($0.257\pm0.006$)\\
            2010-12-02 &  CHARA/CLIMB   & S2-E1-W2  & 1 & HD 32480 ($0.236\pm0.006$)\\
            2010-12-03 &  CHARA/CLIMB   & (S2)-E1-W2  & 1 & HD 32480 ($0.236\pm0.006$), HD 36724 ($0.233\pm0.006$)\\
            2012-10-18 &  CHARA/CLIMB   & S1-E1-W1  & 2 & HD 27777 ($0.204\pm0.006$), HD 34053*\\
            2012-10-19 &  CHARA/CLIMB   & E2-S1-W2  & 4 & HD 27777 ($0.204\pm0.006$), HD 31592*, HD 34053*\\
            2012-10-20 &  CHARA/CLIMB   & S1-W1-W2  & 3 & HD 31592*, HD 34053*\\
            2012-11-27 &  CHARA/CLIMB   & S1-E1-W1  & 5 & HD 32480 ($0.236\pm0.006$), HD 31706 ($0.219\pm0.005$)\\
            2012-11-28 &  CHARA/CLIMB   & S1-E1-E1  & 4 & HD 32480, ($0.236\pm0.006$) HD 31706 ($0.219\pm0.005$), \\
             & & & & {HD 33252} ($0.294\pm0.007$)\\
            2014-11-25 &  CHARA/CLIMB   & E2-S2-W2  & 3 &  {HD 33252}  ($0.294\pm0.007$)\\  
            2014-11-26 &  CHARA/CLIMB   & E2-S2-W2  & 3 &  {HD 33252} ($0.294\pm0.007$)\\  
            \hline
            \noalign{\smallskip}
            2009-10-31 & CHARA/CLASSIC & S1-E1 & 4 &  {HD 32480} ($0.236\pm0.006$)\\
            2009-11-01 & CHARA/CLASSIC & S1-E1  & 3 &  {HD 32480} ($0.236\pm0.006$),  {HD 24365}  ($0.319\pm0.008$)\\
            \hline
            \noalign{\smallskip}
            2011-11-07 &  KI   & FT-SEC  & 5 &  {HD 27777} ($0.204\pm0.006$) \\ \hline
            \noalign{\smallskip} 
            1999-10-09 &  PTI   & NS  & 2 &  {HD 30111} ($0.555\pm0.056$)\\
            1999-11-03 &  PTI   & NS  & 2 &  {HD 28024} ($0.222\pm0.026$),  {HD 27946} ($0.398\pm0.034$), \\
            &&&& {HD 25867} ($0.280\pm0.007$)\\
            1999-11-04 &  PTI   & NS  & 1 &  {HD 28024} ($0.222\pm0.026$),  {HD 32301} ($0.506\pm0.054$), \\
            &&&& {HD 25867} ($0.280\pm0.007$)\\
            1999-12-07 &  PTI   & NS  & 6 &  {HD 27946} ($0.398\pm0.034$),  {HD 30111} ($0.555\pm0.056$)\\
            2000-10-16 &  PTI   & NS  & 4 &  {HD 30111} ($0.555\pm0.056$)\\
            2000-11-13 &  PTI   & NW  & 17 &  {HD 30111} ($0.555\pm0.056$)\\
            2000-11-14 &  PTI   & NW  & 9 & HD  {30111} ($0.555\pm0.056$)\\
            2003-10-16 &  PTI   & SW  & 6 &  {HD 28024} ($0.222\pm0.026$),  {HD 30111} ($0.555\pm0.056$)\\
            2003-10-22 &  PTI   & SW  & 4 &  {HD 28024} ($0.222\pm0.026$),  {HD 30111} ($0.555\pm0.056$), \\
            &&&& {HD 27946} ($0.398\pm0.034$),  {HD 25867} ($0.508\pm0.046$), \\
            &&&& {HD 29645} ($0.507\pm0.035$)\\
            2004-10-01 &  PTI   & NW  & 2 &  {HD 29645} ($0.507\pm0.035$)\\
            2004-10-05 &  PTI   & SW  & 1 &  {HD 29645} ($0.507\pm0.035$)\\
            [1ex] 
            \hline
        \end{longtable}

    \end{landscape}

    \begin{figure*}
        \centering
        \includegraphics[scale=0.4]{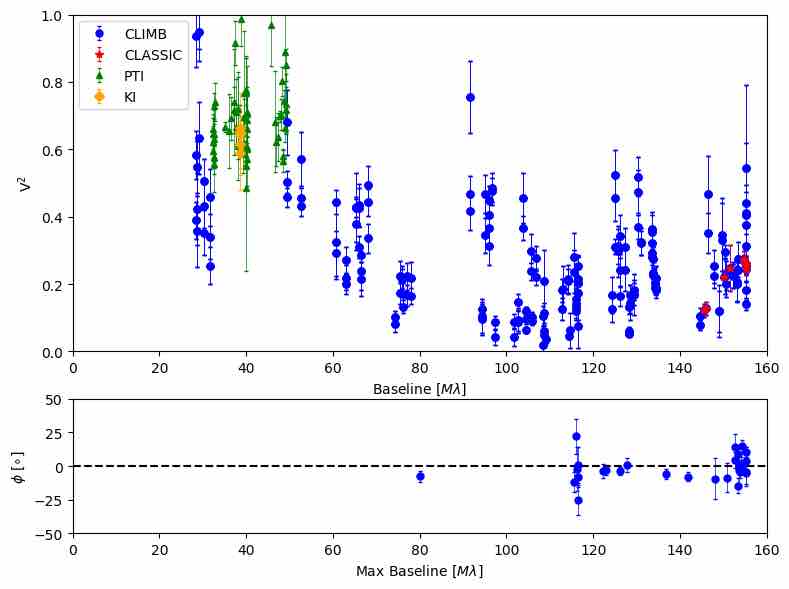}
        \caption[Squared visibility and closure phase measurements of SU\,Aur from CLIMV, CLASSIC, PTI and KI]{Squared visibility and closure phase measurements against the de-projected baseline length of the interferometric data. Blue data points are those from the CLIMB/CHARA instrument. Red data points are from the CLASSIC/CHARA instrument, Green data is that obtained from the PTI instrument and orange is that from the KI instrument.}
        \label{fig:CPVIS_SU1}
    \end{figure*}

    During the data reduction it was found that the calibrators HD\,31592 and HD\,34052 exhibited strong closure phase signals, indicating the presence of close companions around these stars. In order to ensure these binary calibrators could be used to calibrate the primary science target, I cross-calibrated the data on these stars with other calibrators observed during the same nights to determine the binary parameters, as outlined in Section~\ref{appendix1}. Based on the fitted binary parameters, I could then correct the transfer function and use the data for the calibration of SU\,Aur.

    In addition to the new observations from CHARA, archival interferometric data from other facilities was included for the analysis. A small amount of data was available from the Keck Interferometer \citep[KI,][]{Colavita13,Eisner14} from 2011 along a single $84$\,m baseline, while a larger amount of data was also available from the Palomar Testbed Interferometer \citep[PTI,][]{Colavita99} from 1999 to 2004 using a two-telescope beam combiner on 3 different physical baselines between $86$ and $110$\,m. This data was published in \citet{Akeson05}. These additional measurements complement the CHARA observations in the intermediate baseline range; the full uv coverage is shown in Figure\,\ref{fig:uvplane_SU1}.
    
    Both the PTI and KI data were calibrated using the standard method outlined by \citet{Boden98} using the wbCalib software available from NExcScI\footnote{http://http://nexsci.caltech.edu/software/V2calib/wbCalib/}. The calibration pipeline works in conjunction with getCal and the Hipparcos catalogue \citep{Perryman97} for calibrator star astrometry and diameters. This was the same process used by \citet{Akeson05} to extract visibilities from the PTI data. The re-reduction of this data agrees with the results shown in the literature. The fully reduced data from all instruments is shown in Figure\,\ref{fig:CPVIS_SU1}.
    
    As I combine several years worth of data, care was taken to check for time dependencies in the visibilities of baselines of similar length and position angle. Variability in the K band is known to be minimal, so any time dependencies in the visibility amplitudes is likely geometric. However, no significant time dependencies were discovered.

\section{Image reconstruction} \label{ImRecSU}

    Image reconstruction techniques require broad and circular uv coverage along as many baseline lengths as possible. Fortunately, the data from the observations lends itself to this process as the uv plane has been well sampled, though some small gaps remain in the position angle coverage. By comparing visibilities from different instrument at similar baseline lengths and position angles, one can see there is likely very little extended emission in this system and so no correction for instrument field of view is required. This technique is useful for interpretation of non-zero closure phases, indicative of asymmetric distributions, in a model-independent way. The closure phase values are shown in Figure\,\ref{fig:CPVIS_SU1}. There are many different algorithms with which to reconstruct images from interferometric data, but the process described here involved the use of the Polychromatic Image Reconstruction Pipeline (PIRP) which encompasses the $MiRA$ reconstruction algorithm by \citet{MiRA08}. The reconstruction procedure and results are described below.

    \begin{figure*}[b!]
        \centering
        \includegraphics[scale=0.45]{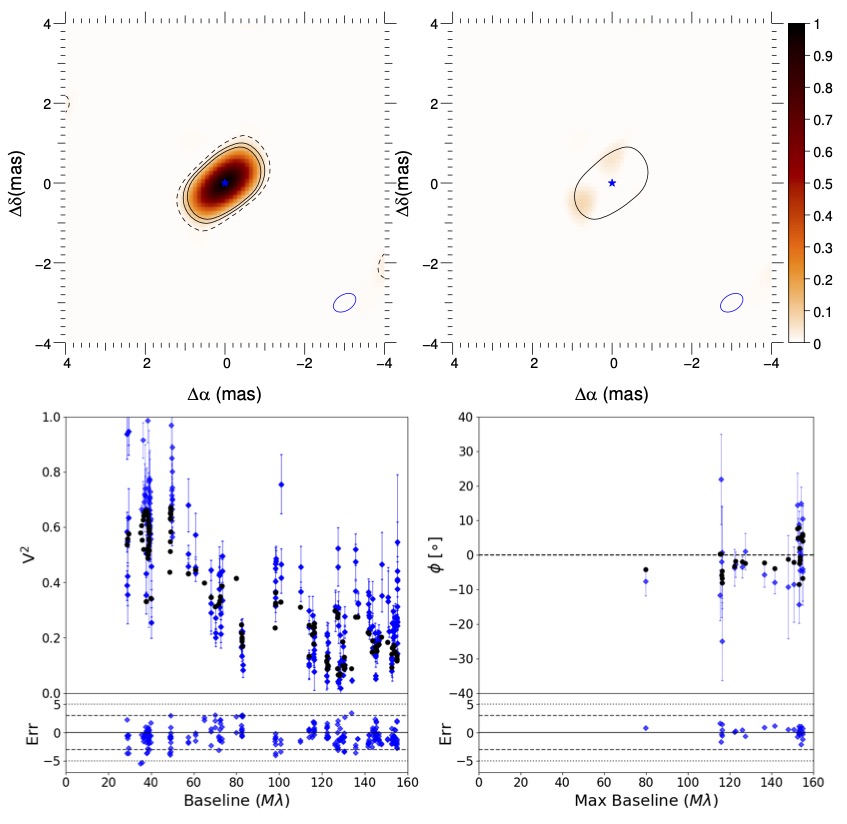}
        \caption[Image reconstruction of SU\,Aur from CLIMB and archival data]{Top left:Resultant reconstructed image showing the $3$ and $5\,\mathrm{\sigma}$ significance levels as solid black lines and the $1\,\mathrm{\sigma}$ level as dashed lines. The beam size is shown in the bottom right. The colour bar is the same for both maps, with the maximum intensity normalised to $1$ for the sake of readability. Top right: Extracted asymmetry in the intensity map shown with $3\,\mathrm{\sigma}$ significance level. Bottom left: Fitted visibilities, data shown in black with synthetic image visibility in blue. Bottom right: Fitted closure phases, data shown in black with synthetic image CP in blue. The residuals normalised to the standard deviation are plotted in the bottom of each graph.}
        \label{fig:newRec_SU1}
    \end{figure*}

    In the $MiRA$ routine, the object is modelled as an unresolved central star with an extended, model-independent, environment \citep{Kluska14}. Both components have different spectral behaviours and so differing spectral indices. Additionally, the type and weight of the regularisation was explored, $MiRA$ allows for either quadratic smoothing or total variation regularisations to be implemented. The regularisation plays the role of the missing information by promoting a certain type of morphology in the image. The quadratic smoothing algorithm aims for the smallest possible changes between pixels to produce a smoother image, it is particularly useful as the quadratic nature means it is less likely to find local minima. On the other hand, total variation aims to minimise the total flux gradient of the image and is useful to describe uniform areas with steep but localised changes. These regularisations are considered to be the best ones for optical interferometric image reconstruction; \citep{Renard11}. 
    The size and number of pixels also plays an important role in image reconstruction. One cannot simply use the maximum number of pixels of the smallest size to obtain better resolution, they have to be chosen to match uv plane sampling. It was found that a quadratic smoothing regularisation with a weight of $1\times10^9$ and $526\times526$ pixels of $0.1$ mas in size provides the best-fit image reconstruction when utilising exact Fourier transform methods. The optimal regularisation parameters were determined using the L-curve method. PIRP also allows for bootstrap iterations \citep{Efron94} starting from a previous best image. This involves a random draw of data points within the dataset to determine the reliable features of the image. The process was carried out 500 times allowing for a pixel by pixel error estimation, see \citet{Kluska16}.

    The final image is shown in Figure\,\ref{fig:newRec_SU1} (upper left panel), which also shows the asymmetry in the intensity map (upper right panel). The contours represent the $1\sigma$ (dashed line), $3\sigma$ and $5\sigma$ (solid lines) uncertainties. The asymmetry is calculated by rotating the image through $180^\circ$ and subtracting it from the un-rotated image. It is this residual flux that produces the non-zero closure phases, highlighting any areas of greater emission within the disk. Using this technique one can see that the disk has greater intensity in the eastern regions, which contains $8\%$ more flux that the western regions with a flux ratio of $1.07$, where the remaining flux is in the central star. By inclining a flared disk with the western region (bottom right of image) towards the observer the nearside of the inner rim becomes self-shadowed by the near-side disk rim, so the eastern region inclined away from the observer appears brighter. The image shows a disk radius of $1.0$~mas, a minor-axis position angle of $41^\circ\pm3$ and an inclination of $51^\circ\pm5$. The fit of the image to the visibility and closure phases in the data is shown in Figure~\ref{fig:newRec_SU1}, with a combined visibility and closure phase reduced $\mathrm{\chi^2_{\mathrm{red}}}$ of $4.57$.
    
\section{Geometric Model Fitting} \label{RAPIDO}
    The next step in interpreting the interferometric observations is the fitting of simple geometric models to the observed quantities. The visibility profile (Figure~\ref{fig:CPVIS_SU1}) reveals a clear drop in visibility through short and intermediate baselines with a possible plateau/second lobe at the longest baselines. 
    In the visibility profile I do not see any evidence for structures on distinct different spatial scales that might indicate the presence of a binary companion or extended halo emission.
    The closure phases are all $<20^\circ$, indicating weak asymmetric features as evidenced in the image reconstruction above. To explore the viewing geometry of the disk, a series of simple intensity distributions were tested against the data.
    All the models tested contained an unresolved point source that was used to represent the central star, a reasonable assumption given the expected angular diameter of the star (see Table.\,\ref{table:Stellar_SU1}). The disk component was modelled as one of three intensity distributions: (i) A Gaussian to simulate a disk with unresolved inner rim with a FWHM free parameter. (ii) A ring to simulate emission from a bright inner rim only with a defined fractional width equal to $20\%$ of the radius with an inner radius ($R_{min}$) free parameter. (iii) A skewed ring model with a diffuse radial profile defined by an inner radius ($R_{min}$), a FWHM ($w$) and an azimuthal brightness modulation, where $c_j$ and $s_j$ are the cosine and sine amplitudes for mode $j$, in an attempt to model disk asymmetries. \citet{Lazareff17} find that the diffuse skewed ring here can be used to successfully model a wide range of YSOs. 
    
    With the exception of the skewed ring, these models are intrinsically axisymmetric, but I project the brightness distribution in order to mimic viewing geometry effects that are parameterised with an inclination angle $i$ (defined with $0^\circ$ as face-on) and a disk position angle $\theta$. I measured disk position angles along the minor axis and follow the convention that position angles are measured from north ($\theta = 0^\circ$) towards east. 
   
    The stellar-to-non-stellar flux ratio can be calculated by comparing K-band photometry \citep{Curi03} with the stellar atmosphere models of \citet{Kurucz04}, which gives a ratio of $1.17$ (using the stellar parameters listed in Table~\ref{table:Stellar_SU1}). However, it cannot be fixed as this has been shown to introduce unreasonable large scale components when fitting long baseline data \citep{Ajay13}. As such the parameter space of the flux ratio was explored step-wise for all models with a range of model parameters. In this way the stellar-to-total flux ratio was constrained for all models. It was found that a ratio of $1.13\pm0.01$ provided the best fit to the data for the ring, skewed ring and TGM models. After determining the star-to-disk flux ratio, the parameter space of the geometric models could be fitted to the observed visibilities and closure phases for each of the observed baselines given initial parameter constraints based on the literature values of \citet{Akeson05} described in Section\,\ref{sec:intro}. I used a bootstrap method to explore the parameter space around these initial values and to compute uncertainties on the individual parameters by fitting Gaussian distributions to parameter histograms.
    
    The best-fit parameters and associated errors for each of the geometric models are shown in Table~\ref{table:geofit_SU1}. All test models agree with respect to the position angle and inclination of the disk very well (Table\,\ref{table:geofit_SU1}). However, none of the models provide good fits to the data as evidence by the $\chi^2_{\mathrm{red}}$ values of between 11.86 (ring model) and $8.57$ (skewed ring). The skewed ring model found a minor axis position angle of $61.0^\circ\pm1.0$ and inclination of $51.2^\circ\pm1.1$. Of the individual models, the skewed ring provides the best fit to both the visibilities and closure phases. Both the Gaussian model and the Skewed Ring are found to be quite diffuse with a FWHM of $2.01~\mathrm{mas}\pm0.02$ and width of $0.96~\mathrm{mas}\pm0.02$ respectively.

\begin{table*}[t!]
        \caption[Geometric modelling parameters of SU\,Aur from CLIMB and archival data]{\label{table:geofit_SU1}Best fit parameters for the simple geometric models investigated. *The closure phase quoted is the achieved when allowing the skewed ring to become asymmetric. While the software did detect an asymmetry, it failed to constrain its location in the disk. $\theta$ is the minor-axis position angle of the disk.}
        \centering
        \begin{tabular}{c c c c c} 
            \hline
            \noalign{\smallskip}
             Parameter & Explored Parameter Space & Gaussian &   Ring & Skewed Ring   \\ [0.5ex]
            \hline
            \noalign{\smallskip}
            $R_\mathrm{{min}}$ [mas] & $0.0 - 10.0$ & ... & $1.0\pm0.13$ & $1.1\pm0.12$   \\
            FWHM & $0.0 - 15.0$ & $2.01\pm0.02$ & ... & ...    \\
            $w$ & $0.01-2.0$ & ... & ... & $0.96\pm0.02$  \\
            $\theta$ [$^\circ$] (PA) & $0.0 - 180.0$ & $60.8\pm1.15$  & $61.36\pm1.2$ & $61.0\pm1.0$ \\
            $i$ [$^\circ$]& $0.0 - 90.0$ & $51.4\pm1.04$  & $50.91\pm0.88$ & $51.2\pm1.1$ \\ 
            \hline
            \noalign{\smallskip}
            $\mathrm{\chi^{2}_{\mathrm{red,Vis}}}$ & ... & $10.97$  & $11.86$ & $8.57$  \\ 
            \noalign{\smallskip}
            $\mathrm{\chi^{2}_{\mathrm{red,CP}}}$ & ... & $1.90$ & $1.90$ & $1.66$* \\ [1ex]
            \hline
        \end{tabular}
    \end{table*}

    Overall, I achieved the best-fit with a skewed ring with an inner radius of $1.10~\mathrm{mas}\pm0.12$ and a ring FWHM of $0.96~\mathrm{mas}\pm0.02$, thus making the ring very diffuse with only a marginally defined inner radius. There was no evidence discovered for any over-resolved extended emission or 'halo' found in many other objects \citep{Monnier06a,Kraus2009}.
    
    In order to model the observed CP signal, with a maximum of $20\pm12^\circ$, which may indicate a slight asymmetry in the disk, I introduced asymmetries to the skewed ring model \citep{Lazareff17}. However, this improved the fit only marginally over the standard ring model from Table~\ref{table:geofit_SU1}. It was found that zero closure phases produced a reduced chi-squared ($\mathrm{\chi^2_{\mathrm{red,CP}}}$) fit of $1.90$, while a sinusoidally modulated asymmetric ring only resulted in a $\mathrm{\chi^2_{\mathrm{red,CP}}}$ of $1.66$. The model visibility curves corresponding to the best-fit skewed ring model are shown in Figure~\ref{fig:IN05}, \ref{fig:THM07} and \ref{fig:BK12} (red curve). The different panels show visibilities towards different position angle bins. 
    
    Of the simple geometric models tested, the skewed ring model provides the best fit. However, none can be said to provide a good fit to the observed data, as evidence by the $\chi^2_{\mathrm{red}}$ values shown in Table\,\ref{table:geofit_SU1}. As such, more complex disk structures are required, such as flared disks, different rim morphologies or disk winds. In the next section, these possibilities are explored in detail using radiative transfer techniques. This allows us to not only explore complex geometries, but to derive physical parameters such as radial density profiles and the disk scale heights. 
    
\section{Radiative Transfer Modelling}\label{MCMC}

    I used the TORUS Monte-Carlo radiative transfer code \citep{Harries00}, allowing for the simultaneous fitting of visibility, closure phase and photometric data to further constrain the geometry and physical dust properties of the SU\,Aurigae circumstellar disk.

    Starting from the disk properties derived by  \citet[][Table~\ref{table:torus_SU1}]{Akeson05}, I explored radiative transfer models with different scale heights (where scale height is that of the gas parameterised at $100~\mathrm{au}$ with flaring index $\beta$) and inner rim shapes. The flaring index $\beta$ describes the radial gradient of the disk scael height, such that $h(r) \propto r^\beta $ \citep{Kenyon87} and is a free parameter. In the TORUS simulations, the dust was allowed to vertically settle where large grains $\geq 1.0\,\mu m$ were settled to $60\%$ of the disk scale height $h$ while small grains $\leq 0.5\,\mu m$ were allowed to inhabit the full disk scale height. The grain size adopted depends on the model adopted as described below. The dust sublimation radius was left as a free parameter, allowing the inner rim radius to define itself based on well-defined rules of the \citet{Lucy99} iterative method to determine the location and the temperature structure of the whole disk. This is implemented whereby the temperature is initially calculated for grid cells in an optically thin disk structure, with dust added iteratively to each cell with a temperature lower than that of sublimation, until the appropriate dust to gas ratio is reached ($0.01$). Once TORUS has converged to radiative equilibrium a separate Monte Carlo algorithm is used to compute images and SEDs based on the optical properties of the dust species implemented. I confirmed that stellar photosphere models of \citet{Kurucz04} using these stellar parameters can reproduce the photometry measurements of SU\,Aur reasonably well across the visible continuum. The grain size distribution used by \citet{Akeson05} is a distribution of astronomical silicate grains up to $1\,\mathrm{mm}$ in size throughout the disk. I adopt a silicate grain species with dust properties and opacities adopted from \citet{Draine84}. The initial density structure of the gas is based upon the $\alpha$-disk prescription of \citet{Shakura73} where the disk density is given as:
    
    \begin{equation}\label{eq:1_SU1}
    \centering
     \rho(r,z) = \frac{\Sigma(r)}{h(r)\sqrt{2\pi}}\exp\bigg\{-\frac{1}{2}\bigg[\frac{z}{h(r)}\bigg]^2\bigg\} .
    \end{equation}
    
    Here, $z$ is the vertical distance from the midplane while the parameters $h(r)$ and $\Sigma(r)$ describe the scale height and the surface density respectively. In the radiative transfer models, I represent the stellar photosphere with a \citet{Kurucz79} model atmosphere using the stellar parameters outlined in Table~\ref{table:Stellar_SU1}.
    The photometric data was obtained from a wide range of instruments from the VizieR database and are compiled in Table~\ref{table:Photometry}. Where multiple observations in the same waveband were present care was taken to minimise the total number of instruments and keep the number of observation epochs as close as possible to minimise any potential variability effects. 

    Visibilities were calculated from synthetic images of the disk system (as shown in Figure~\ref{fig:Timage_SU1}) extracted through application of the van Cittert-Zernicke theorem, applied using a 1D Fourier transforms projected onto the observed baseline position angle. Phases are also extracted from the images and are used to calculate the closure phase, as described in \citet{Davies18}.
    
    The parameter space of the radiative transfer models was explored objectively using the values of \citet{Akeson05} as a starting point. A range of physically realistic values for each parameter was explored in a broad grid of models (as described in Table~\ref{table:torus_SU1}). A $\chi^2$ value was then computed for the visibilities, closure phases and SED fits of each model allowing the grid to be refined around the minimum. The interferometric and photometric data points were fitted simultaneously, with the resulting $\chi^2$ values shown in Table~\ref{table:Chi2}. The silicate feature at $10\,\mathrm{\mu m}$ allows us to place some constraints on the dust sizes, as larger grains produce smaller features. The growth of dust grains and their effects on observed silicate features is described in a review by \citet{Natta07}. The IR flux is controlled by the morphology of the sublimation rim. As the inner radius increases, the amount of circumstellar material emitting in mid-IR wavelengths is reduced, and the IR emission decreases. The shape of the mid-IR excess also describes the degree of flaring present in the disk, where large excess indicates greater flaring. A larger flaring power in a disk leads to an increasing surface intercepting the starlight, and therefore an increase in reprocessed radiation. 
    I adjusted the total dust mass in the model in order to match the millimeter flux. A detailed description of the effect of disk parameters on the SEDs of protoplanetary disks can be found in \citet{Robitaille07}.
    
    Due to the optical depth of the system the inner-rim of the disk appears as the brightest part of the disk at NIR wavelengths. This is because the rest of the disk is shadowed by the inner rim and only rises out of shadow in cooler regions of longer wavelength emission.

    \subsection{Sublimation Rim Model} \label{SubRim}
 
    The curved rim of \citet{Isella05}, henceforth IN05, is based upon a single grain size prescription with a gas density-dependent sublimation temperature. Due to a vertical gas pressure gradient, the sublimation temperature decreases away from the mid-plane, creating a curved rim. The sublimation temperature of the grains follows
    \begin{equation}
        T_{\mathrm{sub}} = G\rho^\gamma(r,z),
    \end{equation}
    where the constant $G=2000$~K, $\gamma = 1.95\times 10^{-2}$, $r$ is the radial distance into the disk and $z$ is the height above the midplane \citep{Pollack94}. 
    
\begin{figure*}[t!]
        \centering
        \includegraphics[scale=0.75]{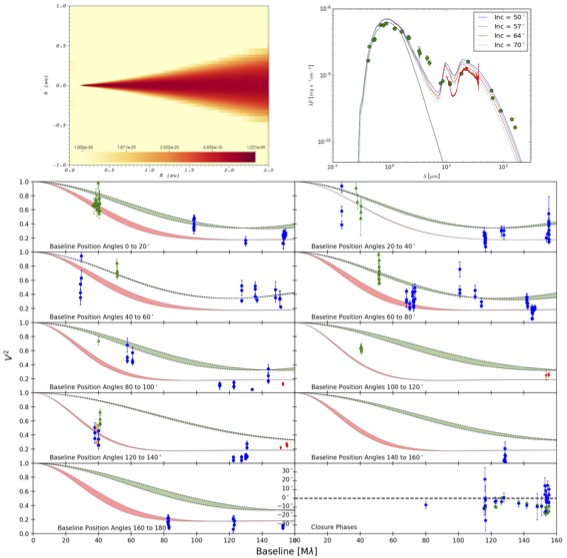}
        \caption[Results of radiative transfer and geometric modelling with the IN05 disk rim prescription]{Results of radiative transfer and geometric modelling with the IN05 disk rim prescription \citep{Isella05} of a gas density-dependent sublimation temperature with a single grain species size $0.1 \mathrm{\mu m}$. TOP LEFT: Disk density cross section of the inner rim. A logarithmic colour scale is used with a minimum density of $1.00\times10^{-30}$ to a maximum of $4.64\times10^{-10}\,\mathrm{g cm^{-3}}$. TOP RIGHT: SED computed with the radiative transfer model. Dark blue curve is the simulated blackbody emission of the central star. Green points are photometric observations while the short red line is the \textit{Spitzer} spectrum. The coloured curves represent the SED at the different inclinations of 50, 57, 64 and $70^\circ$. BOTTOM: Visibility data binned by position angle of observation. The data points are split by instrument consistently with Figures\,\ref{fig:uvplane_SU1} and\, \ref{fig:CPVIS_SU1}, where blue circles are from CHARA/CLIMB, red stars are from CHARA/CLASSIC, green triangles from PTI and orange diamonds from KI. The red curves are the results of the best fit geometric skewed ring model and the green curves are calculated from the radiative transfer image at an inclination of $50^\circ$ and a position angle of $60^\circ$. The dashed bounding lines indicate the minimum and maximum model visibilities for that position angle bin. The very bottom right panel shows the observed CP measurements (black) and the CP computed from the radiative transfer image (green).}
        \label{fig:IN05}
    \end{figure*}

    A curved rim is shown to be a viable disk model by \citet{Flock16a} and \citet{Flock16b}, based upon extensive hydrodynamical simulations. A wide range of disk structure parameters are explored to find the best fit solution to both the interferometric measurements and the SED.
    Figure~\ref{fig:IN05} shows the results of radiative transfer modelling of the IN05 rim prescription. The model SED shows a clear flux deficit in the NIR, with a K-band flux of just 64\% of the 2MASS photometric point, far outside the limited range of variability of SU\,Aur. This is also clear in the visibility curves, shown in green, where the overall shape is a good fit, but the minimum visibility is too high due to a larger than expected stellar contribution.
    
\begin{figure*}[t!]
        \centering
        \includegraphics[scale=0.75]{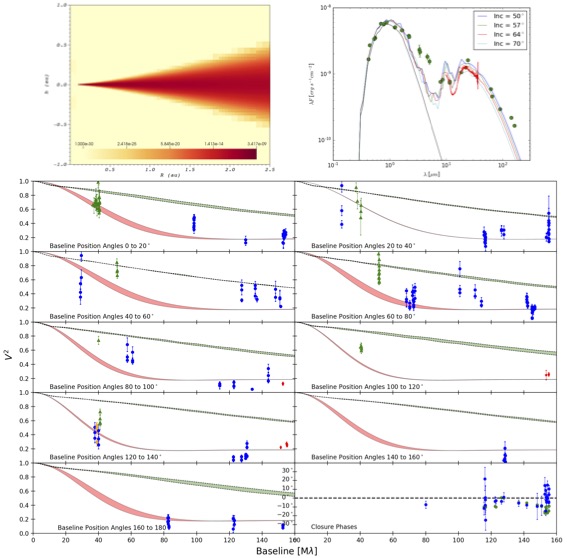}
        \caption[Results of radiative transfer and geometric modelling with the THM07 disk rim prescription]{Results of radiative transfer and geometric modelling with the THM07 disk rim prescription \citep{Tannirkulam07} of a gas density-dependent sublimation temperature with two grain species, a majority larger grains at $1.2 \mathrm{\mu m}$ with fewer smaller grains at $0.1 \mathrm{\mu m}$. The mass of larger grains is fixed at $9$ times the mass of smaller grains. TOP LEFT: Disk density cross section of the inner rim. A logarithmic colour scale is used with a minimum density of $1.00\times10^{-30}$ to a maximum of $1.23\times10^{-09}\,\mathrm{g cm^{-3}}$. TOP RIGHT: SED computed with the radiative transfer model. Dark blue curve is the simulated blackbody emission of the central star. Green points are photometric observations while the short red line is the Spitzer spectrum data. The coloured curves represent the SED at the different inclinations of $50$, $57$, $64$ and $70^\circ$. BOTTOM: Visibility data binned by position angle of observation. The data points are split by instrument consistently with Figures\,\ref{fig:uvplane_SU1} and\, \ref{fig:CPVIS_SU1}, where blue circles are from CHARA/CLIMB, red stars are from CHARA/CLASSIC, green triangles from PTI and orange diamonds from KI. The red curves are the results of the best fit geometric skewed ring model and the green curves are calculated from the radiative transfer image at an inclination of $50^\circ$ and a position angle of $60^\circ$. The dashed bounding lines indicate the minimum and maximum model visibilities for that position angle bin. The very bottom right panel shows the observed CP measurements (black) and the CP computed from the radiative transfer image (green). }
        \label{fig:THM07}
    \end{figure*}

    An alternative sublimation front geometry is proposed by \citet{Tannirkulam07}, henceforth THM07. This model employs a two-grain scenario, were a mixture of small $0.1\,\mathrm{\mu m}$ grains and large $1.2\,\mathrm{\mu m}$ grains has been adopted, with the mass of larger grains fixed at $9$ times the mass of smaller grains. The smaller grains are not allowed to settle, so the scale height is fixed to that of the gas. The larger grains are allowed to settle to 60\% of the scale height of the gas. This combined with the larger grains existing closer to the star, due to more efficient cooling, leads to an elongated and curved sublimation front. Figure~\ref{fig:THM07} shows the results of this modelling. The SED again shows a clear deficit in NIR flux, with a K-band flux of just 68\%, comparable with that of the IN05 prescription. The presence of dust closer to star changes the shape of the visibility curve dramatically, with the first lobe now extending to much longer baselines. The two grain THM07 model proves to be a worse fit than the single grain IN05 model.
    
    The results show that the best fit of the curved-rim disk model of IN05 can be achieved with a single silicate grain species with a single grain size of $0.1\,\mu m$. The addition of larger grain species further reduced near-infrared flux in 1-3\,$\mu m$ region, resulting in a poorer fit to the SED in both the shape and magnitude of the NIR excess. The disk is also found to be highly flared and extending from $0.15$ to $100$\,au, loosely constrained by the long wavelength photometry, while still solving for vertical hydrostatic equilibrium. Importantly, comparison of the stellar atmosphere, represented by the Kurucz model atmosphere \citep{Kurucz04}, with the photometric data can provide the stellar-to-total flux ratio for each waveband. In the K-band this ratio is found to be $1.27$, meaning the circumstellar environment contributes $44\%$ of the total flux. The dust to gas ratio is fixed to $0.01$, as taken from literature values \citep{Akeson05}. The NIR flux deficit results also in a poor fit to the K-band visibilities.

    \clearpage
    
    \begin{figure*}[t!]
        \centering
        \includegraphics[scale=0.75]{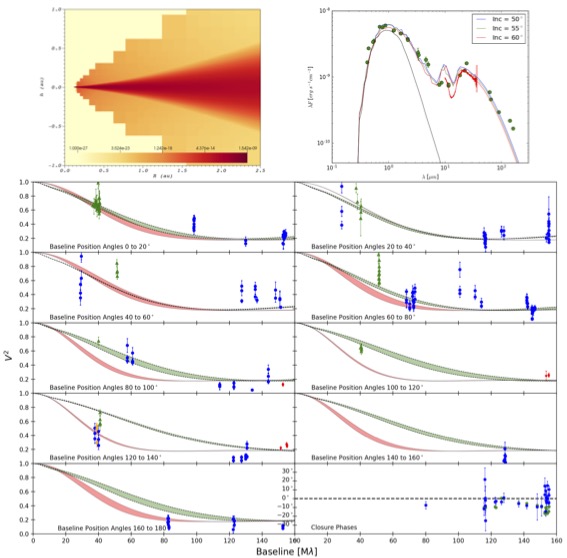}
        \caption[Results of radiative transfer and geometric modelling with the IN05 disk rim prescription and an additional dusty disk wind component]{Results of radiative transfer and geometric modelling with the IN05 disk rim prescription \citep{Isella05} and an additional dusty disk wind (BK12) implemented following \citet{Bans12}. TOP LEFT: Disk density cross section of the inner rim. A logarithmic colour scale is used with a minimum density of $1.00\times10^{-27}\,\mathrm{g cm^{-3}}$ to a maximum of $1.54\times10^{-09}$. TOP RIGHT: SED output of radiative transfer analysis. Dark blue curve is the simulated blackbody emission of the central star. Green points are photometric observations while the short red line is the Spitzer spectrum data. The coloured curves represent the SED at the different inclinations of $50$, $55$ and $60^\circ$. BOTTOM: Visibility data binned by position angle of observation. The data points are split by instrument consistently with Figures\,\ref{fig:uvplane_SU1} and\, \ref{fig:CPVIS_SU1}, where blue circles are from CHARA/CLIMB, red stars are from CHARA/CLASSIC, green triangles from PTI and orange diamonds from KI. The red curves are the results of the best fit geometric skewed ring model and the green curves are calculated from the radiative transfer image at an inclination of $50^\circ$ and a position angle of $60^\circ$. The dashed bounding lines indicate the minimum and maximum model visibilities for that position angle bin. The very bottom right panel shows the observed CP measurements (black) and the CP computed from the radiative transfer image (green).}
        \label{fig:BK12}
    \end{figure*}

    \subsection{Dusty Disk Wind Model}

    In an attempt to increase the NIR flux contributions in the model, I explored a dusty disk wind scenario as set out by \citet{Bans12}, henceforth BK12. This model is described in detail in Section\,2.5.2. The prescription used in TORUS is taken from \citet{Bans12}, and assumes a steady, axisymmetric, effectively cold disk outflow. The wind is launched from the disk surface and is assumed to contained the same dust composition. For the dust distribution, a constant dust-to-gas ratio is assumed to match that of the disk, while this may not a be physical assumption it does allow for estimates of dust quantities in the wind. The wind is populated with dust in the same way as the disk and also converges towards radiative equilibrium with each \citet{Lucy99} iteration. Full details of the disk wind model implemented can be found in \citet{Bans12} and \citet{Konigl11}. A wide parameter search was undertaken, in order to determine the optimum solution (see Table~\ref{table:torus_SU1}). The stellar parameters were fixed to those shown in Table~\ref{table:Stellar_SU1} and the same silicate prescription was adopted for all models, as in the IN05 and THM07 prescriptions The position angle of the disk was fixed to that of the best fit geometric model listed in Table~\ref{table:geofit_SU1}.

    The results of the radiative transfer modelling of the BK12 wind are shown in Figure~\ref{fig:BK12}. The cross section shows the very inner-rim of disk, uplifted dust above and below the mid-plane can clearly be seen. The inner-rim is also curved using the IN05 prescription, although a grain size of $0.4\,\mathrm{\mu m}$ is required to produce the observed excess across the infrared. The graon size was fitted in an integrated way, while producing a grid of mdoels to fit all parameters. A range of grain sizes were tested from $0.1$ to $1.5\,\mathrm{\mu m}$ but $0.4\,\mathrm{\mu m}$ grains provided the best fit to the K-band photometric flux values. 

    \begin{figure}[h!]
        \centering
        \includegraphics[scale=0.35]{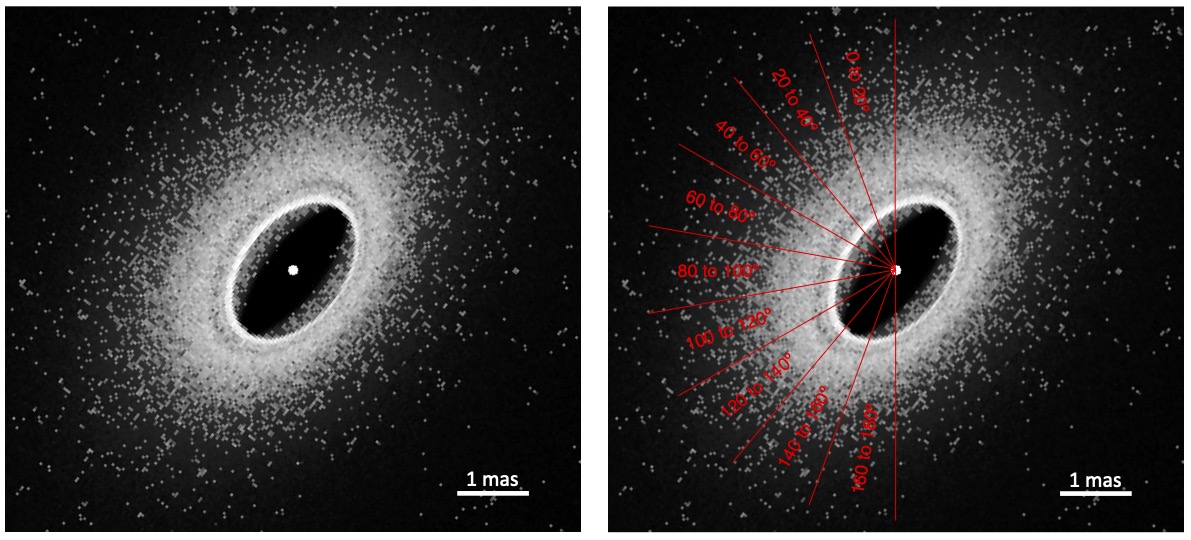}
        \caption[Radiative transfer image of SU\,Aur computed with TORUS]{LEFT: Computed synthetic image from TORUS following the BK12 dusty disk wind prescription. The colour indicates normalised intensity in the K-band. RIGHT: Same image overlaid with the position angle bins used in Figures\,\ref{fig:IN05} \ref{fig:THM07} and \ref{fig:BK12}. The disk minor axis can clear be seen to fall around the 40 to $60^\circ$ and 60 to $80^\circ$ bins, corresponding to a shallower first lobe of the visibility curve.}
        \label{fig:Timage_SU1}
    \end{figure}
    
    \begin{landscape}

    \begin{table*}[t!]
        \caption[Results of radiative transfer fitting of SU\,Aur to CLIMB and archival data]{\label{table:torus_SU1}Best-fit parameters resulting from SED and visibility fitting. (1) \citet{Akeson05}; (2)~\citet{Unruh04}; (3)~\citet{Jeffers14}
        Computed using the TORUS radiative transfer model \citep{Harries00} for the BK12 model scenario. $q_\mathrm{dist}$ is the power law of the grain size distribution. The best fit values are those of the BK12 prescription rather than the IN05 or THM07 models. $\alpha_\mathrm{disk}$ is the incident angle and is fixed at $(\beta_{disk}+1)$ and represent flaring parameters whcih define the angle and pivot point of the flaring. $R_\mathrm{{0min}}$ is the innermost radius of the disk from which the disk wind is launched.
        }
        \centering
        \begin{tabular}{c c c c c} 
            \hline
            \noalign{\smallskip}
            Parameter  & Literature value & Reference & Range explored &  Best fit value    \\ [0.5ex]
            \hline
            \noalign{\smallskip}
            $i$ & $60^\circ$ & (1),(2) & $30 - 80^\circ$ & $50^\circ$    \\
            $R_\mathrm{{inner}}$ & $0.18$ & (1)(3) & $0.1 - 0.6~\mathrm{au}$ & $0.15~\mathrm{au}$    \\
            $R_\mathrm{{outer}}$ & $100~\mathrm{au}$ & (1) & $20.0-120.0~\mathrm{au}$ & $100.0~\mathrm{au}$    \\
            $h_\mathrm{0}$ & ... & ... & $7.0 - 20.0~\mathrm{au}$ & $15.0~\mathrm{au}$     \\
            $\alpha_\mathrm{{disk}}$ & ... & ... & $1.0 - 3.0$ & $2.4$  \\
            $\beta_\mathrm{{disk}}=(\alpha-1)$ & ... & ... & $0.0 - 2.0$ & $1.4$  \\
            $\mathrm{Dust:Gas}$ & $0.01$ & (1) & $0.01 - 0.008$ & $0.01$    \\
            $a_\mathrm{{min}}$ & $0.1\,\mathrm{\mu m}$ & (1) & $0.1-1.4~\mathrm{\mu m}$ & $0.39~\mathrm{\mu m}$ \\  
            $q_\mathrm{{dist}}$ & $3.0$ & (1) & $2.00 - 4.00$ & $3.06$  \\
            $T_\mathrm{{sub}}$ & $1600$ & (1) & $1400- 2000$ & $1600~\mathrm{K}$ \\ [1ex] 
            \hline
            \noalign{\smallskip}
            Dusty disk wind parameter & Literature value & Reference & Range explored & Best fit value \\
            \hline
            \noalign{\smallskip}
            $R_\mathrm{{0min}}$ & ... & ... &  $2.0 - 10.0~\mathrm{R_{\sun}}$ & $4.5~\mathrm{R_{\sun}}$ \\
            $T_\mathrm{{wind}~(near~surface)}$ & ... & ... & $1400 - 2400~\mathrm{K}$ & $1600~\mathrm{K}$ \\
            $\mathrm{Opening~Angle}$ & ... & ... & $25 - 55^\circ$ & $45^\circ$ \\
            $\dot{M}$ & ... & ... & $10^{-6} - 10^{-12}~\mathrm{M_{\odot}yr^{-1}}$ & $10^{-7}~M_{\odot}yr^{-1}$ \\
            \hline
            
        \end{tabular}
        
    \end{table*}

    \end{landscape}

    An into-wind mass outflow rate of $1\times10^{-7}\mathrm{ M_{\odot}yr^{-1}}$ was required to uplift enough material to reproduce the observed excess. Lower into-wind outflow rates or a lower dust-to-gas ratio do not allow enough material to exist exterior to the inner-rim to reproduce the observed K-band excess in the photometry and interferometric data. If one assumes an outflow to accretion ratio of $0.1$ one can predict an accretion rate of $1\times10^{-6}\mathrm{ M_{\odot}yr^{-1}}$. This is unphysically high for an object such as SU\,Aur given the expected age of the system ($5.18\pm0.13~\mathrm{Myr}$), assuming that the rate stayed constant over full period. In addition, this is in disagreement with non-detection of $\mathrm{Br\gamma}$ emission by \citet{Eisner14} which is suggestive of a lower outflow/accretion rate, however, no upper limit is derived. Other free parameters in this parameterisation of a dusty disk wind cannot reproduce the effect of a high into-wind mass accretion rate.
    
    The dust wind scenario has the effect of increasing the amount of the dust close to the star, where temperatures are sufficient for NIR flux contribution. As shown in Figure~\ref{fig:BK12} this model is in agreement with the observed NIR photometry, with a K-band flux of 102\% of the photometric value, well within the range of variability of SU\,Aur. The rest of the SED is still well fitted, as the curvature and shape are based upon the IN05 prescription described above. Another feature of Figure~\ref{fig:BK12} is the 'stair-stepping' structure to the disk wind. This is caused by the adaptive mesh refinement of TORUS, whereby mesh resolution is increased in areas of large density changes. The structure of this 'stair-stepping' was found to have a negligable impact on the results, as the areas of the disk wind that are important to the IR visibilities and SED are close to the disk rim where the disk mesh is sufficiently refined. The visibility curves shown provide a good fit to the data points, matching the lowest visibilities well. However, one cannot negate the potentially unphysically high outflow/accretion rates required to successfully model the data. The introduction of disk wind also had the interesting effect of flattening the 'bump' in the second visibility lobe present in many disk models, including the IN05 prescription. The $\chi^2$ fits for the visibilities and SED for all three models are shown in Table\,\ref{table:Chi2}.

    \begin{table}[b]
        \caption{
        $\chi^2$ values are only reduced by the number of data points due to complexity of the degrees of freedom in TORUS.
        }       
        \centering
        \begin{tabular}{c c c c} 
            \hline
            \noalign{\smallskip}
            Model  &  $\chi^{2}_{V^2}$ & $\chi^{2}_{CP}$ & $\chi^{2}_{SED}$   \\ [0.5ex]
            \hline
            \noalign{\smallskip}
            IN05 & 91.9 & 0.1 & 151.1  \\
            THM07 & 84.2 & 0.1 & 137.7 \\
            BK12 & 35.6 & 0.1 & 121.3 \\ [1ex] 
            \hline
        \end{tabular}
        
    \end{table}

\section{Discussion}\label{Diss}
    
    In investigating the circumstellar environment of SU Aur I have explored the structure and composition of the disk and greatly improved the constraints on the parameters initially taken from literature. The wide variety of techniques used to analyse the interferometric data allow us to precisely define the disk characteristics.
    
    Image reconstruction shows a disk inclined at $52.8^\circ\pm2.2$, this is in agreement with values of $63^{\circ+4}_{-8}$, \textasciitilde$60^\circ$ and \textasciitilde$50^\circ$ found by \citet{Akeson05,Unruh04,Jeffers14} respectively. The minor axis $\theta$ on the other hand was found to be $50.1^\circ\pm0.2$, greater than the literature values of $24^\circ\pm23$,$15^\circ\pm5$ and $\sim15^\circ$ found by \citet{Akeson05,Jeffers14,deLeon15} respectively. This difference is likely due to either: The poor uv coverage and lack of longer baselines in previous interferometric studies, both of which make estimating the position angle and inclination particularly unreliable. The image reconstruction also reveals evidence of slight asymmetries within the disk at $3$ sigma significance level, that are consistent with an inner disk rim seen at an intermediate inclination. As the K-band emission primarily traces the very inner region of the rim; if a disk is inclined the near side of the rim will be partially obscured from view, while the far side of the rim will be exposed to observation. This explanation can successfully account for the over-brightness observed in the eastern disk region in the image reconstruction shown in Fig.\,\ref{fig:newRec_SU1} and can also be seen in the radiative transfer image shown in Fig.\,\ref{fig:newRec_SU1}. Models of the effect of inclined disk on the observed brightness distribution are described by \citet{Jang13}. There are several other scenarios that have been used to explain asymmetries in protoplanetary disks in the past. Two possible scenarios are: Firstly, a shell ejection episode that can carry dust and gas away from the central star can be capable of reproducing the photometric variability in different epochs of observations \citep{Fernandes09,Kluska18}. Also, the presence of a companion embedded within the disk can create dust trapping vortices that capture dust grains substantially altering the aximuthal structure of the disk. Such assymetries, if present in the inner disk woudl substantially alter the visibilities along those position angles. These vortices, however, are known to trap primarily large grains (mm-size), not the small micron-sized grain I observe in the infrared, as shown by \citet{Kraus17,vdMarel13}. I rule out the presence of a companion by undertaking a companion search using the geometric models; a second point source was iterated through the parameter space with a grid size of $100$~mas in steps of $0.1$~mas (see Sec.\,\ref{ImRec}). However, the model fit did not improve significantly by adding an off-centre point source. I therefore favour the explanation of asymmetry arising from an inclined disk. 
    
    The geometric model fits were key in understanding the circumstellar environment of SU\,Aur. It was found that a skewed ring structure is able to fit the data best, which suggests that I trace a diffuse inner disk edge. This finding is consistent with the studies on other YSOs that found bright inner rims, such as the \citet{Lazareff17} survey of 51 Herbig AeBe stars using the PIONIER instrument at the VLTI. They found that over half of the disks could be successfully modelled using a diffuse ring structure. A high optical depth of the circumstellar material causes a bright inner rim, where most of the radiation is absorbed, scattered or re-emitted. The best-fit Skewed ring model suggests that the stellar-to-total flux ratio is $1.13$ and achieves a $\mathrm{\chi^2_{\mathrm{red}}}$ of $8.57$. \citet{Akeson05} find that $44\pm9\%$ of the total flux is in the SED K-band excess, with a $4\%$ of flux in an extended envelope, this is in good agreement with the values found from geometric modelling. The skewed ring model fits a ring of radius $0.17\pm0.02$\,au at an inclination of $51.2^\pm1.1^{\circ}$. These values are in remarkable agreement with both the values derived from image reconstruction in this study and the literature values of \citet{Akeson05} of $0.18\pm0.04$\,au and $62^{\circ+4}_{-8}$, where the slight differences are likely due to difference between the diffuse profile, skewed ring and the standard ring structures employed. The minor axis $\theta$ of $61.0\pm1.0^\circ$ is similar to the image reconstruction value, which is significantly larger than literature values. As above, this is most likely due to the poor uv coverage and short baselines available in previous interferometric studies, making the estimates of position angle and inclination particularly unreliable. The skewed ring also introduces modulated asymmetries into the ring profile.
    The skewed ring geometric model with azimuthal brightness modulation results in an improved fit with $\mathrm{\chi^2_{\mathrm{red}}}=1.66$, where the contrast of the asymmetry is consistent with the one found using image reconstruction techniques. Similarly this can be attributed to inclination effects of the viewing geometry \citep{Jang13}. However, the position angle of the asymmetry is not well constrained in these models.
    
    Radiative transfer modelling of the disk allowed us to fit a physical disk model to the visibility and photometry data simultaneously, meaning the 3-D density distribution of the disk can be explored. In this paper, three different geometries are considered: The single grain curved rim of IN05, the two grain curved rim of THM07 and the addition of a dusty disk wind of BK12 to the single grain curved rim. In the case of the IN05 prescription, I follow the idea that the sublimation temperature is gas-pressure dependent allowing the rim shape to be defined as described in Sec.\,\ref{SubRim}.  This model provides good constraints on the characteristic size of the near-infrared emitting region and the flaring in the colder regions, with a sublimation temperature of $1600$\,K corresponding to an inner radius of $0.12$\,au, slightly smaller than the literature values of $0.18\pm0.04$\,au \citep{Akeson05} and $0.17\pm0.08$\,au \citep{Jeffers14}. The disk was also found to be at an inclination of \textasciitilde$50^\circ$ and position angle of \textasciitilde$45^\circ$, in agreement with both the literature and above mentioned methods. This is shown to fit the photometry well at both shorter and longer wavelengths. However, there is a clear deficit in the IR excess, which also leads to poorly fitted visibilities due to an over-estimation of the stellar-to-total flux ratio. The same issue is obvious in the THM07 prescription, where larger grains are introduced allowing dust to exist closer to the star, though this model also fails to reproduce the visibility curve of the geometric modelling, as the inner radius of the disk is much smaller. A two grain model does not provide a good fit to these observations. 
    
    The dusty disk wind prescription of BK12 was incorporated into the single grain disk model of IN05. Dust flung out from the inner regions of the disk is carried far above and below the mid-plane. This dust is directly exposed to stellar radiation so is hot enough to contribute to NIR flux, whilst also obscuring the direct stellar flux. This was shown to be a physically viable scenario for YSOs, including SU\,Aur, by \citet{Konigl11} and \citet{Petrov19}. The resulting model reveals a disk with an inner radius of $0.15$\,au and a dust-to-gas ratio of $0.01$. The inner radius is in agreement with the literature values discussed above. The flaring parameters $\alpha_{disk}$ and $\beta_{disk}$ were fixed such that $\alpha_{disk} = \beta_{disk}+1$ and found to be $2.4$ and $1.4$ respectively. The dusty disk wind mechanism can directly reproduce the flux ratio in the K-band, allowing for an good visibility fit and an improved SED fit. As the disk wind rises above the mid-plane it also shields the cooler parts of the disk, reducing the longer wavelength flux compared to rim-only models. This was compensated for in the models by increasing the scale height of the disk to $15$\,au at a radius of $100$\,au. However, the implementation of the dusty disk wind in this scenario is not completely physical, owing to the high into-wind outflow rate of $1\times10^{-7}\mathrm{ M_{\odot}yr^{-1}}$ required. A potential solution in the form of an ingoing late infall event onto SU\,Aur is presented in the following chapter. The BK12 model also had the effect of flattening the second visibility lobe, a feature found in other YSOs \citep{Tannirkulam08,Setterholm18} and potentially opens powerful future modelling pathways for these objects. The grain size of the silicate dust species that produced the best fit was found to be $0.4\,\mathrm{\mu m}$ with no evidence of larger grains, as this addition resulted in a worse fit to the shape and magnitude of the NIR excess, particularly the shape of the Silicate feature around $10$\,um. This is in contrast to other inner disk studies where larger $1.2$\,um grains are required \citep{Kama09,Kraus2009,Davies18}.
    
    All the inner rim models investigated differ from the model proposed by \citet{Akeson05} whereby a vertical inner wall was combined with a small optically thick inner gas disk very close to the star aligned with the outer disk. This optically thick gas was implemented through very simple black-body emission models and allowed the author to successfully reproduced the observed NIR bump in excess flux. While TORUS could implement this type of black-body emission, it is unable to simulate the gas emission in a self-consistent physical manner owing to difficulty obtaining optical properties of refractive elements. As such we do not investigate such a model futher given such limited information.

\section{Conclusions}\label{Conc}
    This interferometric study of SU Aurigae has revealed the complex geometry and composition of the disk around SU\,Aurigae. I summarise the conclusions as follows:
    \begin{itemize}
        \item I reconstruct an interferometric image that confirms the inclined disk described in literature.  I see evidence for an asymmetry in the brightness distribution that can be explained by the exposure of the inner-rim on the far side of the disk and its obscuration on the near side due to inclination effects. The data set does not permit the imaging fidelity that would be needed to detect evidence of ongoing planetary formation within the inner disk, such as small-scale asymmetries, gaps or rings.
        \item I see no evidence for a companion, in either the reconstructed images nor in the geometric model fitting procedures.
        \item The simple geometric model fits reveal a disk of inclination $51.2\pm1.2^\circ$ along a minor axis position angle of $61.0\pm1.0^\circ$ and an inner radius of $1.12\pm0.12$\,mas ($=0.17\pm0.02$\,au). The disk is best modelled with a skewed ring which has a Gaussian ring width profile and sinusoidally modulated asymmetry. However, the poor $\chi^2_{\mathrm{red}}$ of this model fit means the uncertainties quoted here are likely not representative of the true range of values.
        \item Radiative transfer modelling shows that simple curved rim disk geometries of IN05 and THM07 cannot effectively model both the SED and visibility data. A deficit of NIR flux is obvious in the failure to reproduce the K-band observations. 
        \item A dusty disk wind scenario can successfully account for both the observed excess in the SED and the observed visibilities. The dusty disk wind scenario described here lifts material above the disk photosphere, thus exposing more dust grains to the higher temperatures close to the star responsible for the NIR excess. However, the high accretion rate required to reproduce the stellar-to-total flux ratio may make this scenario physically invalid.
        \item The best-fit model (dusty disk wind model) suggests that the dust composition in the disk is dominated by medium sized grains ($0.4\,\mathrm{\mu m}$) with a sublimation temperature of $1600$\,K. Introducing larger grains results in a worse fit to the SED shape and NIR excess. The disk is also shown to be highly flared ($15$\,au at $100$\,au).
        \item The dusty disk wind model predicts a rather flat visibility profile at long baselines. This class of models avoids the pronounced visibility 'bounce' that are associated with sharp edges in brightness distributions, as predicted by rim-only models. Therefore, these models may also open a pathway to physically model other YSOs that have been observed with $\gtrsim\,300$\,m infrared long-baseline interferometry, such as AB\,Aur, MWC\,275, and V1295\,Aql \citep{Tannirkulam08,Setterholm18} which all observe very flat long baseline visibility profiles.
    \end{itemize}

\chapter{Imaging the dusty disk wind environment of SU\,Aurigae with MIRC-X}\blfootnote{Large parts of this chapter will form part of a paper that is under preparation for submission to A\&A}
\label{ch:MIRCXSUAur}

\section{Introduction} \label{sec:intro}

    Following my previous study of the object SU\,Aurigae, new observations were obtained with the MIRC-X instrument. Since the publication of \cite{Labdon19} several interesting and relevant studies by other authors were made regarding SU\,Aur, in this introduction I shall introduce those studies and this new study. 

    Spectroscopic and photometric monitoring of SU\,Aur by \citet{Petrov19} has revealed that a dusty disk wind is the potential source of the photometric variability in both SU\,Aur and RY\,Tau. The characteristic time of change in the disk wind outflow velocity and the stellar brightness indicate that the obscuring dust is located close to the sublimation rim of the disk, in agreement with previous theoretical disk wind models \citep{Bans12,Konigl11}. 

    Recent ALMA and SPHERE observations by \citet{Ginski21} reveal a significant disk warp between the inner and out disks of $\sim70 ^\circ$. This misalignment is shown to cause large shadows on the outer disk as it blocks light from the central star. Their observations also suggest that SU\,Aur is currently undergoing a late infall event with significant amounts of material falling inwards from the outermost regions of the disk. Such events have the opportunity to significantly impact the evolution of the disk.

    This study presents one of the first 6-telescope interferometric studies of a YSO to date utilising state of the art observations covering a wider range of baseline position angles and lengths (up to 331\,m) in addition to 50\% greater resolution than previous works. Three different modelling methodologies were used to interpret out data and to provide direct comparisons to \citet{Labdon19}. (i)  Image reconstruction was used to obtain a model-independent representation of the data and to derive the basic object morphology. (ii)  Following this geometric model fitting allowed us to gain an appreciation for the viewing geometry of the disk by fitting Gaussian and ring models to the data. In addition, more complex geometric modelling was used to explore the chromaticity of the data. (iii) Finally, we combine interferometry and photometry to derive physical parameters with radiative transfer analysis, where our focus is on confirming the prescience of a dusty disk wind.

    \section{Observations} \label{Observations}

\begin{figure}
        \centering
        \includegraphics[scale=0.5]{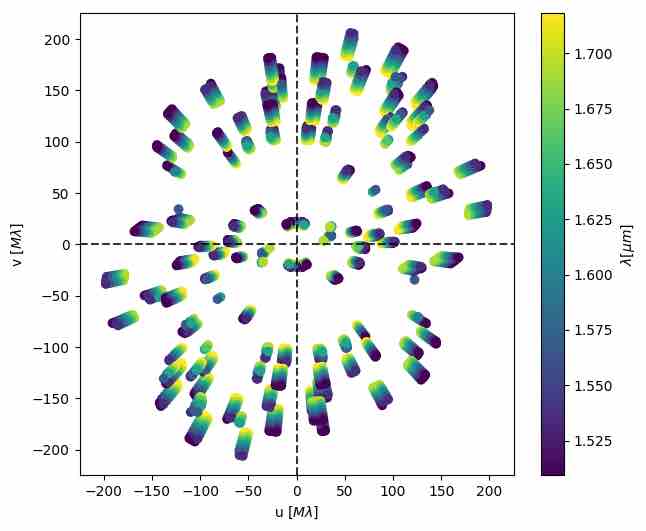}
        \caption[Coverage of the uv plane of the interferometric MIRCX observations of SU\,Aur]{Coverage of the uv plane of the interferometric MIRCX observations obtained with the CHARA array}
        \label{fig:uvplane_SU2}
    \end{figure}

        The CHARA array is a Y-shaped interferometric facility that comprises six $1\,$m telescopes. It is located at the Mount Wilson Observatory, California, and offers operational baselines between $34$ and $331\,$m \citep{Brummelaar05}. The MIRC-X instrument \citep{Kraus18}, a six-telescope beam combiner, was used to obtain observations in the near-infrared H-band ($\lambda=1.63\,\mu m, \Delta\lambda=0.35\,\mu m$) between September and October 2018. We obtained 11 independent pointings of SU\,Aur, using a mixture of 5 and 6-telescope configurations with maximum physical baseline of $331\,$m corresponding to a resolution of $\lambda/(2B) = 0.70\,\mathrm{mas}$ [milliarcseconds], where $\lambda$ is the observing wavelength and $B$ is the projected baseline. Details of our observations, and the calibrator(s) observed for the target during each observing session, are summarised in Table~\ref{table:LOG_SU2}. The uv\,plane coverage that we achieved for the target is displayed in Figure~\ref{fig:uvplane_SU2}. Our data covers an exceptionally wide range of baseline lengths and position angles, making the data ideally suited for image reconstruction.

        The MIRC-X data were reduced using the standard pipeline developed at the University of Michigan by (J.B. le Bouquin, N. Anugu, T. Gardner). The measured visibilities and closure phases were calibrated using interferometric calibrator stars observed alongside the target. Their adopted uniform diameters (UDs) were obtained from JMMC SearchCal \citep{Bonneau06, Bonneau11}, and are listed in Table\,\ref{table:LOG_SU2}. 

        Considering the short timescale over which the observations were taken the effect of time dependencies/variability of the object is thought to be minimal. However, care was taken to check for time dependencies in the visibilities of baselines of similar length and position angle. Variability in the NIR is known to be minimal \citep{Akeson05}, so any time dependencies in the visibility amplitudes is likely geometric. However, no significant time dependencies were discovered.

    \begin{landscape}
        \renewcommand*{\arraystretch}{0.6}
        \begin{longtable}{c c c c c } 
        \caption[MIRC-X observing log of SU\,Aur]{\label{table:LOG_SU2}Observing log from 2018 from the CHARA interferometer. All uniform disk (UD) diameters quoted obtained from \citet{Bourges14}.}
        \label{table:LOG_SU2}
        \endfirsthead
        \endhead
        
          \endlastfoot
          \hline
            \noalign{\smallskip}
            Date  &  Beam Combiner &   Stations  & Pointings & Calibrator (UD [mas]) \\ [0.5ex]
            \hline
            \noalign{\smallskip}
            2018-09-13 &  CHARA/MIRC-X   & S1-S2-E1-E2-W1-W2   & 2 & {HD 34499} ($0.256\pm0.007$) \\
            
            2018-09-16 &  CHARA/MIRC-X   & S1-S2-E1-E2-W1-W2   & 1 & {HD 28855} ($0.303\pm0.008$) \\
            
            2018-09-17 &  CHARA/MIRC-X   & S1-S2-E1-W1-W2   & 2 & {HD 40280} ($0.599\pm0.051$) \\
            
            2018-10-26 &  CHARA/MIRC-X   & S1-S2-E1-E2-W1-W2   & 6 & {BD+31 600} ($0.391\pm0.011$), \\ 
            & & & & {BD+44 1267} ($0.317\pm0.008$), \\
            & & & &{BD+43 1350} ($0.318\pm0.008$), \\
            & & & &{HD 28855} ($0.303\pm0.008$)\\
            [1ex] 
            \hline 
          \end{longtable}
          \end{landscape}

\section{Image Reconstruction} \label{ImRec}
\begin{figure*}[h!]
        \centering
        \includegraphics[scale=0.35]{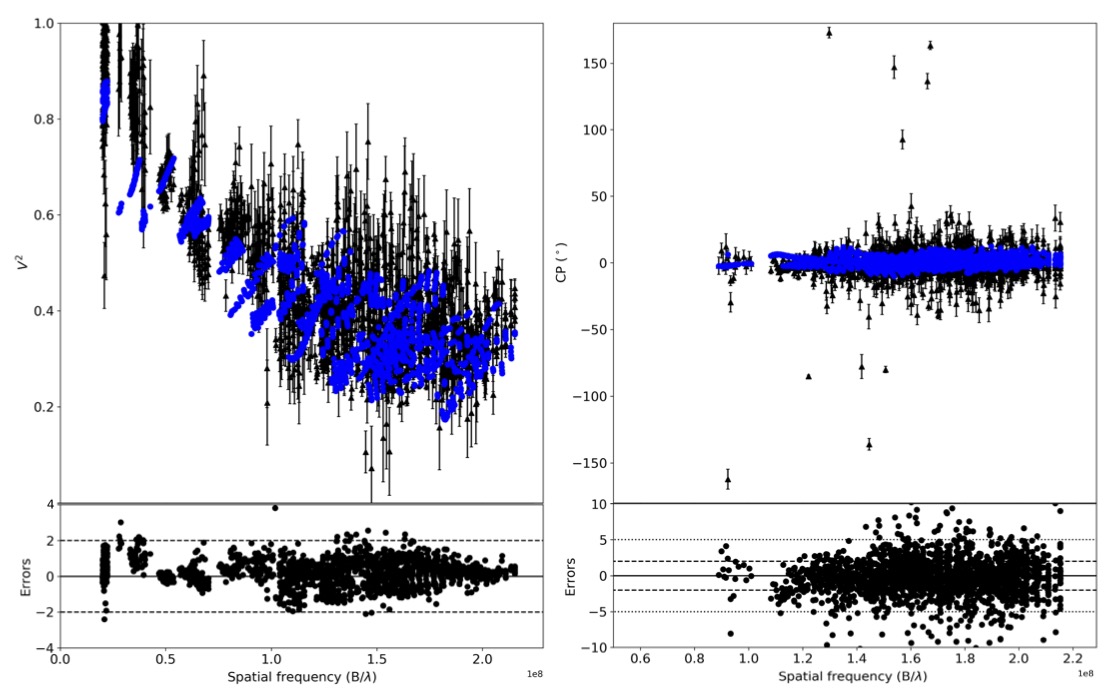}
        \caption[Visibilities and closure phases of the image reconstruction of SU\,Aur MIRC-X data]{Visibilities and closure phases of the image reconstruction. Black triangles with error bars are the original calibrated observables (squared visibilities on the left and closure phases on right), over plotted as blue circles is the model observables of the reconstructed image. Below each plot is the fit residuals normalised by the standard deviation as black circles. }
        \label{fig:ImRec_SU2_V2CP_SU2}
    \end{figure*}

    The 6-telescope observations from MIRC-X lends itself to the process of image reconstruction as the uv plane has been well sampled, though some small gaps remain in the position angle coverage. Of particular improtance is the large number of closure pahse measurements, which are shown in Figure\,\ref{fig:ImRec_SU2_V2CP_SU2}. There are many different algorithms with which to reconstruct images from interferometric data, but the process described here involved the use of the Polychromatic Image Reconstruction Pipeline (PIRP) which encompasses the $MiRA$ reconstruction algorithm by \citet{MiRA08}. For a detailed description of the $MiRA$ image reconstruction algorithm and its implementation, see Section\,6.3.

    \begin{figure*}
        \centering
        \includegraphics[scale=0.70]{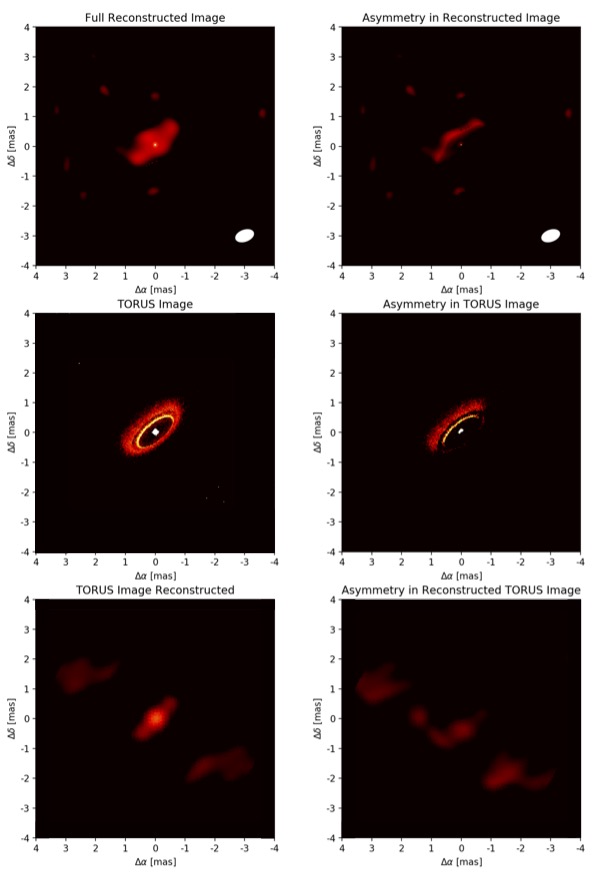}
        \caption[Image reconstruction and radiative transfer image of SU\,Aur computed from MIRC-X observation of SU\,Aur]{TOP LEFT: Image reconstruction resultant bootstrapped image, including beam size and orientation. MIDDLE LEFT: Radiative transfer image produced using TORUS including a dusty disk wind. BOTTOM LEFT: Reconstruction of simulated data created using the best fit TORUS image above, reconstruction parameters are equivalent to those of top left image. RIGHT COLUMN: Asymmetry maps of the left column, created by rotating the image $180^\circ$ and subtracting from original image. Colours are normalised intensity.}
        \label{fig:ImRec_SU2}
    \end{figure*}

    
    The final image is shown in Figure\,\ref{fig:ImRec_SU2} (top left panel), which also shows the asymmetry in the intensity map (top right panel). The asymmetry is calculated by rotating the image through $180^\circ$ and subtracting it from the un-rotated image. It is this residual flux that produces the non-zero closure phases, highlighting any areas of greater emission within the disk. The inclination of the disk appears to be greater than that found by \citet{Labdon19} with a similar minor-axis position angle. There also appears to be a central bulge along the minor disk axis likely caused by the over brightness of the star along this axis. The asymmetry map clearly shows a bright 'ribbon' along the south-west of the outer disk, parallel to the major axis of the disk. This is consistent with the asymmetry found by \citet{Labdon19} and is indicative of a highly inclined disk where the far side of the inner rim in directly exposed to the observer, while the nearside is obscured by flaring in the outer disk. 
    
    The visibility and closure phase fits of the image reconstruction are shown in Figure\,\ref{fig:ImRec_SU2_V2CP_SU2} (top panels) along with the residuals of the fit (bottom panels). The combined visibility and closure phase reduced chi-squared $\chi^2_{\mathrm{red}}$ of the image reconstruction was found to be $4.38$.

\section{Geometric Modelling} \label{GeoMod}
    In order to understand the geometry of the system one must consider the application of simple geometric models. In this section we explore several different approaches to modelling our data with both non-chromatic 'grey' models and techniques which explore the chromaticity.

    \subsection{Basic Geometric Models}
    
    The fitting of Gaussian and ring like distributions to the interferometric variables allows highly accurate estimations of the characteristic size, inclination and position angle of the object. In all models the central star is modelled as a point source, which is an acceptable assumption given the expected angular diameter of the star. The disk parameters are then fitted in the RAPIDO (Radiative transfer and Analytic modelling Pipeline for Interferometric Disk Observations) framework which utilises the Markov Chain Monte Carlo (MCMC) sampler corner to produce a fit and error estimate \citep{ForemanMackey16}. Three disk models were employed, a standard Gaussian brightness distribution which is characterised by its full-width-half-maximum (FWHM). Along with two ring models, a sharp ring with a width fixed to 20\% of the disk radius ($R$) and a 'skewed' ring with a more diffuse radial profile produced by convolving with a Gaussian with a FWHM, which is also a free parameter. The skewed ring is also capable of modelling azimuthal modulation or disk asymmetries, a detailed description of this model can be found in \citet{Lazareff17}. In addition to the model specific parameters, we also fitted in inclination ($INC$), minor-axis position angle ($PA$) and disk-to-total flux ratio ($f_{\mathrm{disk}}$). As we see no evidence of time variability in the data we are able to fit all data simultaneously. The results from the simple geometric model fitting are shown in Table\,\ref{table:geomodels_SU2}.

    \begin{table*}
        \caption[Geometric modelling results of fitting MIRC-X observations of SU\,Aur]{\label{table:geomodels_SU2}. Best fit parameters for the simple geometric models investigated. (*) The closure phase quoted is the achieved when allowing the skewed ring to become asymmetric. While the software did detect an asymmetry, it failed to constrain its location in the disk. PA is the minor-axis position angle of the disk and is measured from north ($PA = 0^\circ$) towards east.   
        }
        \centering
        \begin{tabular}{c c c c c} 
            \hline
            \noalign{\smallskip}
            Parameter & Explored Parameter Space & Gaussian & Ring & Skewed Ring    \\ [0.5ex]
            \hline
            \noalign{\smallskip}
            $R$ [mas] & $0.0 - 10.0$ & -- & $0.83\pm0.01$ & $0.17\pm0.16$\\
            $FWHM$ [mas] & $0.0 - 15.0$ & $1.52\pm0.01$ & -- & $0.75\pm0.04$ \\
            $INC$ [$^\circ$] & $0.0 - 90.0$ & $56.9\pm0.4$ & $57.4\pm0.4$ & $56.9\pm0.5$   \\
            $PA$ [$^\circ$] & $0.0 - 360.0$ & $55.9\pm0.5$ & $56.8\pm0.4$ & $55.8\pm0.5$ \\
            $f_{\mathrm{disk}}$ & $0.0 - 1.0$ & $0.43\pm0.01$ & $0.32\pm0.01$ & $0.43\pm0.01$ \\
            \hline
            \noalign{\smallskip}
            $\chi^2_{\mathrm{vis}}$ & & $11.63$ & $13.87$ & $11.62$ \\
            $\chi^2_{\mathrm{cp}}$ & & $6.05$ & $6.05$ & $6.01*$ \\
            \hline
            
        \end{tabular}
    \end{table*} 
    
    Out of the geometric models tested, the Gaussian model is considered to be the best fit. Even though the skewed ring produced a slightly small $\chi^2$ value for the closure phase and visibility measurements, we do not consider this significant given the additional complexities in the model. The Gaussian model finds a disk of FWHM $1.52\pm0.01\,\mathrm{mas}$ which is inclined at $56.9\pm0.4^\circ$ and a minor-axis position angle of $55.9\pm0.5^\circ$. In addition, we find that $43\pm1\%$ of the total flux originates from the disk in the H band. This is consistent with measurements based on the infrared excess of the spectral energy distribution (SED) \citep{Labdon19}. 
    
\begin{figure*}
        \centering
        \includegraphics[scale=0.7]{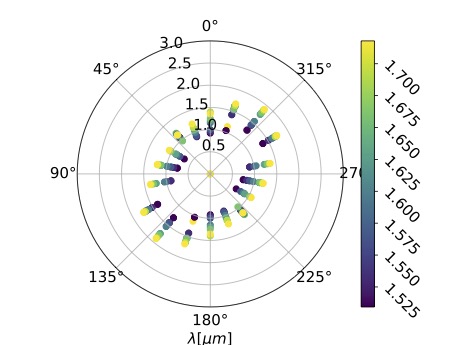}
        \caption[Results of fitting spectrally independent ellipses to MIRC-X data of SU\,Aur]{Geometric modelling Gaussian fit to the individual PA bins and spectral channels to highlight wavelength dependence. Colours represent MIRCX spectral channels as shown in the colour bar. Gaussian FWHM found for each position angle bin. Colour scheme is shared across plots.}
        \label{fig:RingPALamFIT_SU2}
    \end{figure*}

    The primary limitation of the simple geometric models described above is that they are intrinsically 'grey' in nature. Meaning they contain no spectral information, in other words, all 6 spectral channels of MIRCX are modelled using the same geometry, hence the large $\chi^2$ values obtained in the fitting process. In order to better model the spectral dependency of the visibility there are two geometric methods available. The first, and simplest is to model each spectral channel with a separate grey model. Figure\,\ref{fig:RingPALamFIT_SU2} shows the best fit Gaussian model for each spectral channel for each $20$ degree position angle bins. For these models the $INC$ and $PA$ were fixed to $0$ as position angles binned data can be considered orientation independent. The elongation of the object can clearly be seen, indicating a minor axis position angle of around $50 \mathrm{to} 60^\circ$. The wavelength dependence of the visibilities is also obvious, with short wavelengths corresponding to smaller ring radii. This is as expected given the hotter temperatures found at smaller radii with blackbody peaks at shorter wavelengths. The true temperature gradient of the inner disk can be found by applying more complex temperature gradient models that are able to account for observing wavelength.

    \subsection{Temperature Gradient Models} \label{sec:TGMs}
    
    The separation of the spectral channels for separate modelling is a tool of limited use. A more physically correct model can be applied by considering the temperature gradient of the disk.  A temperature gradient model (TGM) allows for the simultaneous fitting of interferometric and photometric observables. It is built up by several rings extending from an inner radius $R_{\mathrm{in}}$ to an outer radius $R_{\mathrm{out}}$. Each ring is associated with temperature and hence flux. Therefore, a model SED can be computed by integrating over the resulting blackbody distributions for each of the concentric rings. Such a model allows us to not only build up a picture of the temperature profile, but also approximate the position of the inner radius. The TGM is based upon $T_R = T_0(R/R_0)^{-Q}$ where $T_0$ is the temperature at the inner radius of the disk $R_0$, and $Q$ is the exponent of the temperature gradient \citep{Kreplin20,Eisner11}. A TGM represents an intrinsically geometrically thin disk. A point source is used at the centre of each model to represent an unresolved star, which is a reasonable approximation given the expected angular diameter of $0.05\,\mathrm{mas}$ \citep{Perez20}.

    \begin{figure}
        \centering
        \includegraphics[scale=0.6]{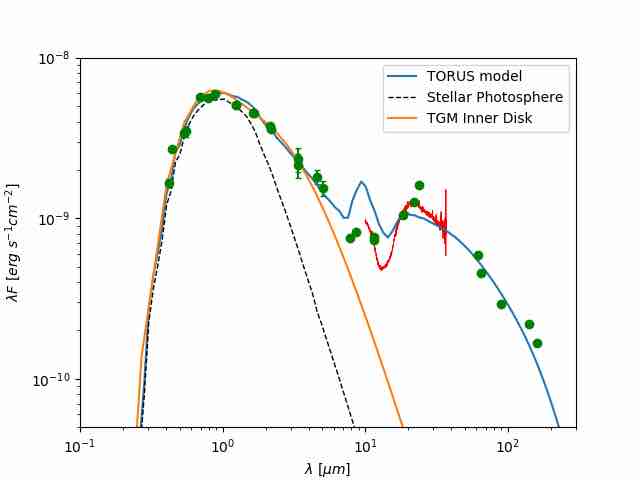}
        \caption[Spectral energy distribution of SU\,Aur and best fit temperature gradient and radiative transfer models]{Spectral energy distribution of SU\,Aurigae. Green points are photometric data from a variety of instruments (see Appendix\,B). Red line is Spitzer IR data. Black dashed line is direct radiation from the stellar photosphere. Blue line is the best TORUS computed radiative transfer model inclined at $56^\circ$. Orange line is the SED computed from the simple temperature gradient models described in Section\,\ref{sec:TGMs}.}
        \label{fig:SED_SU2}
    \end{figure}
    
\begin{figure}[b!]
        \centering
        \includegraphics[scale=0.3]{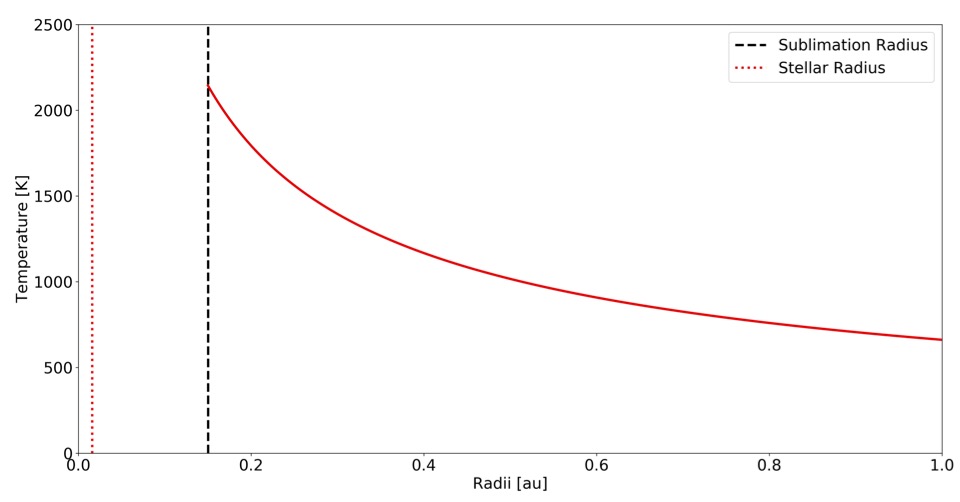}
        \caption[Derived temperature gradient of SU\,Aur from photometry and MIRC-X visibilities]{Temperature gradient profile of the inner disk of SU\,Aur. The red dashed line represents the stellar radius at 0.016\,au, the black dashed line represents the inner edge of the disk model at 0.15\,au which is equivalent to $9.2\,R_*$.}
        \label{fig:Gradient_SU2}
    \end{figure}

    The inclination and position angle of the disk are maintained at fixed values of $56.9\pm0.4^\circ$ and $55.9\pm0.5^\circ$ respectively, from the fitting of the Gaussian distribution. This was done to reduce the number of free parameters in the model. The fitting was undertaken using all of visibility data shown in Figure\,\ref{fig:ImRec_SU2} and all the SED points described in Appendix\,B simultaneously. The fitting and error computation was once again done using the MCMC sampler corner \citep{ForemanMackey16}. The temperature gradient modelling finds an inner disk radius of $0.15\pm0.04\,\mathrm{au}$ where the temperature is equivalent to $2100\pm200\,\mathrm{K}$ and decreases with an exponent of $Q=0.62\pm0.02$. These results are shown graphically in Figure\,\ref{fig:Gradient_SU2}. 
    
\section{Radiative Transfer} \label{RadTrans}

    We used the TORUS Monte-Carlo radiative transfer code \citep{Harries00}, allowing for the simultaneous fitting of visibility, closure phase and photometric data to further constrain the geometry and physical dust properties of the SU\,Aurigae circumstellar disk.
    
    The models adopted here are based on the disk models used by \citet{Labdon19}, adapted to account for the higher inclination and different observing wavelength. In these TORUS simulations, the dust was allowed to vertically settle and the dust sublimation radius was left as a free parameter, allowing the inner rim radius to define itself based on well-defined rules of the \citet{Lucy99} iterative method to determine the location and the temperature structure of the whole disk. For a full description of the disk model within TORUS see Section\,6.5.

     

    The dusty disk wind model is adapted from \citet{Bans12}. This mechanism is based on the presence of a large-scale, ordered magnetic field which threads the disk along which disk material is flung out. The high magnetic pressure gradient above the disk surface accelerates the material which is then collimated through the azimuthal and poloidal field components \citep{Bans12}. These centrifugally driven winds are highly efficient at distributing density above and below the plane of the disk, carrying angular momentum away from the disk surface. A full description of the implementation within the TORUS radiative transfer code can be found in \citet{Labdon19}.
    
    The disk model adopted follows the curved inner rim prescription of \citet{Isella05} with a density dependent sublimation radius whereby grains located in the disk midplane are better shielded due to higher densities and so can exist closer to the central star than grains in the less dense upper layers. A full summary of the disk parameters can be found in Table\,\ref{table:torus}. The key difference in the models described here compared to \citep{Labdon19} is the grain size adopted. Here we adopt a smaller grain size of $0.1\,\mathrm{\mu m}$, which in turn leads to a slightly smaller inner radius of $0.13\,\mathrm{au}$ and a hotter inner rim temperature of $2000\,K$. In order to improve the SED fit at longer wavelengths we also adopt a smaller scale height of $9.0\,\mathrm{au}$ at $100\,\mathrm{au}$.
    
    The presence of additional IR flux is once again required to fit both the visibilities and the SED, which is implemented in the form of a dusty disk wind. The absence of a disk wind fails to reproduce IR excess across both the H and K bands, with insufficient NIR disk flux. A disk wind is required to eject more hot dust above the midplane of the disk where it is directly exposed to stellar radiation which is reprocessed as an IR excess. As seen in the SED in Figure\,\ref{fig:SED_SU2} the current model overestimates the flux in the mid/far-IR, it is thought that this might be related to the disk warp predicted by \citet{Ginski21} and further modelling is required.

\begin{figure*}[h!]
        \centering
        \includegraphics[scale=0.53]{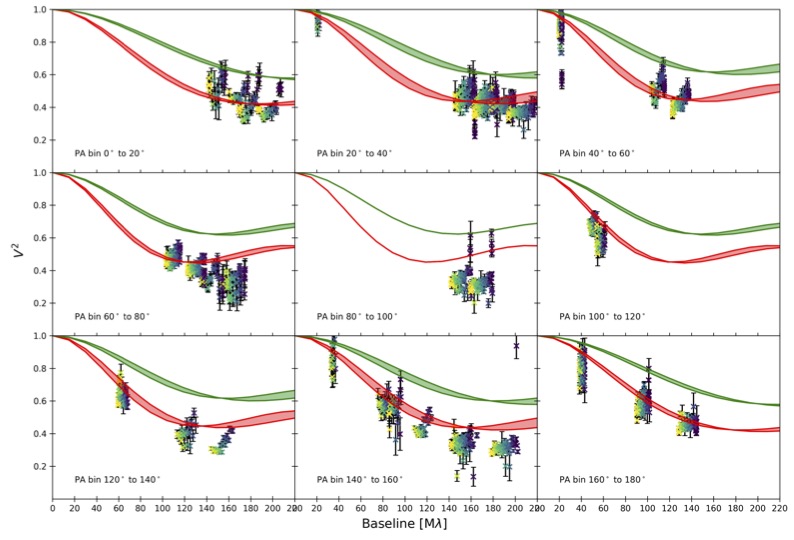}
        \caption[Results of radiative transfer modelling of MIRC-X data of SU\,Aur]{Results of radiative transfer modelling of MIRC-X data of SU\,Aur. Data split by baseline position angle into $20^\circ$ bins, colours of points represents wavelength on the same scale as other plots in this work, with darker colours representing short wavelengths. Red shaded areas are the visibility curve for the best fit TORUS radiative transfer model. The green shaded areas are the same model, but without the presence of a dusty disk wind.}
        \label{fig:PA_Split_SU2}
    \end{figure*}

    The detailed model fits are shown in Figure\,\ref{fig:PA_Split_SU2}, where the visibilities are split by position angle in $20^\circ$ bins. The green curves show the basic disk model without a dusty disk wind component, while the red curves show the same disk with the added disk wind component described above. The disk wind model provides a far superior fit to the observations, being able to successfully reproduce the NIR excess. The adopted disk wind parameters are the very similar as those used in \citet{Labdon19}, including the into-wind accretion rate of $10^{-7}\,\mathrm{M_\odot yr^{-1}}$. Considering a typical on-to-star accretion rate to into-wind accretion ratio of $0.01$, this level of transport is perhaps unphysically high given the age of the star. 

    \begin{landscape}
        \renewcommand*{\arraystretch}{0.6}
        \begin{longtable}{c c c c c} 
        \caption[Results of radiative transfer modelling of MIRC-X data of SU\,Aur]{\label{table:torus}Best-fit parameters resulting from SED and visibility fitting. (*) Computed using the TORUS radiative transfer model \citep{Harries00}. $q_\mathrm{{dist}}$ is the power law of the grain size distribution. $\mathrm{\alpha_{\mathrm{disk}}\,\,is\,\,fixed\,\,at\,\,(\beta_{\mathrm{disk}}+1)}$. (1) \citet{Akeson05}; (2)~\citet{Jeffers14}; (3)~\citet{Labdon19}}
        \label{table:LOG_Rad2}
        \endfirsthead
        \endhead
       
          \endlastfoot
          \hline
            \noalign{\smallskip}
            Parameter  & Literature value & Reference & Range explored &  Best fit value    \\ [0.5ex]
            \hline
            \noalign{\smallskip}
            $R_\mathrm{{inner}}$ & $0.18~\mathrm{au}$ & (1)(2) & $0.1 - 0.6~\mathrm{au}$ & $0.13~\mathrm{au}$ \\
            $R_\mathrm{{outer}}$ & $100~\mathrm{au}$ & (1) & $20.0-120.0~\mathrm{au}$ & $100.0~\mathrm{au}$    \\
            $h_\mathrm{0}$ & $15.0~\mathrm{au}$ & (3) & $7.0 - 20.0~\mathrm{au}$ & $9.0~\mathrm{au}$     \\
            $\alpha_\mathrm{{disk}}$ & $2.4$ & (3) & $1.0 - 3.0$ & $2.3$  \\
            $\beta_\mathrm{{disk}}=(\alpha-1)$ & $1.4$ & (3) & $0.0 - 2.0$ & $1.3$  \\
            $\mathrm{Dust:Gas}$ & $0.01$ & (1) & $0.01 - 0.008$ & $0.01$    \\
            $a_\mathrm{{min}}$ & $0.1\,\mathrm{\mu m}$ & (1) & $0.1-1.4~\mathrm{\mu m}$ & $0.09~\mathrm{\mu m}$ \\  
            $q_\mathrm{{dist}}$ & $3.0$ & (1) & $2.00 - 4.00$ & $3.06$  \\
            $T_\mathrm{{sub}}$ & $1600~\mathrm{K}$ & (1) & $1400- 2200$ & $2000~\mathrm{K}$ \\ [1ex] 
            \hline
            \noalign{\smallskip}
            Dusty disk wind parameter & Literature value & Reference & Range explored & Best fit value \\
            \hline
            \noalign{\smallskip}
            $R_\mathrm{{0min}}$ & $4.5$ & (3) &  $2.0 - 10.0~\mathrm{R_{\sun}}$ & $6.5~\mathrm{R_{\sun}}$ \\
            $T_\mathrm{{wind}~(near~surface)}$ & $1600~\mathrm{K}$ & (3) & $1400 - 2400~\mathrm{K}$ & $2000~\mathrm{K}$ \\
            $\mathrm{Opening~Angle}$ & $45^\circ$ & (3) & $25 - 55^\circ$ & $45^\circ$ \\
            $\dot{M}$ & $10^{-7}~M_{\odot}yr^{-1}$& (3) & $10^{-6} - 10^{-12}~\mathrm{M_{\odot}yr^{-1}}$ & $10^{-7}~M_{\odot}yr^{-1}$ \\
            \hline
          \end{longtable}
          \end{landscape}

\begin{figure}[h!]
        \centering
        \includegraphics[scale=0.5]{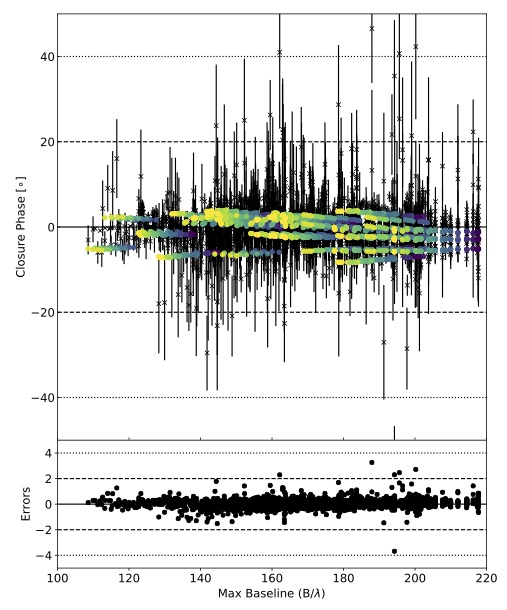}
        \caption[Closure phases of MIRC-X data of SU\,Aur and radiative transfer model fit]{Black crosses are the closure phase data obtained with MIRCX. Overlaid as coloured points are the TORUS model closure phases, the colours represent the wavelength of the spectral channels and follows the same convention as other plots in this work. Below in black points are the normalised residual errors of the fit. }
        \label{fig:Closure_TORUS_SU2}
    \end{figure}

    The final computed image is shown in Figure\,\ref{fig:ImRec_SU2} (middle left panel) and shows the clear asymmetry originating from the inclination in the asymmetry map in the same figure (middle left panel). In order to approximate what this computed image would look like if observed at the same resolution as the original observations, we computed synthetic visibility and closure phases based on the radiative transfer images. Artificial noise and error bars were computed to be representative of the original data and to ensure an accurate representation. These synthetic observable were then reconstructed in the same manner as the original data, as described in Section\,\ref{ImRec}. Care was taken to ensure the constancy of the reconstruction parameters for both the real and synthetic observables. The reconstructed TORUS image is also shown in Figure\,\ref{fig:ImRec_SU2} (bottom right panel), and shows clear similarities with both the original TORUS image and the image reconstructed from the original data. A comparison between the observed and model closure phases is shown in Figure\,\ref{fig:Closure_TORUS_SU2}, the TORUS model is shown to recreaste the scale of the closure phases (considering the error bars) well with relatively small residuals. 

    \begin{figure*}[h!]
        \centering
        \includegraphics[scale=0.4]{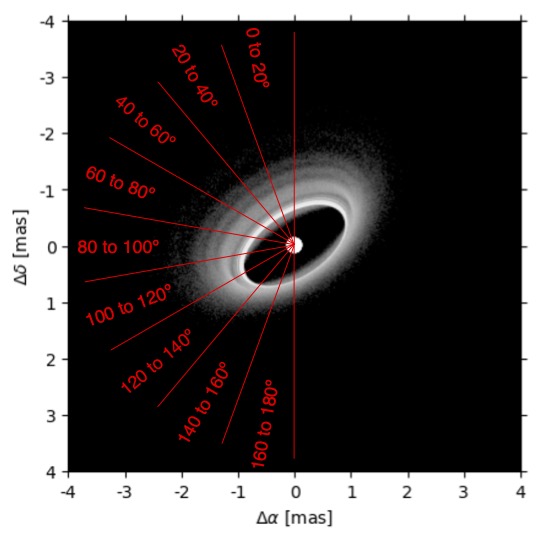}
        \caption[Annotaed radiative transfer image of SU\,Aur]{Radiative transfer image of SU\,Aur created with TORUS. Overlaid with position angle bins plotted in Figure\,\ref{fig:PA_Split_SU2}.} 
        \label{fig:Timage_SU2}
    \end{figure*}
    
\section{Discussion} \label{Discuss}

    Our extensive observations and analysis of the circumstellar environment of SU\,Aurigae have revealed the details of the inner disk in unprecedented detail. The wide variety of techniques used to analyse our interferometric data allows us to precisely define the disk characteristics. 
    
    Image reconstruction is a crucial, model independent, method of analysis which is ideally suited to our dataset with extensive uv and baseline coverage. Our analysis reveals a highly elliptical shape, indicative of an object with a high inclination. There appears to be a central bulge to the disk, however this feature is not thought to be physical but rather a manifestation of the brightness of the central star combined with the width of the disk at this point. The thinner 'arms' of the image are thought to be a depiction of the far-side of the disk rim which is un-obscured by the outer disk. This is confirmed by the asymmetry map shown alongside the image is Figure\,\ref{fig:ImRec_SU2}. There is a significant asymmetric feature in the form of a thin brightness on the south-eastern edge of the disk. The unique shape of this feature indicates that this is again caused by the high inclination obscuring the nearside disk rim. The effect of an inclined disk on the observed brightness distribution is described extensively by \cite{Jang13} and accurately describes the observations here. The scale and shape of the image is similar to that of \citet{Labdon19}, with a slightly higher inclined viewing angle. We consider the images of this work to be the more accurate depiction of SU\,Aur given the higher quality observations, taken over a much shorter timescale, with the added detail and resolution this entails.

    Although geometric modelling is much more constrained in the geometries it can explore, in does provide a more quantitative view of the disk. It was found that the model which best fit our data was a simple Gaussian distribution with a point source representing the star. A Gaussian model is consistent with other work, both on this object by \citet{Labdon19}, but also in other YSO studies such as the survey by \citet{Lazareff17} who find that little under half of their 51 objects can be modelled by a Gaussian structure. Although the data quality of this work is variable, with many objects only modelled on a sparse uv coverage, meaning more objects may appear Gaussian in nature than truly are. The Gaussian fitted in this work has a FWHM of $1.52\pm0.01\,\mathrm{mas}$ ($0.239\pm0.002\,\mathrm{au}$) at an inclination of $56.9^\circ\pm0.4$ at a minor axis position angle of $55.9^\circ\pm0.5$ with a stellar-to-total flux ratio of $0.57\pm0.01$. The reduced $\chi^2$ value for the visibilities is $11.63$ and $6.05$ for the closure phases, which are equivalent to $0^\circ$ for this centro-symmetric model. These values are in agreement with the literature values of \citet{Akeson05} who find a K band radius of $0.18\pm0.04$\,au and an inclination of $62^{\circ+4}_{-8}$. Similar values for the inclination in literature are \textasciitilde$60^\circ$ and \textasciitilde$50^\circ$ found by \citet{Unruh04,Jeffers14} respectively. The minor axis position angle derived here is significantly greater than the literature values of $24^\circ\pm23$ and $15^\circ\pm5$ found by \citet{Akeson05,Jeffers14}. This difference is likely due to either: The poor uv coverage and lack of longer baselines in previous interferometric studies, both of which make estimating the position angle and inclination particularly unreliable. Other non-interferometric studies focus on the outer disk, rather than the inner au-scale regions. 
    
\begin{figure}[h!]
        \centering
        \includegraphics[scale=0.4]{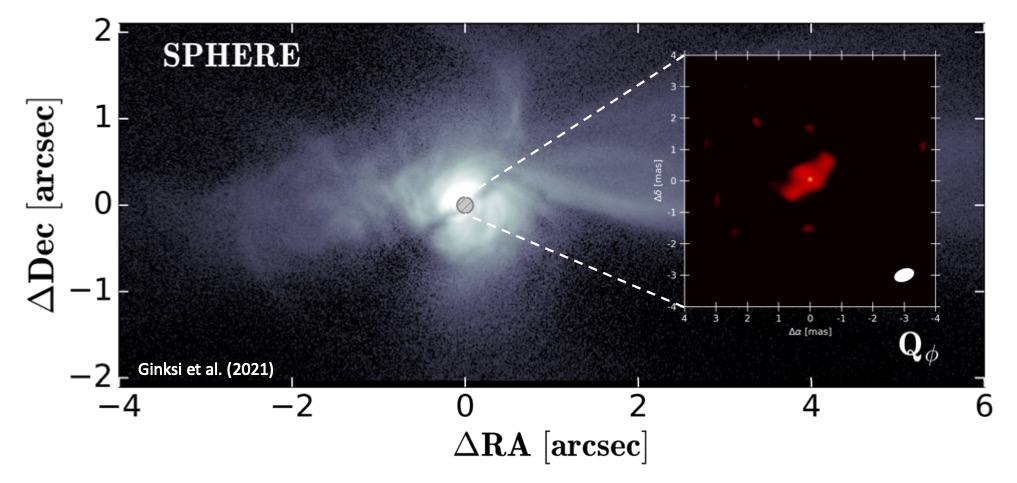}
        \caption[SPHERE image of SU Aur inlaid with MIRC-X image]{SPHERE/IDRIS image of SU\,Aurigae system, the H band $\mathrm{Q_\phi}$ polarised light signal, a dark shadow is visible along the centre of disk which is aligned with the major axis position angle of the MIRC-X reconstructed image insert. SPHERE image taken from \citet{Ginski21}.}
        \label{fig:GinksiComp}
    \end{figure}

    The geometric modelling results are broadly similar to those presented in our previous work \citep{Labdon19} where an inclination of $50.9\pm1.0^\circ$ and minor axis position angle of $60.8\pm1.2^\circ$ were found and the data were marginally better described by a ring-like brightness distribution. The values and models presented here are considered to be more accurate due high precision observations and significantly smaller potential for temporal variations. These values are also consistent with observations of the outer disk by \citet{Ginski21} where dark shadows are observed in scattered light origination from a significant disk warp between the inner and outer regions. On larger scales the near-side of disk is seen to the north-east, in our observations it is seen to the south west. Figure\,\ref{fig:GinksiComp} shows the comparison between the SPHERE image of \citet{Ginski21} and the reconstructed image from this work. The SPHERE image probes scattered polarised IR light on significantly larger scales. 
    
    In modelling the temperature gradient of SU\,Aur we can gain an appreciation for spectral dependence of our interferometric variables across the 6 spectral channels of MIRCX. Our modelling finds a disk which extends down to $0.15\pm0.04\mathrm{au}$ where the temperature is equivalent to $2100\pm200\,K$. The presence of such hot dust is unlikely for silicate and graphite domianted disks and potentially suggests the prescence of refractory grain species, likely at a much lower dust to gas ratio. However, these results cannot confirm this, but simply find evidence of emission conistent with such temperatures. The temperature gradient and decreases with an exponent of $0.62\pm0.02$. The outer edge of this temperature regime was found to be $0.20\pm0.03\mathrm{au}$, showing this this prescription only covers the very innermost regions of the disk. Modelling outer regions of the disk is beyond the scope of this paper, as our NIR interferometric data does not cover emission from these regions. The temperature gradient exponent found here lies between two established models from literature. That of \citet{Pringle81} who find that a steady state, optically-thick accretion disk heated by viscous processes will exhibit an exponent of $0.75$ and of \citet{Kenyon87,Dullemond04} who show that a flared disk heated by reprocessed stellar radiation alone will exhibit an exponent of $0.5$ or less \citep{Chiang97}. As such, we infer that the circumstellar environment of SU\,Aurigae is not heated by stellar radiation alone, but additional heating processes must also be present in the inner disk. One potential issue with these interpritations is that the NIR regions probed in this work are only sensitive to the upper layers of the disk, as the disk is optically thick in the NIR. As physical disks are not vertically isothermal the tempersature between the midplane and upper layers will likely differ. However, \citet{Adams87,Kenyon87} find that for the innmost regions of a disk the discrepancy between midplane and surface temperature gradients is very small.

    The radiative transfer modelling presented in this paper is heavily based on LA19. For a detailed discussion of the motivation behind certain choices, particularly in relation to the shape of the inner rim, we recommend the reader see there. In this work we are able to achieve a similar SED fit to LA19, even with the adoption of a smaller $0.1\,\mu m$ grain size. The smaller grain size is in line with previous radiative transfer work of SU\,Aur by \citet{Akeson05}. The smaller grain size results in a smaller and hotter inner rim which now extends down to $0.13\,\mathrm{au}$ at a temperature of $2000\,K$. This is within the uncertainties of values predicted by the temperature gradient modelling and is roughly consistent with older literature values of $0.18\pm0.04\,\mathrm{au}$ and $0.17\pm0.08\,\mathrm{au}$ by \citet{Akeson05} and \citet{Jeffers14} respectively. The flaring parameters $\alpha_{\mathrm{disk}}$ and $\beta_{\mathrm{disk}}$ were fixed such that $\alpha_{\mathrm{disk}} = \beta_{\mathrm{disk}} + 1$ and found to be $2.3$ and $1.3$ respectively. Similarly to LA19, a dusty disk wind is required in order to fit both the SED across the NIR and visibilities as shown in Figure\,\ref{fig:PA_Split_SU2}. A dusty disk wind allows material to be listed from the disk surface along magnetic field lines where it can reprocess stellar radiation to contribute to the NIR excess. However, the implementation of the dusty disk wind in this scenario is not completely physical, owing to the high into-wind outflow rate of $1\times10^{-7}\mathrm{ M_{\odot}yr^{-1}}$ required. If one assumes an outflow to accretion ratio of $0.1$ the resulting onto-star-accretion rate is greater than those typically found in T\,Tauri stars. In particular spectroscopic measurments of the accretion rate of SU\,Aur by \citet{Calvet04} find a moderate rate of $0.5-0.6\times10^{-8}\mathrm{ M_{\odot}yr^{-1}}$ However, the suggestion that SU\,Aur is undergoing a late stage infall event could be a potential explanation for such a high level of mass transport through the system.  
    
    The simplest way to directly compare the analytical methods employed in this work, is through the images produced. Figure\,\ref{fig:ImRec_SU2} shows a collage of the images produced by the geometric modelling, image reconstruction and radiative transfer methods. It clearly shows the similarities between the images all with very similar inclinations and position angles with asymmetries oriented in the same direction. The asymmetries all appear to be the result of the high inclination of the disk causing the near-side inner rim to be shadowed by the flared outer disk. The lowest panel of Figure\,\ref{fig:ImRec_SU2} shows an image reconstructed from synthetic visibilities and closure phases obtained from the radiative transfer image. Care was taken to ensure the observables were matched in baseline length and position angle and detector mode, while the image reconstruction process was exactly that described in Section\,\ref{ImRec}

\section{Conclusions}

This interferometric study of SU Aurigae has revealed the complex geometry and composition of the disk around SU\,Aurigae. We summarise our conclusions as follows:
    
    \begin{itemize}
        \item We reconstruct an interferometric image that confirms the inclined disk described in literature.  We see evidence for an asymmetry in the brightness distribution that can be explained by the exposure of the inner-rim on the far side of the disk and its obscuration on the near side due to inclination effects. 
        
        \item Our simple geometric model fits show that the circumstellar environment is best modelled as a Gaussian distribution with a disk of inclination $56.9\pm0.4^\circ$ along a minor axis position angle of $55.9.0\pm0.5^\circ$ and an FWHM of $1.52\pm0.01\,\mathrm{mas}$ ($0.239\pm0.002\,\mathrm{au}$). Such geometry is consistent with strong disk shadows observed in the outer disk originating from a disk warp. 
        
        \item We model the radial temperature profile of the inner disk and find a disk which extends down to $0.15\pm0.04\mathrm{au}$ where the temperature is equivalent to $2100\pm200\,K$ and decreases with an exponent of $0.62\pm0.02$. 
        
        \item A dusty disk wind scenario is still required to account for both the observed excess in the SED and the observed visibilities. The dusty disk wind scenario described here lifts material above the disk photosphere, thus exposing more dust grains to the higher temperatures close to the star responsible for the NIR excess. The high accretion rate required to reproduce the stellar-to-total flux ratio could be explained by a late stage infall event. 
        
        \item Our best-fit model (dusty disk wind model) suggests that the dust composition in the disk is dominated by small grains ($0.1\,\mathrm{\mu m}$) with a sublimation temperature of $2000$\,K. Introducing larger grains results in a worse fit to the SED shape and NIR excess. The disk is also shown to be highly flared with a scale height of $15$\,au at $100$\,au.
        
    \end{itemize}

\chapter{Viscous Heating and Boundary Layer Accretion in the Disk of an Outbursting Young Star}
\blfootnote{Large parts of this chapter are published in \citet{Labdon20}}
\label{ch:FUOri}

FU\,Orionis is the archetypal FUor star, a subclass of YSO that undergo rapid brightening events, often gaining 4-6 magnitudes on timescales of days. This brightening is often associated with a massive increase in accretion; one of the most ubiquitous processes in astrophysics from planets and stars to super-massive black holes. We present multi-band interferometric observations of the FU\,Ori circumstellar environment, including the first J-band interferometric observations of a YSO. We investigate the morphology and temperature gradient of the inner-most regions of the accretion disk around FU\,Orionis. We aim to characterise the heating mechanisms of the disk and comment on potential outburst triggering processes.
   
Recent upgrades to the MIRC-X instrument at the CHARA array allowed the first dual-band J and H observations of YSOs. Using baselines up to 331\,m, we present high angular resolution data of a YSO covering the near-infrared bands J, H, and K. The unprecedented spectral range of the data allows us to apply temperature gradient models to the innermost regions of FU\,Ori.
   
We spatially resolve the innermost astronomical unit of the disk and determine the exponent of the temperature gradient of the inner disk to $T\propto r^{-0.74\pm0.02}$. This agrees with theoretical work that predicts $T\propto r^{-0.75}$ for actively accreting, steady state disks, a value only obtainable through viscous heating within the disk, assuming it is flared. We find a disk which extends down to the stellar surface at $0.015\pm0.007\,\mathrm{au}$ where the temperature is found to be $5800\pm700\,\mathrm{K}$ indicating boundary layer accretion as the disk is continuous down to the stellar surface. We find a disk inclined at $32\pm4^\circ$ with a minor-axis position angle of $34\pm11^\circ$. We demonstrate that J-band interferometric observations of YSOs are feasible with the MIRC-X instrument at CHARA. The temperature gradient power-law derived for the inner disk is consistent with theoretical predictions for steady-state, optically thick, viscosly heated accretion disks.

\section{Introduction} \label{sec:intro}

    Accretion onto astronomical objects is one of the most fundamental processes in astrophysics and facilitates mass transport onto a wide range of astrophysical objects, from planets and stars to super-massive black holes \citep{Lin96}. The mass transport proceeds through accretion disks, where viscosity transports angular momentum outwards, thus enabling the mass infall \citep{Pringle72}. A key prediction is that the disk viscosity should actively heat the disk, where the radial temperature profile has been predicted as far back as the 1970’s \citep{Shakura73,Shibazaki75,Hartmann85}.
    
    The radial temperature gradient of circumstellar disks are determined by the heating mechanisms that power them. The two primary heating mechanisms in protoplanetary disks are the reprocessing of stellar radiation and viscous heating \citep{DAlessio05}. Stellar radiation is reprocessed through absorption, re-emission, and scattering of photons and is most effective in the outer layers of the disk owing to the optical depth of material\citep{Natta01}. Viscous heating on the other hand is thought to be most prevelant in the mid-plane, owing to higher densities of material, where it is thought to be driven by turbulence and instabilities in the disk material, such as magneto-rotation instabilities (MRI) \citep{Balbus98}. However, the presence and nature of viscosity is highly debated and relatively unconstrained by current observational data. 

    The appreciation of accretion and viscosity processes in young stellar objects (YSOs) is vital to the understanding of both star- and planet-formation mechanisms. Active accretion disks have been observed around a wide range of YSO classes, however YSOs are known to be 10-100 times less luminous than expected from steady-state accretion scenarios. Particularly given the accretion rates of the order $10^{-7}$ to $10^{-8}\,\mathrm{M_\odot yr^{-1}}$ observed around many YSOs. This raises the possibility that accretion is not consistent across the early stages of stellar evolution, but is episodic \citep{Kenyon95,Evans09}. Such scenarios may manifest in outbursting events.


    
    Many stars are known to undergo episodic accretion events on sifferent timescales, such as EX\,Lupi objects which undergo short outbursts on regular periods. However, it is FU\,Orionis (FUor) stars that are perhaps the best known. FUors are characterised by rapid brightening events \citep{Audard14} followed by a protracted period of dimming, on the order of decades to centuries to return to the quiescence state  \citep{Hartmann96,Herbig07,kra16}. In such an outburst the accretion rates can increase
    to $\sim 10^{-4}\,\mathrm{M_\odot yr^{-1}}$. It is now thought that most YSOs stars exhibit episodic accretion, undergoing at least one or more outburst events throughout their lifetime \citep{Hartmann96,Audard14}.

    The triggering mechanism of outbursts in FUor-type stars is not well understood. Various scenarios have been proposed to explain the massive increase in accretion rate seen in these objects. \citet{Vorobyov05,Vorobyov06} propose gravitational instabilities on large scales cause the disk to fragment and for clumps of material to fall onto the central star. \citet{Bell94} suggest that a thermal instability in the very inner regions ($<0.1$\,au) could be sufficient to cause outbursts of this magnitude. \citet{Bonnell92} propose that a binary companion on a highly eccentric orbit could perturb the disk and cause repeated outburst of accretion. Similarly, \citet{Reipurth04} suggest that FUor stars are newly created binary systems, where the two stars become bound following the breakup of larger multiple systems. Such a scenario leads to the ejection of companions and the rapid infall of material. However, given the limited number of known FUors there is little consensus on the FUor outburst triggering mechanism.
    
    FU\,Orionis (FU\,Ori) is the archetypal FUor object located in the Lambda-Orionis star forming region at a distance of $416\pm8\,\mathrm{pc}$ \citep{BailerJones18}, obtained from GAIA parallax measurements. FU\,Ori is known to be a binary system with an actively accreting companion located $0.5"$ (200\,au at 416\,pc) to the South. Despite being \textasciitilde$4$\,mag fainter (in the V band), FU\,Ori~S is thought to be the more massive object ($1.2\,\mathrm{M_\odot}$) \citep{Beck12}, as it is highly embedded, hence only appearing fainter and less massive. Meanwhile the mass of the Northern component has been estimated to $0.3\pm0.1\,\mathrm{M_\odot}$ based on modelling of the Spectral Energy Distribution (SED) \citet{Zhu07}. In 1937 FU\,Ori~N underwent a rapid brightening event whereby its brightness increased from $15.5^m\pm0.5$ to $9.7^m\pm0.1$ \citep{Herbig66,Clarke05} in the photographic system (similar spectral response to Johnson B filter). Since reaching a peak shortly after outburst its magnitude has decreased steadily at around $0.0125^{m}$ per year in B band ($540\,\mathrm{nm}$).
    
    FU\,Ori is one of the best-studied YSOs due to its current brightness and unique nature. Scattered light observations by \citet{Takami14} and \citet{Laws20} with Subaru and GPI respectively reveal a large spiral arm structure in the north-west of the disk extending around 200\,au that has been attributed to a gravitational instability in the outer disk. Additionally, they reveal a stripe/shadow in the Northern disk and a diffuse outflow extending to the east, which is tentatively attributed to a jet in polar direction. \citet{Eisner11} observed FU\,Ori using the Keck interferometer in high resolution K-band mode ($\lambda = 2.15\,-\,2.36$, $R=2000$). They found a temperature gradient of $T\propto R^{-0.95}$ for the inner disk using a single baseline. Additionally, \citet{Liu19} presented medium resolution GRAVITY observations ($\lambda = 2.0\,-\,2.45$, $R=500$) which they model as an off-centre face-on Gaussian object. They also highlight $\mathrm{H_2O}$ and CO absorption features that are consistent with a $\Dot{M}\sim10^{-4}\,\mathrm{M_\odot yr^{-1}}$ accretion disk model. ALMA observations taken by \citet{Perez20} show a disk inclined at $\sim37^\circ$ at a position angle of $134\pm2^\circ$, both values are aligned to a disk detection around the southern component. Using methods outlined by \citet{Zhu07}, \citet{Perez20} were able to constrain the stellar parameters of the northern component based on the disk geometry. They find a stellar mass of $0.6\,M_\odot$, with a mass accretion rate of $3.8\times10^{-5}\,\mathrm{M_\odot yr^{-1}}$. They go further to derive an inner disk radius of $3.5\,R_\odot$ or $0.016\,\mathrm{au}$. Mid-infrared (MIR) interferometry in the N-band ($8-13\,\mathrm{\mu m}$) and SED analysis by \citet{Quanz06} derive a temperature gradient for the outer disk ($>3\,\mathrm{au}$) which is in good agreement to what can be found for vertically isothermal flared disks in theoretical work \citep{Kenyon87}.

    In non-outbursting low-mass YSO (T\,Tauri stars) the mass-infall is believed to proceed through magnetospheric accretion columns operating between the inner disk and the photosphere \citep{Bouvier07}.  It is not yet understood how the accretion geometry differs in highly-accreting FUors, which also limits our understanding of the outburst-triggering mechanisms. In order to further our understanding of these processes I have conducted the first tri-waveband NIR interferometric study of a YSO at the CHARA array using MIRC-X in the J and H bands. These include simultaneous J- and H-band observations using MIRC-X, the first of their kind. This data is complemented with observations in the K-band from the "CLassic Interferometry with Multiple Baselines" (CLIMB) beam combiner. The observations are introduced in Section\,\ref{Observations}. The modelling techniques and temperature gradient analysis are shown in Section\,\ref{GeoMod}. The details of the companion search are outlined in Section\,\ref{Comp} and I discuss the implications and draw conclusions in Sections\,\ref{discussion} and \ref{Conclusion}, respectively.

    \section{Observations} \label{Observations}
    
\begin{figure}[h]
        \centering
        \includegraphics[scale=0.35]{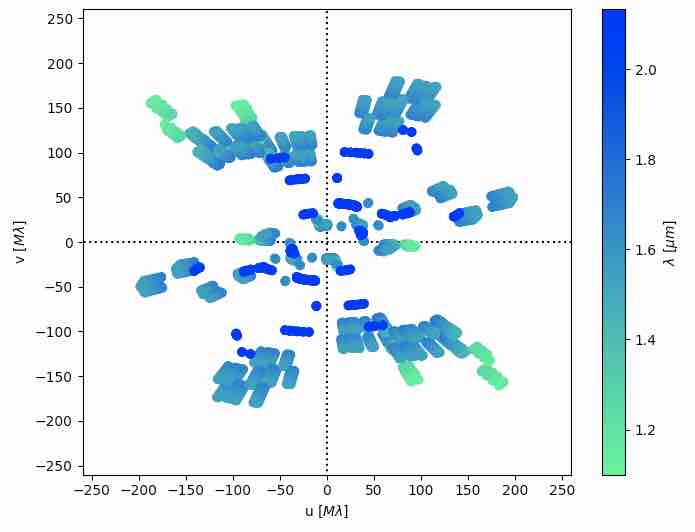}
        \caption[uv-plane of all interferometric observations of FU\,Ori]{The uv-plane of all observations detailed in Table\,\ref{table:LOG_FU}. The H+J dual-band observations were obtained with CHARA/MIRC-X. The H-band only observations were obtained with CHARA/MIRC-X with $R=50$ spectral resolution and with VLTI/PIONIER in free mode, while the K band observations were obtained with CHARA/CLIMB and PTI.}
        \label{fig:uv_map_FU}
    \end{figure}

    To collect the data for this multi-wavelength study, a variety of instruments operating at a range of wavelengths were employed. A summary of the observations is provided in Table\,\ref{table:LOG_FU} and the resultant uv-coverage is shown in Figure\,\ref{fig:uv_map_FU}.

    \subsection{CHARA/MIRC-X H-band Observations}
    
     The primary instrument used in this study is the "Michigan InfraRed Combiner-eXeter" (MIRC-X), which is a six-telescope beam combiner instrument located at the CHARA array.
     The CHARA array is a Y-shaped interferometric facility that comprises six $1\,$m telescopes. It is located at the Mount Wilson Observatory, California, and offers operational baselines between $34$ and $331\,$m \citep{Brummelaar05}. The MIRC-X instrument \citep{Monnier04,Kraus18,Anugu20} is recording data routinely in H-band ($\lambda=1.63\,\mathrm{\mu m}, \Delta\lambda=0.35\,\mathrm{\mu m}$ with $R=\lambda/\Delta\lambda=50$) since late 2017 \citep[first results in][]{Kraus20}.
     
     Overall I obtained 4 independent H-band pointings of FU\,Ori with MIRC-X, using a mixture of 5 and 6-telescope configurations in bracketed calibrator-science concatenation sequences. A maximum physical baseline of $331\,$m was used, corresponding to a maximum resolution of $\lambda/(2B) = 0.50\,\mathrm{mas}$ [milliarcseconds], where $\lambda$ is the observing wavelength and $B$ is the projected baseline.
     
     The data were reduced using the MIRC-X data reduction pipeline v1.2.0\footnote{https://gitlab.chara.gsu.edu/lebouquj/mircx\_pipeline} to produce calibrated squared visibilities and closure phases. The MIRC-X detector is susceptible to bias in the bispectrum estimation as pointed out by \citet{Basden04}. The pipeline corrects for this with a method similar to Appendix C of that paper. The measured visibilities and closure phases were calibrated using interferometric calibrator stars observed alongside the target; the calibrators are listed in Table\,\ref{table:LOG_FU}. The adopted uniform diameters (UDs) were obtained from JMMC SearchCal \citep{Bonneau06, Bonneau11}.
    
    \subsection{CHARA/MIRC-X J+H band Observations}
    
    Recent developments have allowed the wavelength of MIRC-X to be extended to allow for simultaneous H and J-band observations through the implementation of longitudinal dispersion compensators (LDCs) to correct for atmospheric dispersion between the two bands \citep{Berger03}. LDCs consist of a wedge of SF-10 glass which is moved across the beam to increase or decrease the thickness of glass depending on the total airpath of the interferometric system. At the time, the LDCs were tracked manually resulting in sub-optimal fringe contrast in the J band, however a consistent approach was used across calibrator and science stars to ensure accurate calibration could be obtained. Such observations were conducted for the first time in November 2019, with automated LDC control now in the late stage of development \citep{Anugu20}.
    
    The data presented in this paper represents the first successful J-band interferometric observations of a YSO. These dual-band observations correspond to 14 spectral channels across wavelengths $1.08\,\mathrm{to}\,1.27\,\mathrm{\mu m}$ and $1.41\,\mathrm{to}\,1.73\,\mathrm{\mu m}$. The gap in the band pass is due to the presence of the CHARA metrology laser at around $1.3\,\mathrm{\mu m}$, this was removed using a narrow-band 'notch' filter. 
    
    I obtained two J+H band pointings on FU\,Ori with MIRC-X in 2019, using a 4-telescope configuration in CAL-SCI concatenation sequences. Only a 4-telescope configuration corresponding to the lowest spatial frequency fringes can be used for dual-band observations. Recording with 5 or 6 telescopes would result in the highest spatial frequency being undersampled (sub-Nyquist) on the detector. Of the available baselines, a maximum physical baseline of $280\,$m was used corresponding to a maximum resolution of $0.34\,\mathrm{mas}$.
    
    These data were reduced using an adapted version of the MIRC-X data reduction pipeline v1.2.0. The UDs of the calibrator stars were obtained from JMMC SearchCal \citep{Bonneau06, Bonneau11}. The spectral dependence of the UDs between the J and H bands is small enough to be considered negligible by the calibration pipeline.

    \subsection{VLT/PIONIER H-band Observations}

    FU\,Ori was recorded with the PIONIER instrument \citet{PIONIER11}. PIONIER is a four telescope beam combiner operating in the H-band ($\lambda = 1.64\,\mathrm{\mu m}$) at the VLTI. Data was obtained in December 2017 without a dispersive element (FREE mode) and reduced using the standard PNDRS pipeline \citep{JBLB2011}.

    \subsection{VLTI/CLIMB K-band Observations}
    
    I present observations obtained with the CLIMB instrument \citep{Brummelaar13}, also located at the CHARA array. CLIMB is a three telescope beam combiner that was used to obtain near-infrared K-band data ($\lambda=2.13\,\mathrm{\mu m}, \Delta\lambda=0.35\,\mathrm{\mu m}$) between November 2009 and October 2011. The CLIMB data were reduced using pipelines developed at the University of Michigan that are optimised for recovering faint fringes from low visibility data.
    
    Archival K-band data were also available from the Palomar Testbed Interferometer \citep[PTI,][]{Colavita99} from 1998 to 2008 using a two-telescope beam combiner on 3 different physical baselines between $86$ and $110$\,m. The data on FU\,Ori was published in \citet{Malbet98,Malbet05}. These additional measurements complement the CHARA observations in the short to intermediate baseline range; the full uv coverage is shown in Figure\,\ref{fig:uv_map_FU}.
    
    \begin{landscape}
    
    \renewcommand*{\arraystretch}{0.6}
    \begin{longtable}{c c c c c c} 
    \caption[Observing log of FU\,Ori from all instruments]{Observing log of all instruments, data spanning 10 years from 1998 to 2019.} 
    \label{table:LOG_FU}
    \endfirsthead
    \endhead
        \hline
        \noalign{\smallskip}
        Date  &  Beam Combiner & Filter &   Stations  & Pointings & Calibrator (see Appendix\,C) \\ [0.5ex]
        \hline
        \noalign{\smallskip}
        2019-11-07 &  CHARA/MIRC-X  & J+H & S1-S2-W1-W2 & 2 & HD\,64515 ($0.47\pm0.01$)  \\
        \hline
        \noalign{\smallskip}
        2018-11-27 &  CHARA/MIRC-X & H  & S1-S2-E1-E2-W1-W2   & 2 & HD\,246454 ($0.60\pm0.01$) , HD\,38164 ($0.43\pm0.01$)  \\
        2019-11-06 &  CHARA/MIRC-X  & H  & S1-S2-E1-W1-W2    & 1 & HD\,28855 ($0.30\pm0.01$)  \\
        \hline
        \noalign{\smallskip}
        2017-12-25 &  VLTI/PIONIER  & H  & D0-G2-J3-K0   & 1 & HD\,37320 ($0.19\pm0.01$), HD\,39985 ($0.19\pm0.01$) \\
        \hline
        \noalign{\smallskip}
        2016-11-24 & VLTI/GRAVITY & K  & D0-G2-J3-K0 & 2 &  HD\,36814 ($0 .75\pm0.06$), HD\,38494 ($0.71\pm0.06$) \\
        2016-11-25 & VLTI/GRAVITY & K  & D0-G2-J3-K0 & 2 & HD\,38494 ($0.71\pm0.06$) \\
        \hline
        \noalign{\smallskip}
        2010-10-02 &  CHARA/CLIMB  & K  & S2-E2-W2 & 3 & HD\,38164 ($0.43\pm0.01$)  ,HD\,42807 ($0.45\pm0.01$) \\
        2010-11-30 &  CHARA/CLIMB  & K  & S1-E1-W1 & 2 & HD\,38164 ($0.43\pm0.01$)   \\
        2010-12-04 &  CHARA/CLIMB  & K  & S2-E1-W2 & 5 & HD\,28527 ($0.43\pm0.03$)\\
        2011-10-27 &  CHARA/CLIMB  & K  & S2-E2-W2 & 1 & HD\,38164 ($0.43\pm0.01$)   \\
        2011-10-29 &  CHARA/CLIMB  & K  & S1-E1-W1 & 1 & HD\,38164 ($0.43\pm0.01$)   \\[1ex] 
        \hline
        \noalign{\smallskip}
        1998-11-14 &  PTI  & K  & N-S   & 2 & HD\,37147 ($0.62\pm0.01$) , HD\,35296 ($0.62\pm0.06$), \\
        & & & & & HD\,37147 ($0.62\pm0.01$), HD\,28910 ($0.54\pm0.05$) \\
        1998-11-16 &  PTI  & K  & N-S   & 2 & HD\,37147 ($0.62\pm0.01$), HD\,35296 ($0.62\pm0.06$), \\
        & & & & &  HD\,28910 ($0.54\pm0.05$) \\
        
        1998-11-17 &  PTI  & K  & N-S   & 12 & HD\,37147 ($0.62\pm0.01$), HD\,35296 ($0.62\pm0.06$), \\
        & & & & &  HD\,37147 ($0.62\pm0.01$) \\
        
        1998-11-19 &  PTI  & K  & N-S   & 5 & HD\,42807 ($0.45\pm0.01$), HD\,35296 ($0.62\pm0.06$), \\
        & & & & &  HD\,32301 ($0.50\pm0.06$), HD\,37147 ($0.62\pm0.01$) \\
        
        1998-11-22 &  PTI  & K  & N-S   & 9 & HD\,42807 ($0.45\pm0.01$), HD\,35296 ($0.62\pm0.06$), \\
        & & & & &  HD\,37147 ($0.62\pm0.01$), HD\,32301 ($0.50\pm0.06$) \\
        
        1998-11-23 &  PTI  & K  & N-S   & 4 & HD\,42807 ($0.45\pm0.01$), HD\,35296 ($0.62\pm0.06$), \\
        & & & & &  HD\,37147 ($0.62\pm0.01$) \\
        1998-11-24 &  PTI  & K  & N-S   & 6 & HD\,42807 ($0.45\pm0.01$), HD\,35296 ($0.62\pm0.06$), \\
        & & & & & HD\,37147 ($0.62\pm0.01$) \\
        1998-11-25 &  PTI  & K  & N-S   & 6 & HD\,42807 ($0.45\pm0.01$), HD\,35296 ($0.62\pm0.06$), \\
        & & & & & HD\,37147 ($0.62\pm0.01$) \\
        1998-11-26 &  PTI  & K  & N-S   & 5 & HD\,42807 ($0.45\pm0.01$), HD\,37147 ($0.62\pm0.01$) \\
        
        1998-11-27 &  PTI  & K  & N-S   & 7 & HD\,42807 ($0.45\pm0.01$), HD\,37147 ($0.62\pm0.01$) \\
        
        1999-11-24 &  PTI  & K  & N-S   & 11 & HD\,42807 ($0.45\pm0.01$), HD\,37147 ($0.62\pm0.01$) \\
        
        1999-11-25 &  PTI  & K  & N-S   & 1 & HD\,42807 ($0.45\pm0.01$), HD\,32301 ($0.50\pm0.06$), \\
        & & & & &  HD\,50635 ($0.50\pm0.06$), HD\,32923 ($0.99\pm0.11$) \\
        
        1999-11-26 &  PTI  & K  & N-S   & 6 & HD\,42807 ($0.45\pm0.01$), HD\,37147 ($0.62\pm0.01$),  \\
        & & & & & HD\,50635 ($0.50\pm0.06$), HD\,32923 ($0.99\pm0.11$) \\
        
        1999-11-27 &  PTI  & K  & N-S   & 16 & HD\,42807 ($0.45\pm0.01$), HD\,37147 ($0.62\pm0.01$), \\
        & & & & &  HD\,32301 ($0.50\pm0.06$), HD\,50635 ($0.50\pm0.06$),  \\
        & & & & & HD\,32923 ($0.99\pm0.11$) \\
        
        1999-11-28 &  PTI  & K  & N-S   & 21 & HD\,42807 ($0.45\pm0.01$), HD\,37147 ($0.62\pm0.01$), \\
        & & & & &  HD\,32301 ($0.50\pm0.06$), HD\,50635 ($0.50\pm0.06$), \\
        & & & & &  HD\,32923 ($0.99\pm0.11$) \\
        
        2000-11-19 &  PTI  & K  & N-S   & 9 & HD\,42807 ($0.45\pm0.01$), HD\,32923 ($0.99\pm0.11$), \\
        & & & & &  HD\,37147 ($0.62\pm0.01$), HD\,46709 ($1.57\pm0.17$) \\
        
        2000-11-20 &  PTI  & K  & N-W   & 14 & HD\,42807 ($0.45\pm0.01$), HD\,32923 ($0.99\pm0.11$), \\
        & & & & &  HD\,37147 ($0.62\pm0.01$) \\
        
        2000-11-22 &  PTI  & K  & N-W   & 3 & HD\,42807 ($0.45\pm0.01$), HD\,32923 ($0.99\pm0.11$), \\
        & & & & & HD\,37147 ($0.62\pm0.01$) \\
        
        2000-11-26 &  PTI  & K  & N-W   & 13 & HD\,42807 ($0.45\pm0.01$), HD\,32923 ($0.99\pm0.11$),  \\
        & & & & & HD\,43042 ($0.68\pm0.08$), HD\,37147 ($0.62\pm0.01$) \\
        
        2003-11-19 &  PTI  & K  & N-W   & 3 & HD\,43042 ($0.68\pm0.08$), HD\,42807 ($0.45\pm0.01$), \\
        & & & & & HD\,46709 ($1.57\pm0.17$) \\
        2003-11-20 &  PTI  & K  & N-S   & 2 & HD\,43042 ($0.68\pm0.08$), HD\,42807 ($0.45\pm0.01$),  \\
        & & & & & HD\,46709 ($1.57\pm0.17$) \\
        2003-11-21 &  PTI  & K  & S-W   & 2 & HD\,43042 ($0.68\pm0.08$), HD\,42807 ($0.45\pm0.01$),  \\
        & & & & & HD\,46709 ($1.57\pm0.17$) \\
        2003-11-27 &  PTI  & K  & S-W   & 1 & HD\,43042 ($0.68\pm0.08$), HD\,42807 ($0.45\pm0.01$),  \\
        & & & & & HD\,46709 ($1.57\pm0.17$) \\
        2004-10-12 &  PTI  & K  & N-W   & 3 & HD\,43042 ($0.68\pm0.08$), HD\,42807 ($0.45\pm0.01$),  \\
        & & & & & HD\,35956 ($0.36\pm0.01$), HD\,46709 ($1.57\pm0.17$) \\
        2004-11-12 &  PTI  & K  & S-W   & 6 & HD\,42807 ($0.45\pm0.01$), HD\,43042 ($0.68\pm0.08$),  \\
        & & & & & HD\,35956 ($0.36\pm0.01$), HD\,46709 ($1.57\pm0.17$) \\
        2004-12-12 &  PTI  & K  & N-W   & 6 & HD\,42807 ($0.45\pm0.01$), HD\,43042 ($0.68\pm0.08$),  \\
        & & & & & HD\,35956 ($0.36\pm0.01$), HD\,46709 ($1.57\pm0.17$) \\
        2008-10-25 &  PTI  & K  & N-W   & 5 & HD\,42807 ($0.45\pm0.01$), HD\,37147 ($0.62\pm0.01$),  \\
        & & & & & HD\,32923 ($0.99\pm0.11$) \\
        2008-11-14 &  PTI  & K  & N-S   & 1 & HD\,42807 ($0.45\pm0.01$), HD\,37147 ($0.62\pm0.01$),  \\
        & & & & & HD\,32923 ($0.99\pm0.11$), HD\,32301 ($0.50\pm0.06$) \\
        \hline
    \end{longtable}
    \end{landscape}

    \subsection{Photometric Observations}
    
    Photometric observations were collected from a variety of sources in order to build up the spectral energy distribution (SED) of FU\,Ori. Where possible, care was taken to minimize the time difference between observations and the number of instruments used. In particular, photometric data taken during the 1998 to 2019 period of the interferometric observations was considered whenever possible. A full list of the photometric observations used can be seen in Appendix\,\ref{appendix3}.

\section{Modelling and Results}\label{Modelling}

    \subsection{Presentation of Results} \label{Results}
    
\begin{figure}
        \centering
        \includegraphics[scale=0.31]{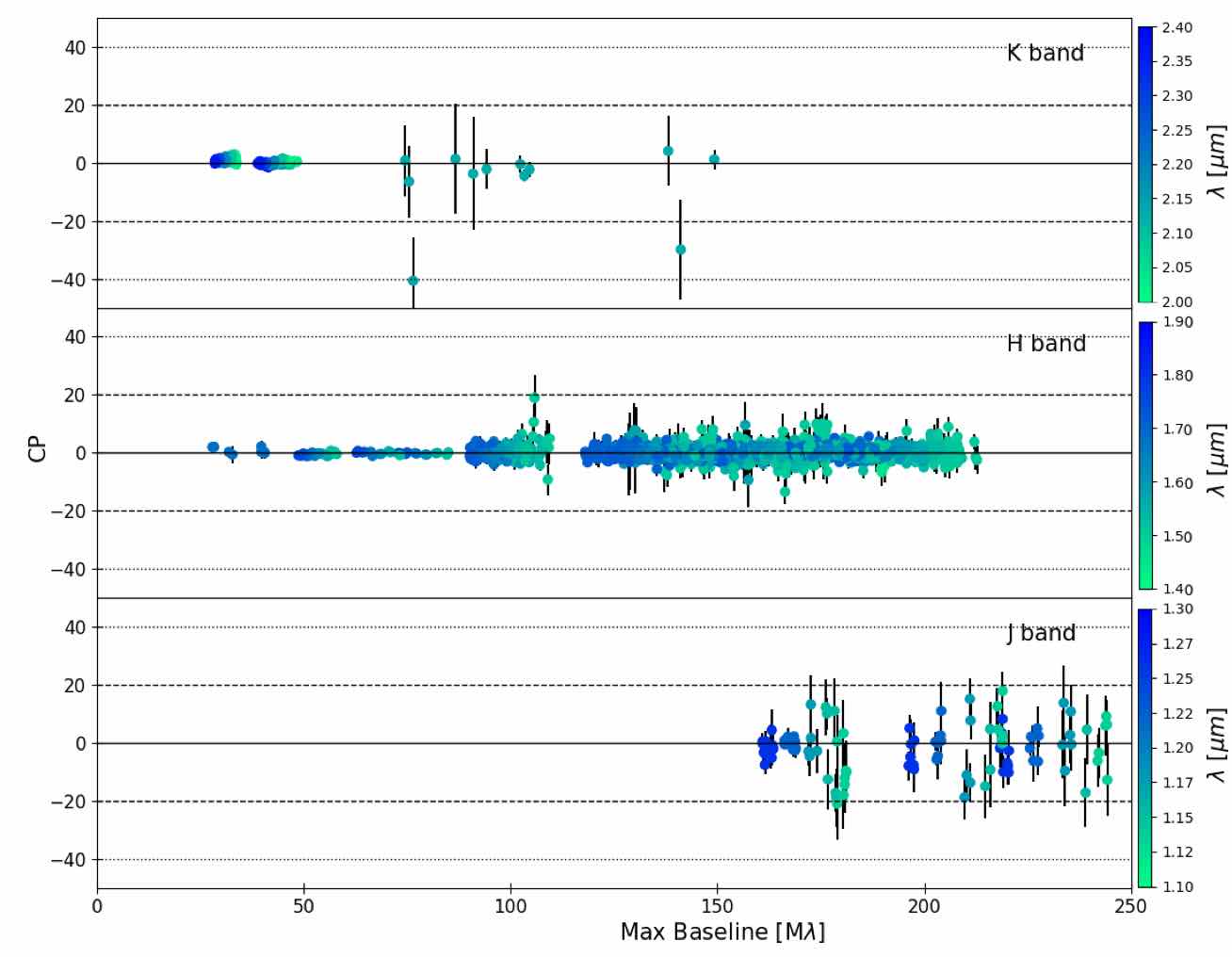}
        \caption[Closure phases from all observations of FU\,Ori]{Closure phases for each waveband plotted against maximum baseline length, coloured according to observing wavelength. Shorter wavelengths in each band correspond to lighter green colours while longer wavelengths in each band are represented as dark blue.}
        \label{fig:CPs_FU}
    \end{figure}

\begin{figure*}
        \centering
        \includegraphics[scale=0.24]{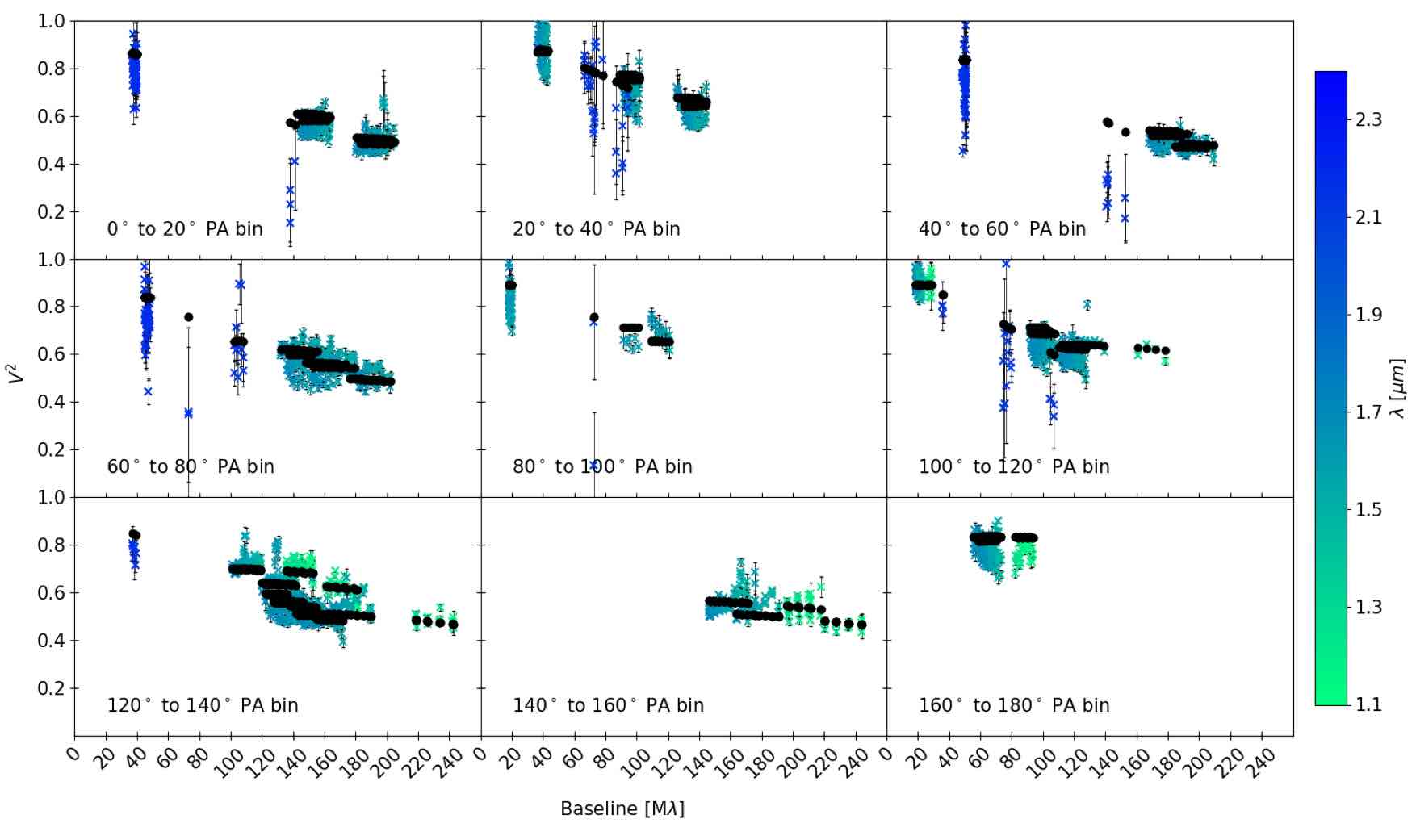}
        \caption[Squared visibilities of FU\,Ori and resultant temperature gradient model fit, split by position angle]{Squared visibilities plotted against spatial frequency, split by baseline position angle into $20^\circ$ bins. The blue/green crosses represent the interferometric observations across all instruments, coloured according to observing wavelength: dark blue is K-band data, light blue is H-band and green is J-band data. The black circles are the model visibilities of the best fit temperature gradient model corresponding to each data point.}
        \label{fig:PAmodel_FU}
    \end{figure*}

     The closure phases and squared visibilities obtained for all instruments are shown in Figures\,\ref{fig:CPs_FU} and \ref{fig:PAmodel_FU}, respectively. The visibilities are split into position angle bins of $20^\circ$ and coloured according to wavelength.

     The majority of the observations are contained within the MIRC-X, PIONIER and GRAVITY data which were taken over the relatively short period of 3 years. Hence the effect of photometric variability in the NIR of the object is thought to be minimal. On the other hand, the CLIMB and PTI observations were taken over a significantly longer timescale, were a small amount of photometric variability may be expected. Based on the established decreasing trend in the magnitude of 0.0125 per year in the B band, the expected drop change in the visibility based on the changing stellar-to-disk flux ratio between 1998 and 2019 is 0.045. However, I am confident that the large error bars caused by the poor signal-to-noise of these observations will successfully account for variability on the visibility and closure phase measurements. Even so, care was taken to check for time dependencies in the visibilities of baselines of similar length and position angle. K-band squared visibilities were binned according to the year of the observation and compared with each other. Each night of H-band data was compared separately to other nights. While J-band data from MIRC-X could not be compared directly, the H-band data taken simultaneously could be compared with other nights. None of these checks revealed any time variability in the data beyond the noise level, hence all interferometric data can be studied together. Also, I fitted the model to the post-2009 interferometric data alone and obtain values that are consistent on the $1.3\sigma$-level with those obtained from fitting the complete data. This confirms that any potential temporal variability does not affect the results significantly.

     The closure phases (Fig\,\ref{fig:CPs_FU}) are consistent with $0^\circ$ within the error bars, indicating a centro-symmetric brightness distribution. The reduced chi-square for the closure phase measurements \citep[$\chi^2_{\mathrm{red-cp}}$, see definition in][]{Kraus2009} for a centro-symmetric model (i.e.\ with closure phases of $0^\circ$ along all triangles) is $3.65$.
     
     \subsection{Simple Geometric Models} \label{GaussMods}

     As a first step for interpreting the recorded interferometric observables I fitted simple geometric models to the data. This allows for the characteristic size, inclination and position angle of the object to be derived. I employ a Gaussian model within the RAPIDO (Radiative transfer and Analytic modelling Pipeline for Interferometric Disk Observations) framework. RAPIDO utilises the Markov chain Monte Carlo (MCMC) sampler {\it emcee} \citep{emcee13} to produce a fit and error estimate. The Gaussian model employed is a 'grey' model, in that it contains no spectral information. As such the three wavebands of the observations were fitted separately. In addition to a Gaussian model a secondary unresolved, extended component was also required. The free parameters of the model were the full-width-half-maximum (FWHM) of the Gaussian, the inclination (INC), minor axis position angle (PA) and the flux of the unresolved, extended component ($\mathrm{F_{bg}}$), which is measured as a percentage of the total flux in the model. 
     
     Table\,\ref{table:GuassFit_FU} summarises the results of the Gaussian fitting. The inclination ($\sim37^\circ$) and position angle ($\sim40^\circ$) of the object are consistent across all three wavebands as expected. The size of the disk as characterised by the FWHM of the Gaussian shows an increase in size with increasing wavelength from 0.38\,mas in the J-band to 0.60\,mas in the K-band. This is expected given that longer wavelengths probes cooler regions of the disk, found at a larger distance from the central star. The flux contained within the extended component also shows a spectral dependency with a lower flux contrast at shorter wavelengths of only $2.5\%$ in the J band compared to $8.6\%$ in the K band. 
     
     A Gaussian model is intrinsically centro-symmetric and as such has a closure phase of $0^\circ$ across all triangles. This is a very good approximation to the observations, were very small closure phase signals are measured. A reduced $\chi^2_{red-cp}$ value for the closure phases of $3.65$ was calculated for a Gaussian model fitted to all data. 

      \begin{table}
    \caption[Geometric modelling results of FU\,Ori from all observations]{\label{table:GuassFit_FU} Best fit parameters for the Gaussian models to each of the three wavebands, independently. $\mathrm{F_{bg}}$ is the flux present in the background as a percentage of the total flux in the field of view. $\chi^2_{red}$ is the reduced chi-squared value of the best fit model for the visibilities \citep[see definition in][]{Kraus2009}. } 
    \centering
    \begin{tabular}{c c c c c c} 
        \hline
        \noalign{\smallskip}
        Band  &  FWHM & INC & PA & $\mathrm{F_{bg}}$ & $\chi^2_{red}$ \\ [0.5ex]
         & [mas] & [$^\circ$] & [$^\circ$] & [\%] & \\
        \hline
        \noalign{\smallskip}
        J & $0.38^{+0.03}_{-0.04}$ & $38.6^{+5.5}_{-9.0}$ & $37.6^{+4.0}_{-7.2}$ & $2.4^{+1.7}_{-1.2}$ & 0.22 \\
        \noalign{\smallskip}
        H & $0.41^{+0.02}_{-0.02}$ & $37.0^{+0.5}_{-0.5}$ & $41.4^{+0.7}_{-0.7}$ & $3.8^{+0.5}_{-0.5}$ & 1.68 \\
        \noalign{\smallskip}
        K & $0.60^{+0.02}_{-0.02}$ & $44.6^{+2.8}_{-3.3}$ & $32.0^{+4.2}_{-3.8}$ & $8.6^{+0.5}_{-0.4}$ & 0.76 \\
        \noalign{\smallskip}
        \hline
        \noalign{\smallskip}
    \end{tabular}
    \end{table}

     \subsection{Disk Temperature Structure and Geometry} \label{GeoMod}
     
     The primary limitation of the Gaussian models employed is the lack of spectral information within the intrinsically 'grey' model. As the interferometric data covers three wavebands it is vital to account for the spectral dependency, as each wavelength channel probes a different temperature regime and hence a different disk radius.
    
     A temperature gradient model (TGM) allows for the simultaneous fitting of interferometric and photometric observables. It is built up by several rings extending from an inner radius $R_{\mathrm{in}}$ to an outer radius $R_{\mathrm{out}}$. Each ring is associated with temperature and hence flux. Therefore, a model SED can be computed by integrating over the resulting blackbody distributions for each of the concentric rings. Such a model allows us to not only to build up a picture of the temperature profile, but also approximate the position of the inner radius. The TGM is based upon $T_R = T_0(R/R_0)^{-Q}$ where $T_0$ is the temperature at the inner radius of the disk $R_0$, and $Q$ is the exponent of the temperature gradient \citep{Kreplin20,Eisner11}. While a model lacks any flaring, in this case it traces the surface layers of the disk as observed in the infrared. A point source is used at the centre of each model to represent an unresolved star, which is a reasonable approximation given the expected diameter (from the mass estimate) of the star of $4.3\,R_\odot$ resulting in an angular diameter of $0.05\,\mathrm{mas}$ \citep{Perez20}. 
     
    \begin{table}
    \caption[Best fit parameters of the temperature gradient model to FU\,Ori observations]{\label{table:TGM_fit_FU} Best fit parameters of the temperature gradient model. Inner disk parameters are derived in this work. Outer disk parameters ($>3\,\mathrm{au}$) are taken from \citet{Quanz06}.} 
    \centering
    \begin{tabular}{c c} 
        \hline
        \noalign{\smallskip}
        Parameter  &  Best Fit Value \\ [0.5ex]
        \hline
        \noalign{\smallskip}
        Inner Disk & $ < 3\,\mathrm{au}$\\
        \hline
        \noalign{\smallskip}
        $R_{in}$ & $0.015\pm0.007\,\mathrm{au}$ \\
        $R_{out}$ & $0.76\pm0.35\,\mathrm{au}$ \\
        $T_{in}$ & $5800\pm700\,K$ \\
        $Q$ & $0.74\pm0.02$\\
        $PA$ & $34\pm11^\circ$\\
        $INC$ & $32\pm4^\circ$\\
        \hline
        \noalign{\smallskip}
        Outer Disk & $ > 3\,\mathrm{au}$\\
        \hline
        \noalign{\smallskip}
        $T_{3\,\mathrm{au}}$ & 340\,K \\
        $Q$ & 0.53 \\
        $R_{outer}$ & 7.7\,au\\
        \hline
        \noalign{\smallskip}
    \end{tabular}
    \end{table}
    
\begin{figure*}
        \centering
        \includegraphics[scale=0.24]{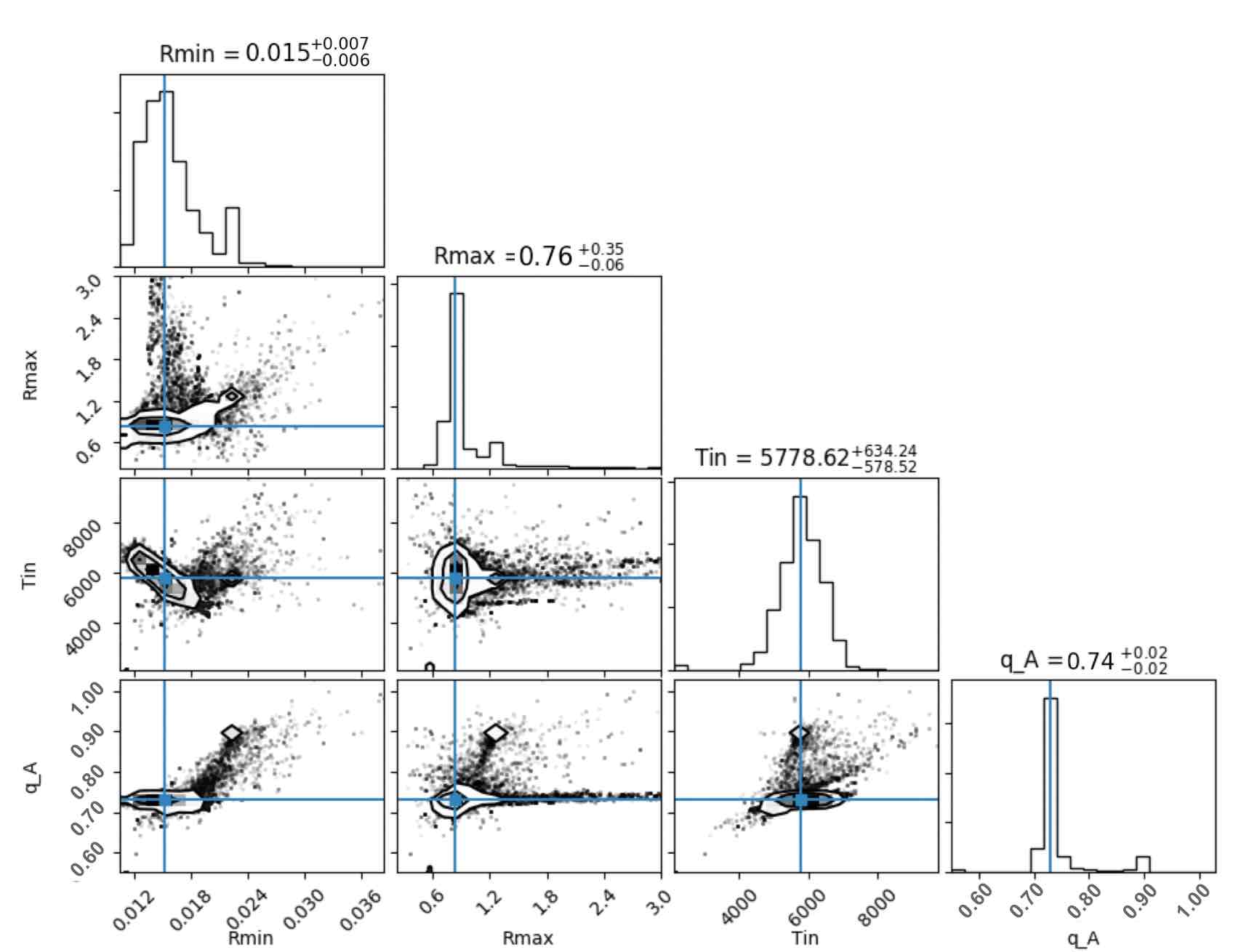}
        \caption[Posterior distributions of the parameters of the Temperature Gradient Model to FU\,Ori data]{Posterior distributions of the parameters of the Temperature Gradient Model, produced using the corner package \citep{ForemanMackey16}. The position angle and inclination of the disk are fixed by to those obtained through Gaussian modelling. }
        \label{fig:Corner_FU}
    \end{figure*}

    In the previous section I was able to reliably constrain the inclination and position angle of the disk for the first time. The geometric modelling finds an inclination of $32\pm4^{\circ}$ and a minor axis position angle of $34\pm11^{\circ}$ measured East from North. These values were fixed in the fitting of temperature gradient model in order to reduce the number of free parameters. The fitting was undertaken using all the visibility data, shown in Figure\,\ref{fig:PAmodel_FU} and all of the SED shown in Figure\,\ref{fig:SED_FU}, which is constructed from the photometry detailed in Appendix\,\ref{appendix3}. Interstellar extinction was taken as $E_{(B-V)} = 0.48\,(A_V = 1.4)$ as described by \citet{Pueyo12} and was used to redden the model results during fitting. The SED of FU\,Ori is unusual in the context of YSOs, in that it is almost completely dominated by disk flux even across the visible where the central star only contributes $1-2\%$. This is due to the lack of gap between the star and disk resulting in very hot material close to the star, which dominates the emission from the relatively small central star. The fitting and error computation is done using the MCMC sampler corner \citep{ForemanMackey16} to produce corner plots from which parameter degeneracies and errors can be analysed. This is ideal for fitting many parameters simultaneously in a consistent manner. Fig.~\ref{fig:Corner_FU} shows the posterior distributions, local minima can be seen in several parameters, the signifcance of these were explored by limiting the range of the search, however, the minima were ruled to be insigificant sompared to the global minima. The final results of fitting the temperature gradient are shown in Table\,\ref{table:TGM_fit_FU}.

\begin{landscape}
    \begin{figure*}
        \centering
        \includegraphics[scale=0.35]{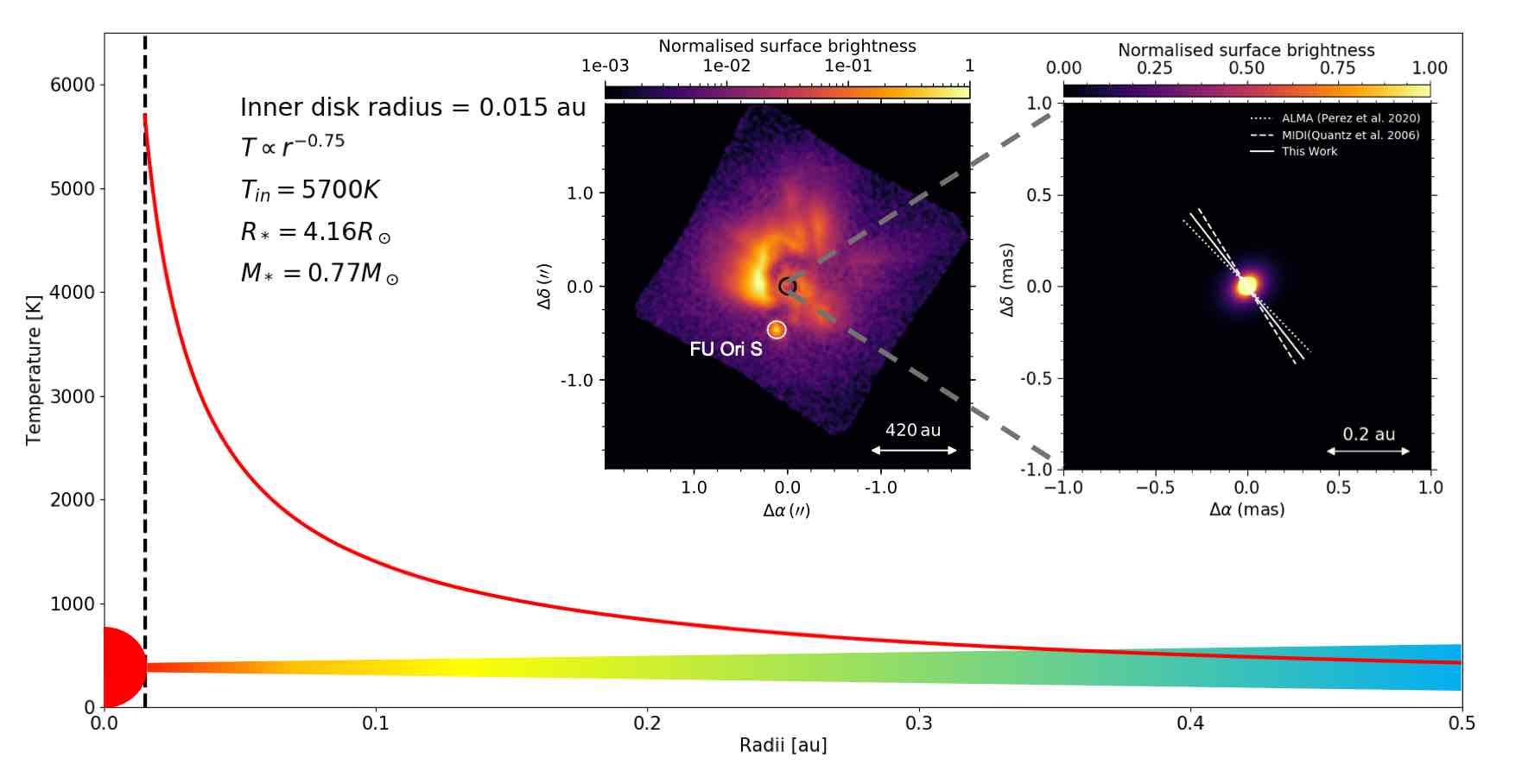}
        \caption[Temperature gradient of FU\,Ori over the inner disk and simulated brightness distribution]{Temperature gradient of FU\,Ori over the inner disk. The temperature and location of the inner radius of the disk are $5800\,\mathrm{K}$ and $0.015\,\mathrm{au}$ respectively. The dashed line represents the inner disk edge. The vertical profile of the disk schematic is not to scale, nor does it represent the vertical temperature structure of the disk. {\it INSET-left:} Linearly-polarized intensity, scattered light image obtained with the Gemini Planet Imager (GPI) in the J-band \citep{Laws20}. The image is centered on FU\,Ori\,N within the black circle. FU\,Ori\,S is also visible to the South-West. {\it INSET -right:} The brightness distribution of the TGM disk model in the J-band, as derived in this work. In both panels, I plot normalised surface brightness with a logarithmic colour scale. The white lines in the image indicate the position angle of the disk minor-axis derived from other observations in literature and in this work. The central star is included in the image as the disk extends down to the photosphere.}
        \label{fig:gradient_FU}
    \end{figure*}
    \end{landscape}

    Figure\,\ref{fig:gradient_FU} shows the best-fit temperature gradient model which corresponds to an inner disk radius of $0.015\pm0.007\,\mathrm{au}$ with a temperature of $5800\pm700\,\mathrm{K}$ and an exponent $Q = 0.74\pm\,0.02$. By design, the brightness distribution computed from the model of a geometrically thin disk with a radial temperature gradient is intrinsically centro-symmetric, meaning all closure phase measurements are equivalent to $0^\circ$. There is a slight discrepency between the temperature gradient of the inner disk derived here and the disk beyond $3\,\mathrm{au}$ taken from \citet{Quanz06}. At $3\,\mathrm{au}$ there is a temperature difference of $\sim20\,\mathrm{K}$ with the inner disk regieme producing a slightly cooler temperature, but this is within the error margins.
    
    \begin{figure}
        \centering
        \includegraphics[scale=0.55]{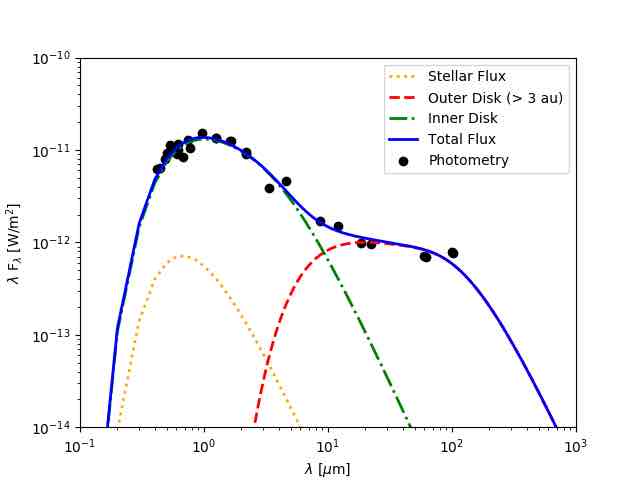}
        \caption[Spectral Energy Distribution of FU\,Ori and fitted model SED]{Spectral Energy Distribution of FU\,Ori. The black data represents the photometry points compiled in Apppendix\,\ref{appendix3}. The yellow dotted line indicates the stellar flux contribution, the red dashed line the outer ($>3\,\mathrm{au}$ disk as determined by \citet{Quanz06}. The green dash-dot line represents the contribution from the inner disk, described here as a temperature gradient model. The blue line is the total flux, a sum of all components.}
        \label{fig:SED_FU}
    \end{figure}

    \subsection{Companion Constraints} \label{Comp}
    
    In addition to disk modelling I undertook a companion search, in order to search for a putative, previously-undetected companion within the field of view of the observations. FU\,Ori is a known binary system, however the southern component, located at a separation of $0.5"$ is outside the field of view of all interferometric observations. The companion search was undertaken using the RAPIDO model-fitting code and a companion finder extension described in \citet{Davies18}. This module computes a grid of model, where a point source is added to the best-fit model described in section\,\ref{GeoMod}. The resulting $\chi^2_{red}$ detection significance map allows us to estimate the detection significance, or to derive an upper detection limit if no significant companion is detected.
    
    A companion search was undertaken separately on the two epochs of MIRC-X H-band data taken in November 2018 and November 2019. Only these dates were chosen as they offer good uv-coverage over the shortest possible time period (1 night), making it ideal to conduct companion searches upon. On the other hand, the supplementary data (MIRC-X J-band data, CLIMB K-band data, and PIONIER H- band data) is taken over many years and on fewer baselines, so individual epochs contain sparse uv-coverage.
    
    In order for a detection to be considered significant the p-value must be greater than $5\sigma$. This search finds that the non-zero flux solution for a companion is not significant with a p-value of $3.04$. Following the non-detection the upper limits to the flux of any companion can be calculated for each (x, y) positions in the field-of-view (grid search process described in \citet{Davies18}). The maximum flux contribution from any companion that could remain undetected by the observations is $1.3\%$ of FU Ori’s total flux in H-band within $0.5$ to $50$\,mas. Assuming a standard mass luminosity relation of $L\propto M^4$ for low mass stars, this flux contrast corresponds to a maximum companion mass of $0.12\,M_\odot$.

\section{Discussion} \label{discussion}

    I have presented the first J-band interferometric observations of a young stellar object, thus demonstrating the feasibility of such observations, particularly in the context of multi-waveband interferometry. The J-band has  been a relatively untapped resource in interferometry, and the feasibility of these observations is of great interest to the wider scientific community. The J-band has the potential to not only be used in YSO studies to examine the sublimation rims and the potential for optically-thick gas inside the sublimation radius, but also in stellar photosphere studies, as a waveband which is relatively free from molecular opacities.
    
    By investigating the circumstellar environment of FU\,Ori I have explored the morphology and temperature gradient of the inner disk and greatly improved the constraints on the parameters of both the star and the disk. Geometric modelling finds a disk inclined at $32\pm4^\circ$ with a minor axis position angle of $34\pm11^\circ$. This inclination estimate is significantly more face-on than earlier estimates of $50^\circ$ and $60^\circ$ that were based on a variety of techniques, including SED analysis and near/mid-IR interferometry  \citet{Malbet05,Zhu08,Quanz06}. \citet{Calvet91} derived an inclination of $20-60^\circ$ based on the CO line-width  and \citet{Liu19} found FU\,Ori to be face-on based on NIR closure phases with GRAVITY. However, previous interferometric studies were based on very limited uv-coverage consisting solely of baselines below 100\,m and at a single wavelength, making these estimates less accurate compared to the comprehensive uv-coverage. The limitations in uv-coverage of earlier interferometric studies resulted likely also in the wide spread of minor axis position angle estimates that range from $19^\circ\pm12$ \citep{Quanz06}, $47^{\circ+7}_{-11}$ \citep{Malbet05}. In addition, many literature values are at odds with the tentative detection of a jet/outflow detected on larger scales by \citet{Takami18} and \citet{Laws20} (features C and D respectively). The putative jet/outflow feature was detected in scattered light imagery obtained with Subaru and GPI, respectively. Assuming the minor disk axis is aligned with the stellar polar axis and hence the jet position angle, a position angle of $\sim-25^\circ$ would be expected. Compared to $34\pm11^\circ$ measured here, I find a discrepancy of almost $60^\circ$. I see no evidence of such a jet in the continuum imaging on smaller scales.
    
    
    Temperature gradient models were used in order to fit the spectral dependence of the data. These models allow for the simultaneous fitting of the interferometric data and the SED. Application of these models finds a disk that extends down to the stellar surface at $0.015\pm0.007\,\mathrm{au}$ (stellar radius of $4.3\,R_\odot$ corresponds to $0.0019\,\mathrm{au}$), where the temperature of disk is $5800\pm700\,\mathrm{K}$. This is expected of an object that is actively accreting with such a high rate of $10^{-4}\,\mathrm{M_\odot yr^{-1}}$, and is consistent with estimates by \citet{Zhu07,Zhu08} where the disk temperature peaks at around $6000\,\mathrm{K}$. Such a high temperature would be unexpected and perhaps unfeasible in a standard protoplanetary disk, as standard silicate or graphite based dust grain species will sublimate at such high temperatures. However, this is not unexpected in FU\,Ori objects where the disk extends down to the stellar photosphere, indeed one of the first works deriving temperatures of FU\,Ori by \citet{Kenyon88} finds a maximum disk temperature of $7200\,\mathrm{K}$ based on SED and spectral line fitting, significantly hotter than this work. The existence of dust at such temperature is possible through highly refractive grain species such as Mg and Ti which are known to be present \citet{Kenyon88}. In addition, strong gas emission could appear as continuum emission for such low resolution observations. This highlights how these objects cannot be thought of as simple star-disk systems, but as one continuous object. The derived maximum temperature of the disk is comparable to estimates of the efefctive temperature of FU\,Ori\,N of $\sim5000-6500\,\mathrm{K}$ \citep{Kenyon87,Zhu07,Zhu08}.

    The inner disk radius is equivalent to that of the star indicating boundary layer accretion directly from the disk onto the central star. Beyond the inner radius, the temperature falls off with the power-law $T\propto r^{-0.74\pm0.02}$ to an outer radius cut-off at $0.76\pm0.35\,\mathrm{au}$. The determined power-law index is consistent with the predicted temperature profile for a steady state, optically-thick accretion disk \citep{Pringle81}. A temperature gradient of this profile is only possible if viscous heating processes are present in the inner disk. Heating of flared disks by reprocessed stellar radiation alone is shown to produce temperature exponents of $q <= 0.5$ \citep{Kenyon87,Dullemond04}. Only through viscous heating can the observed temperature profile be replicated.
    
    The derived inner disk temperature gradient is in agreement with MIR work conducted with the MIDI instrument by \citet{Quanz06}, who also found a value of $Q = 0.75$ for the inner disk, although it is unclear whether their estimate was constrained mainly by interferometry or SED data. This contrasts with the outer ($>3$\,au) disk model they derive, which is also adopted here. In order to successfully fit the long wavelength SED and N-band MIDI interferometry, a temperature gradient of $T\propto r^{-0.53}$ is adopted, in good agreement to what can be found for vertically isothermal flared disks \citep{Kenyon87}. In order to test this result, more comprehensive MIR N-band observations with the MATISSE instrument \citet{MATISSE14} at VLTI are required. 
    
    It has been proposed that FUor stars may be newborn binaries that have become bound when a small non-hierarchical multiple system breaks up \citep{Bonnell92}. In such a scenario \citet{Reipurth04} predict a close companion ($<10$\,au). Accordingly, for the FU\,Ori systems, such models would predict a third component orbiting FU\,Ori\,N in the inner 20\,mas. \citet{Malbet98} found tentative evidence for a companion located at $\sim1$\,au ($2.4$\,mas), however their data can be equally well interpreted as a circumstellar disk, which was confirmed in later work. No other studies have previously detected a companion in the inner few astronomical units around FU\,Ori\,N. I conducted a companion search based on the MIRC-X H-band visibilities and closure phases, I derive an upper limit to the flux contrast of 1.3\%, which corresponds to a maximum companion mass of 0.12\,$M_{\sun}$ in the separation range between 0.5 and 50.0 mas. Therefore, the observations do not support this scenario.

\section{Conclusions} \label{Conclusion}
    This multi-wavelength study has probed the inner disk geometry of FU\,Ori at the highest angular resolution yet. Furthermore, I put for the first time tight constraints on the disk temperature structure in the inner astronomical unit.
    
    I summarise my conclusions as follows:
    \begin{itemize}
    
    
    \item I believe this first-of-its-kind study demonstrates a powerful and exciting new technique in the study of accretion and circumstellar disks. This allows us to directly test long-posited theoretical work with observational evidence and provide clues to the true nature of accretion and viscosity processes. 
    
    \item Temperature gradient models find an inner disk that extends down to the stellar photosphere at $0.015\pm0.007\,\mathrm{au}$ where the temperature reaches $5800\pm700\,\mathrm{K}$. This is in agreement with a heavily accreting star such as outbursting FUors and indicates boundary layer accretion processes. 
    
    \item The temperature of the inner disk falls off with a power-law $T\propto r^{-0.74\pm0.02}$. This is consistent with theoretical work for steady state, optically-thick accretion disks. Such a temperature profile is only possible if viscous heating processes are present in the inner disk.
    
    \item The inclination and position angle of the disk are tightly constrained, providing a significant improvement over literature values. An inclination of $32\pm4^\circ$ and a minor axis position angle of $34\pm11^\circ$ are found from geometric modelling of the on sky brightness distrubution.
    
    \item The minor axis position angle is around $60^\circ$ mis-aligned with the detection of a jet/outflow detected in scattered light images, assuming that the jet is perpendicular to the disk.
    
    \item No significant companion is detected within the field of view of $0.5$ to $50$\,mas. I place an upper limit of $1.3\%$ of the total $H$-band flux on an potential companion in this separation range.

    \end{itemize}
    
    This study demonstrates the potential of combined J and H-band interferometry to constrain the temperature structure on milliarcsecond scales. This technique enables exciting new studies on a broad range of science applications, from characterising the disks around young stellar objects to stellar surface imaging.

\chapter{Conclusions and Future Work} \label{ch:conclusions}

\epigraph{``The mystery of life isn't a problem to solve, but a reality to experience.''}{--- Frank Herbert, Dune \hphantom{spacing}}

The aim of my PhD was to explore the innermost astronomical unit of protoplanetary disks with optical interferometry in order to search for evidence of planet formation processes. Although no direct evidence of ongoing planet formation was discovered, each of the studies presented in this thesis unveil new and exciting characteristics of protoplanetary disks which have potentially significant impacts for planet formation theory. In addition, the technical work undertaken opens up a new frontier in optical interferometry, which will significantly impact our understanding. In this concluding chapter, I shall summarise the key results from this thesis and their impacts in addition to discussing the principal skills and experience gained along the way. Finally, I shall touch on the future work planned to build upon the techniques and results of this work.

\section{Conclusions}

	A key aspect of my PhD has been my involvement in the engineering and instrumentation work with the CHARA array and MIRC-X instrument. Firstly, my development of the CHARA baseline solution (Section\,5.2), has led to increasingly productive and efficient routine observing for users of all beam combiners. Searching with delay lines for fringes previously required scanning over $\pm15\,\mathrm{mm}$ taking up precious time. Since this development and subsequent updates the required scan is reduced to $\pm0.5\,\mathrm{mm}$ saving time and making observation of faint/low contrast objects significantly easier. The other aspect of instrumentation work concerns the expansion of the operating wavelength of MIRC-X to include the J band ($\sim1.2\,\mathrm{\mu m}$) (Section\,5.1). This involved the modelling and correcting for both atmospheric dispersion and internal instrument birefringence using wedges of glass. I successfully obtained the first science observations in the new dual JH-band observing mode in November 2019, highlighting the immense possibilities of the J band. It will allow access to the photosphere in giant and super-giant stars relatively free from opacities of molecular bands which are not associated. In addition, the J band traces the warmest and smallest scales of protoplanetary disks, where accretion and viscosity processes are strongest and hence detectable. Finally, the J band allows for higher resolution observations than near and mid-IR allowing us to probe the smallest scales of astrophysical objects.  

	The first science conducted during my PhD was in regard to the young star SU\,Aurigae, these are described in detail in chapters 6 \& 7. In the first of these studies, I interpreted interferometric data from a variety of different instruments. Using this I was able to derive significantly better constraints on the Gaussian-like geometry of the system than previous studies. By investigating radiative transfer models in the context of both visibilities and photometry I was able to explore the shape of the dust sublimation rim. It was found that a classic gas density dependent sublimation temperature which creates a curved rim was insufficient to recreate the observed stellar-to-total flux ratio in the interferometry or SED. In order to explain the dearth of flux in the NIR, I invoked a dusty disk wind scenario where material is lifted out of the disk along inclined magnetic field lines allowing for an increased reprocessing of stellar radiation. Although the into-wind transfer rate is potentially unphysically high, this scenario can recreate the observables of SU\,Aur very well.

	Follow-up observations of SU\,Aur were conducted with the MIRC-X instrument, many of which were conducted by myself at the CHARA array during several observing runs over several years. Such observing has given great insight into the interferometric process and a deeper understanding of the physics behind the observables. These new observations provided increased resolution and more comprehensive baseline coverage than my previous study. From this data I was able to reconstruct a detailed image of the innermost region to reveal strong, inclination induced, asymmetries where the far-side of the disk rim is illuminated, while the near-side of the rim is obscured by its own flaring. Interpretation of the geometry of SU\,Aurigae in the context of SPHERE observations emphasises the discovery of a large warp between the inner and outer disk of $\sim70^\circ$, creating large shadows on the outer disk. 

	The first science observations in the new dual JH band mode were conducted on the outbursting young star FU\,Orionis. FU\,Ori underwent a massive, accretion fueled, outburst over 80 years ago and has remained at a heightened luminosity ever since. This object, with a very hot disk extending down to the stellar photosphere, was the ideal testbed for a multi-wavelength observing campaign to explore the temperature gradient across the inner disk. I discovered that the temperature gradient exponent of $0.74\pm0.02$ was consistent with theoretical models of a viscously heated disk in contrast to a disk heated by reprocessed stellar radiation alone. The presence of viscous heating within the disk is a major milestone as the presence and extent of such a process has been much debated and relatively unconstrained by observations. It remains to be seen is viscous heating is ubiquitous across FUors and the wider disk population.

\section{Future Work}

	In the author's opinion, the most exciting work, with the greatest potential in this thesis is the expansion of the MIRC-X wavelength range and the subsequent multi-wavelength study of FU\,Orionis. It is for this reason, I have chosen to continue this work in the form of a fellowship at the European Southern Observatory (ESO). The temperature gradient of FUors and the implications for viscous heating and accretion will be the focus of the study. Over four years, I aim to conduct multi-wavelength observations using both CHARA and VLTI on both FUor and non-FUor YSOs. This will allow for studies of the temperature gradients across the population of FUors and comparisons with non-outbursting stars. The key questions will be approached:

	\begin{itemize}
	\item How widespread are viscous heating processes in circumstellar environments? I will determine if viscous heating is common across the FUor object population and in Herbig Be disks.

	\item How far through the disk does the actively accreting/viscously heated region extend? Is viscous heating only prevalent in the innermost regions, or does it extend further through the disk?

	\item What would be the associated disk depletion times for steady-state and variable accretion scenarios? Conclusions can be drawn about the disk depletion and mass infall rate from the outer disk reservoir. A range of near and mid-IR wavelength observations will be crutial to answer such questions.

	\item How do these timescales compare to the age and expected lifetime of the disk? Does the disk depletion timescale exceed or under-predict the expected disk lifetime? 

	\item What are the driving forces behind FUor accretion outbursts? Search for evidence of undetected companions and large asymmetric disk clumps. 
	\end{itemize} 

	Furthermore, these multi-wavelength interferometric observations will allow me to investigate possible triggering mechanisms behind outbursting events. Various competing models have been proposed including gravitational instabilities and disk-companion interactions. Both of these can be directly explored by searching for point source companions or larger disk asymmetric clumps. In order build a complete picture of the disks, I have a range of techniques in my arsenal of expertise. Image reconstruction algorithms allows for model independent analysis of disk structures. More complex disk models can be explored using radiative transfer techniques, specifically using the \emph{TORUS} code with which I have much experience. This code also allows for coupling with hydrodynamical disk simulations, which can explore the impact of reservoirs or companions a disk over time and their potential for causing accretion outbursts.

\chapter{Publications}
\label{ch:Publications}

\textbf{The orbit and stellar masses of the archetype colliding-wind binary WR 140}; Thomas, J. D., Richardson, N. D., Eldridge, J. J., Schaefer, G. H., Monnier, J. D., Sana, H., Moffat, A. F. J., Williams, P., Corcoran, M. F., Stevens, I. R., Weigelt, G., Zainol, F. D., Anugu, N., Le Bouquin, J.-B., ten Brummelaar, T., Campos, F., Couperus, A., Davies, C. L., Ennis, J., Eversberg, T., Garde, O., Gardner, T., Fló, J. G., Kraus, S., \textbf{Labdon, A.}, Lanthermann, C., Leadbeater, R., Lester, T., Maki, C., McBride, B., Ozuyar, D., Ribeiro, J., Setterholm, B., Stober, B., Wood, M., Zurmühl, U. (2021), MNRAS, 504, 5221

\textbf{$\nu$ Gem: a hierarchical triple system with an outer Be star}; Klement, R., Hadrava, P., Rivinius, T., Baade, D., Cabezas, M., Heida, M., Schaefer, G. H., Gardner, T., Gies, D. R., Anugu, N., Lanthermann, C., Davies, C. L., Anderson, M. D., Monnier, J. D., Ennis, J., \textbf{Labdon, A.}, Setterholm, B. R., Kraus, S., ten Brummelaar, T. A., le Bouquin, J.-B. (2021), arXiv, arXiv:2105.13437

\textbf{Viscous heating in the disk of the outbursting star FU Orionis}; \textbf{Labdon, A.}, Kraus, S., Davies, C. L., Kreplin, A., Monnier, J. D., Le Bouquin, J.-B., Anugu, N., ten Brummelaar, T., Setterholm, B., Gardner, T., Ennis, J., Lanthermann, C., Schaefer, G., Laws, A. (2021), A\&A, 646, A102

\textbf{ARMADA. I. Triple Companions Detected in B-type Binaries {\ensuremath{\alpha}} Del and {\ensuremath{\nu}} Gem}; Gardner, T., Monnier, J. D., Fekel, F. C., Schaefer, G., Johnson, K. J. C., Le Bouquin, J.-B., Kraus, S., Anugu, N., Setterholm, B. R., \textbf{Labdon, A.}, Davies, C. L., Lanthermann, C., Ennis, J., Ireland, M., Kratter, K. M., Ten Brummelaar, T., Sturmann, J., Sturmann, L., Farrington, C., Gies, D. R., Klement, R., Adams, F. C. (2021), AJ, 161, 40

\textbf{CHARA/MIRC-X: a high-sensitive six telescope interferometric imager concept, commissioning and early science}; Anugu, N., Le Bouquin, J.-B., Monnier, J. D., Kraus, S., Schaefer, G., Setterholm, B. R., Davies, C. L., Gardner, T., \textbf{Labdon, A.}, Lanthermann, C., Ennis, J., ten Brummelaar, T., Sturmann, J., Anderson, M., Farrington, C., Vargas, N., Majoinen, O. (2020), SPIE, 11446, 114460N

\textbf{A new frontier for J-band interferometry: dual-band NIR interferometry with MIRC-X}; \textbf{Labdon, A.}, Monnier, J. D., Kraus, S., Le Bouquin, J.-B., Setterholm, B. R., Anugu, N., ten Brummelaar, T., Lanthermann, C., Davies, C. L., Ennis, J., Gardner, T., Schaefer, G. H., Sturmann, L., Sturmann, J. (2020), SPIE, 11446, 114460H

\textbf{MIRC-X polarinterferometry at CHARA}; Setterholm, B. R., Monnier, J. D., Le Bouquin, J.-B., Anugu, N., \textbf{Labdon, A.}, Ennis, J., Johnson, K. J. C., Kraus, S., ten Brummelaar, T. (2020), SPIE, 11446, 114460R

\textbf{MIRC-X: A Highly Sensitive Six-telescope Interferometric Imager at the CHARA Array}; Anugu, N., Le Bouquin, J.-B., Monnier, J. D., Kraus, S., Setterholm, B. R., \textbf{Labdon, A.}, Davies, C. L., Lanthermann, C., Gardner, T., Ennis, J., Johnson, K. J. C., Ten Brummelaar, T., Schaefer, G., Sturmann, J. (2020), AJ, 160, 158

\textbf{A triple-star system with a misaligned and warped circumstellar disk shaped by disk tearing}; Kraus, S., Kreplin, A., Young, A. K., Bate, M. R., Monnier, J. D., Harries, T. J., Avenhaus, H., Kluska, J., Laws, A. S. E., Rich, E. A., Willson, M., Aarnio, A. N., Adams, F. C., Andrews, S. M., Anugu, N., Bae, J., ten Brummelaar, T., Calvet, N., Curé, M., Davies, C. L., Ennis, J., Espaillat, C., Gardner, T., Hartmann, L., Hinkley, S., \textbf{Labdon, A.}, Lanthermann, C., LeBouquin, J.-B., Schaefer, G. H., Setterholm, B. R., Wilner, D., Zhu, Z. (2020), Sci, 369, 1233

\textbf{Optical interferometry and Gaia measurement uncertainties reveal the physics of asymptotic giant branch stars}; Chiavassa, A., Kravchenko, K., Millour, F., Schaefer, G., Schultheis, M., Freytag, B., Creevey, O., Hocdé, V., Morand, F., Ligi, R., Kraus, S., Monnier, J. D., Mourard, D., Nardetto, N., Anugu, N., Le Bouquin, J.-B., Davies, C. L., Ennis, J., Gardner, T., \textbf{Labdon, A.}, Lanthermann, C., Setterholm, B. R., ten Brummelaar, T. (2020), A\&A, 640, A23

\textbf{Dusty disk winds at the sublimation rim of the highly inclined, low mass young stellar object SU Aurigae}; \textbf{Labdon, A.}, Kraus, S., Davies, C. L., Kreplin, A., Kluska, J., Harries, T. J., Monnier, J. D., ten Brummelaar, T., Baron, F., Millan-Gabet, R., Kloppenborg, B., Eisner, J., Sturmann, J., Sturmann, L. (2019), A\&A, 627, A36

\textbf{Simultaneous Spectral Energy Distribution and Near-infrared Interferometry Modeling of HD 142666}; Davies, C. L., Kraus, S., Harries, T. J., Kreplin, A., Monnier, J. D., \textbf{Labdon, A.}, Kloppenborg, B., Acreman, D. M., Baron, F., Millan-Gabet, R., Sturmann, J., Sturmann, L., Ten Brummelaar, T. A. (2018), ApJ, 866, 23

\textbf{The MIRC-X 6-telescope imager: key science drivers, instrument design and operation}; Kraus, S., Monnier, J. D., Anugu, N., Le Bouquin, J.-B., Davies, C. L., Ennis, J., \textbf{Labdon, A.}, Lanthermann, C., Setterholm, B., ten Brummelaar, T. (2018), SPIE, 10701, 1070123

   \begin{appendices}
\chapter{Binary fit for CLIMB calibrator stars} \label{appendix1}

The two calibrators HD\,31952 and HD\,34053 were found to be binary systems, based on strong non-zero closure phase signals. As both of these stars were needed for the calibration for 3 nights of our data, a binary fit was undertaken in order to recalculate the transfer function with which our data was calibrated. The software package \textit{LITpro} was used to construct the fits within a search radius of $10$\,mas.  The parameters fitted are: angular separation, position angle and the uniform disk diameters (UDD) of the primary and secondary components. Two uniform disks were used to represent the stars and $\chi^2$ maps were constructed to find the best-fit location of the secondary star. The parameters are displayed in Table~\ref{table:binaries}.

\begin{landscape}

\begin{table*}
    \centering
    \caption[Binary fit parameters for the two calibrator stars HD\,31952 and HD\,34053]{\label{table:binaries} Binary fit parameters for the two calibrator stars HD\,31952 and HD\,34053. Flux ratio is given as primary/secondary, the position angle is taken from north to east and UDD is the uniform disk diameter of the individual stars. }
    \begin{tabular}{c c c c c c c}
        \hline
        \noalign{\smallskip}
        Star & Obs. Date & Flux Ratio & Separation [mas] & Position Angle [$^\circ$] & $\mathrm{UDD_{pri}}$ [mas] & $\mathrm{UDD_{sec}}$ [mas]  \\
        \hline
        \noalign{\smallskip}
        HD\,31952 & 2012-10-(19,20) & $3.76\pm0.07$ & $7.03\pm0.12$ & $95.55\pm2.35$ & $0.15\pm0.02$ & $0.15\pm0.03$\\
        HD\,34053 & 2012-10-(18,19,20) & $1.17\pm0.05$ & $1.66\pm0.23$ & $85.82\pm3.27$ & $0.14\pm0.02$ & $0.16\pm0.04$\\ [1ex]
        \hline
    \end{tabular}

\end{table*}

\end{landscape}

\chapter{Photometric catalogue of SU\,Aurigae} \label{appendix2}

Table of photometry used in the SED fitting procedure as described in Chapter\,5.
    \begin{table*}
        \centering
        \caption[Photometric values used to construct the SED of SU\,Aur]{\label{table:Photometry} Photometric values used to construct the SED of SU\,Aur}
        \begin{tabular}{c c c} 
            \hline
            Wavelength [$\mathrm{\mu m}$] & Flux [Jy] & Reference \\
            \hline
            0.15&1.31E-04&\citet{Bianchi11}\\
            0.23&0.00219&\citet{Bianchi11}\\
            0.42&0.233&\citet{Ammons06}\\
            0.44&0.402&\citet{Anderson12}\\
            0.53&0.6&\citet{Ammons06}\\\
            0.69&1.32&\citet{Morel78}\\
            0.79&1.47&\citet{Davies14}\\
            0.88&1.75&\citet{Morel78}\\
            1.24&2.08&\citet{Roser08}\\
            1.25&2.12&\citet{Ofek08}\\
            1.63&2.47&\citet{Ofek08}\\
            2.17&2.71&\citet{Roser08}\\
            2.19&2.62&\citet{Ofek08}\\
            3.35&2.6&\citet{Cutri14}\\
            3.40&2.44&\citet{Bourges14}\\
            4.50&1.75&\citet{Esplin14}\\
            4.60&2.78&\citet{Cutri14}\\
            5.03&2.58&\citet{Bourges14}\\
            7.88&1.99&\citet{Esplin14}\\
            8.62&2.36&\citet{Abrahamyan15}\\
            11.57&2.83&\citet{Cutri14}\\
            11.60&3.52&\citet{Abrahamyan15}\\
            18.40&6.47&\citet{Abrahamyan15}\\
            22.11&9.24&\citet{Cutri14}\\
            23.90&12.8&\citet{Abrahamyan15}\\
            61.89&12.2&\citet{Abrahamyan15}\\
            65.04&9.89&\citet{Toth14}\\
            90.06&8.8&\citet{Toth14}\\
            140.10&10.2&\citet{Toth14}\\
            160.11&8.88&\citet{Toth14}\\
            849.86&0.074&\citet{Mohanty13}\\
            887.57&0.071&\citet{Andrews13}\\
            1300.90&0.03&\citet{Mohanty13}\\
            1333.33&0.0274&\citet{Andrews13}\\ [1ex]
            \hline
        \end{tabular}
    \end{table*}
\chapter{Photometric catalogue of FU\,Orionis} \label{appendix3}

\begin{table}[h]
    \caption[Photometric data used in the SED of FU\,Ori]{\label{table:Phot} Table of photometric data points used in the construction of the spectral energy distribution of FU\,Ori. Data is plotted, along with the best fitting model in Figure\,\ref{fig:SED}. Care was taken to use synchronous data wherever possible to reduce the effect of photospheric variability on the SED, the method and fitting procedure are described in Section\,\ref{GeoMod}.}
    
    \centering
    \begin{tabular}{c c c} 
        \hline
        \noalign{\smallskip}
        Wavelength [$\mathrm{\mu m}]$ & Flux [Jy] & Reference \\ [0.5ex]
        \hline
        \noalign{\smallskip}
        0.42    &   0.13    &   \citet{HIP97}\\
        0.44    &   0.02    &   \citet{Saunders00}  \\
        0.48    &   0.25    &   \citet{Henden15}    \\
        0.50    &   0.35    &   \citet{GAIA2phot}   \\
        0.53    &   0.57    &   \citet{HIP97}   \\
        0.54    &   0.00    &   \citet{Page12}  \\
        0.55    &   0.64    &   \citet{HIP97}   \\
        0.60    &   0.55    &   \citet{McDonald17}  \\
        0.61    &   0.75    &   \citet{Chambers16}  \\
        0.62    &   0.68    &   \citet{GAIA2phot}   \\
        0.67    &   0.70    &   \citet{GAIA2phot}   \\
        0.74    &   1.36    &   \citet{Chambers16}  \\
        0.77    &   1.20    &   \citet{GAIA2phot}   \\
        0.96    &   2.78    &   \citet{Chambers16}  \\
        1.24    &   3.89    &   \citet{Cutri14} \\
        1.25    &   3.97    &   \citet{Cutri03} \\
        1.63    &   5.46    &   \citet{Cutri03} \\
        1.65    &   5.52    &   \citet{Cutri14} \\
        2.16    &   5.83    &   \citet{Cutri14} \\
        2.19    &   5.64    &   \citet{Cutri03} \\
        3.35    &   3.91    &   \citet{Cutri14} \\
        4.60    &   5.82    &   \citet{Cutri14} \\
        8.61    &   4.77    &   \citet{Abrahamyan15}    \\
        11.65   &   4.16    &   \citet{Cutri14} \\
        11.59   &   5.95    &   \citet{Gezari93}    \\
        12.00   &   5.95    &   \citet{Saunders00}  \\
        18.39   &   6.02    &   \citet{Abrahamyan15}    \\
        22.09   &   6.95    &   \citet{Cutri14} \\
        23.88   &   14.09   &   \citet{Gezari93}    \\
        25.00   &   14.09   &   \citet{Saunders00}  \\
        60.00   &   14.29   &   \citet{Saunders00}  \\
        61.85   &   14.29   &   \citet{Gezari93}    \\
        100.00  &   26.18   &   \citet{Saunders00}  \\
        101.95  &   26.18   &   \citet{Gezari93}    \\
        \hline
    \end{tabular}
    \end{table}
   \end{appendices}

   \cleardoublepage

   \phantomsection

   \addcontentsline{toc}{chapter}{Bibliography}
   \bibliographystyle{apalike}
   \bibliography{thesis} 
   \cleardoublepage

   \phantomsection

   \addcontentsline{toc}{chapter}{Index}
   \printindex

\end{document}